\documentclass[12pt]{article}
\usepackage{savesym}
\usepackage{cite}
\usepackage{wasysym}
\usepackage{amscd}
\usepackage{graphicx}
\usepackage{pifont}
\usepackage{float}
\usepackage{subfig}
\usepackage[all]{xy}
\usepackage{multicol}
\usepackage{amsfonts}
\usepackage{picture}
\usepackage{amssymb}
\savesymbol{iint}
\savesymbol{iiint}
\usepackage{amsmath}
\usepackage{amsthm}
\restoresymbol{TXF}{iint}
\restoresymbol{TXF}{iiint}
\usepackage{color}
\usepackage{clay}
\usepackage{hyperref}
\hypersetup{colorlinks=true}
\hypersetup{linkcolor=black}
\hypersetup{citecolor=black}
\hypersetup{urlcolor=black}
\usepackage{setspace}
\usepackage{multirow}
\usepackage{wrapfig}
\usepackage{verbatim}
\usepackage{dsfont}
\usepackage[vcentermath]{youngtab}

\numberwithin{equation}{section}
\usepackage{float}
\restylefloat{table}
\allowdisplaybreaks
\usepackage{accents}

\def\tilde{\widetilde}

\def\hat{\widehat}

\def\1{{\mathds 1}}

\usepackage{datetime}

\begin{document}

\date{November, 2017}

\institution{IAS}{\centerline{${}^{1}$School of Natural Sciences, Institute for Advanced Study, Princeton, NJ, USA}}
\institution{Princeton}{\centerline{${}^{2}$Physics Department, Princeton University, Princeton, NJ, USA}}

\title{ Global Symmetries, Counterterms, and Duality in Chern-Simons Matter Theories with Orthogonal Gauge Groups }

\authors{Clay C\'{o}rdova\worksat{\IAS}\footnote{e-mail: {\tt claycordova@ias.edu}},   Po-Shen Hsin\worksat{\Princeton}\footnote{e-mail: {\tt phsin@princeton.edu}}, and  Nathan Seiberg\worksat{\IAS}\footnote{e-mail: {\tt seiberg@ias.edu}}}

\abstract{  

We study three-dimensional gauge theories based on orthogonal groups.  Depending on the global form of the group these theories admit discrete $\theta$-parameters, which control the weights in the sum over topologically distinct gauge bundles.  We derive level-rank duality for these topological field theories.  Our results may also be viewed as level-rank duality for $SO(N)_{K}$ Chern-Simons theory in the presence of background fields for discrete global symmetries.  In particular, we include the required counterterms and analysis of the anomalies.  We couple our theories to charged matter and determine how these counterterms are shifted by integrating out massive fermions.  By gauging discrete global symmetries we derive new boson-fermion dualities for vector matter, and present the phase diagram of theories with two-index tensor fermions,  thus extending previous results for $SO(N)$ to other global forms of the gauge group.

 }

\maketitle

\setcounter{tocdepth}{2}
\tableofcontents

\section{Introduction}

In this paper we study various gauge theories with a gauge group based on the Lie algebra $\frak{so}(N)$.  These include $SO(N)$, $Spin(N)$, $O(N)$, and $Pin^\pm (N)$ gauge theories.  (There are also other such groups that we will not study here; e.g.\ $SO(N)/\mathbb{Z}_2$.)  Depending on the gauge group the Lagrangians of these theories can include various Chern-Simons couplings and discrete $\theta$-parameters, which can also be viewed as more subtle Chern-Simons terms.  We will discuss them and show how they affect the behavior of the theory.  We will also couple these gauge theories to bosonic or fermionic matter fields in various representations.  

The novelty in our analysis will be the careful attention to the global aspects of the gauge group and the dependence on these discrete $\theta$-parameters.  This understanding will lead to a rich spectrum of dualities including level-rank dualities between topological quantum field theories (TQFTs), and various dualities between Chern-Simons-Matter (CSM) theories.

Beyond the intrinsic interest in these theories, there are several motivations for our investigations.  

Consider for example the $Spin(N)$ gauge theory with matter fields in the vector representation.  This theory has a $\mathbb{Z}_2$ one-form global symmetry \cite{Kapustin:2014gua,Gaiotto:2014kfa} associated with the center of the gauge group.  This means that we can explore the behavior of Wilson loops in a spinor representation of $Spin(N)$ and learn about confinement.  This has been done in many papers and our discussion here will follow the line of investigation of \cite{Aharony:2013hda,Kapustin:2014gua,Gaiotto:2014kfa} in $4d$ and of \cite{Aharony:2013kma,Komargodski:2017keh,Gomis:2017ixy} in $3d$.  Specifically, we will focus on the one-form global symmetry as a clear diagnostic of confinement. 

Our second motivation is associated with the role of global symmetries.  Consider for example the $SO(N)$ theories.  They have a $\mathbb{Z}_2$ charge conjugation symmetry $\mathcal{C}$ and a $\mathbb{Z}_2$ magnetic symmetry $\mathcal{M}$, which we will discuss in detail below.  In addition, depending on the matter fields, the Chern-Simons couplings and the value of $N$ they can have time-reversal symmetry and one-form global symmetry. (More subtle phenomena associated with time-reversal symmetry will be discussed in \cite{Cordova:2017kue}.)  We will couple these global symmetries to background gauge fields and analyze their t' Hooft anomalies, and Chern-Simons counterterms \cite{Closset:2012vg,Closset:2012vp}.    Once these counterterms are understood, it is straightforward to gauge these symmetries, i.e.\ to promote the background fields to dynamical fields.

Another motivation for studying these theories is the presence of discrete $\theta$-parameters. It is often the case that the configuration space of a gauge theory breaks into distinct sectors labelled by $\nu$ with a well defined partition function in each sector $Z_\nu$.  The total partition function
\begin{equation}
\label{Znu}
Z=\sum_\nu a_\nu Z_\nu
\end{equation}
depends on the coefficients $a_\nu$.  These can be viewed as partition functions of some other theory that couples to our gauge theory.  In the special case where all the coefficients $a_\nu$ are phases, $a_\nu$ are the partition functions of invertible topological quantum field theories \cite{Freed:2004yc}.  But we can also have situations where some of the coefficients $a_\nu$ vanish; i.e.\ some sectors are absent in the sum.  Examples of that were presented in \cite{Seiberg:2010qd,Freed:2017rlk}.  Here we will see more examples of this phenomenon.

As is clear from these points, these systems are concrete examples of the interplay between topology and symmetries.  The recent interest in this interplay was formalized in \cite{EtingofNOM2009,Barkeshli:2014cna}.  Our systems are an explicit laboratory allowing exploration of these phenomena.

We will start our discussion with a review of known facts about the various groups that we will study, $SO(N)$, $Spin(N)$, $O(N)$, and $Pin^\pm (N)$.  We will also review how gauge theories based on these groups are constructed.  We will be particularly interested in three cases.

$Spin(N)$ gauge theories are special because they do not involve a sum over topological sectors as in \eqref{Znu}.  In the other extreme, $O(N)$ gauge theories include all the bundles of interest to us.  In addition to the regular Chern-Simons level $K\in\mathbb{Z}$, these theories are labeled by discrete parameters associated with  distinct consistent ways to sum over these bundles as in \eqref{Znu}.  We will denote them as $O(N)^r_{K,L}$ with $r=0,1$ and $L=0,1,...,7$.  (The label $r=0,1$ was denoted as $O(N)_\pm$ in \cite{Aharony:2013kma}.)
  
$SO(N)$ gauge theories are the ones with the largest (zero-form) discrete symmetry.  They have a charge conjugation $\mathbb{Z}_2$ and a magnetic $\mathbb{Z}_2$ symmetries.  We will couple them to background gauge fields $B^{\mathcal{C}}$ and $B^{\mathcal{M}}$.  When these gauge fields are made dynamical the $SO(N)$ gauge theory becomes the gauge theory of a different group.

We will then turn to a careful analysis of level-rank duality in topological field theories based on Chern-Simons theories of these gauge groups.  For unitary gauge groups the level-rank duality relations are \cite{Naculich:1990pa,Mlawer:1990uv, Witten:1993xi,Douglas:1994ex,Naculich:2007nc,Hsin:2016blu}
\begin{eqnarray}\label{SUlevelranki}
SU(N)_{K}&\quad\longleftrightarrow\quad& U(K)_{-N,-N}~, \\
U(N)_{K,K\pm N}&\quad\longleftrightarrow\quad &U(K)_{-N,-N \mp K}~, \nonumber
\end{eqnarray}
where $U(N)_{K,K'} = (SU(N)_K\times U(1)_{NK'})/\mathbb{Z}_N$.\footnote{Throughout most of this paper we label TQFTs by the corresponding Chern-Simons gauge group and its level.  A quotient as in this expression is interpreted from the $2d$ RCFT as an extension of the chiral algebra \cite{Moore:1988ss} and from the $3d$ Chern-Simons theory as a quotient of the gauge group \cite{Moore:1989yh}.  More abstractly, it can be interpreted as gauging a one-form global symmetry of the TQFT \cite{Kapustin:2014gua,Gaiotto:2014kfa}. This quotient is referred to in the condensed matter literature as ``anyon condensation''  \cite{Bais:2008ni}.  Occasionally we will meet a TQFT that is easier to describe not by the Chern-Simons gauge group, but by starting with another TQFT and performing such a quotient by a ``magnetic one-form symmetry.''  Such TQFTs can be described by a Chern-Simons group, but that description is more complicated.  We will alert the reader whenever we use this more abstract notation.} For orthogonal groups they are \cite{Hasegawa:1989741, Verstegen:1990at, Naculich:1990pa, Mlawer:1990uv,  Aharony:2016jvv}
\begin{equation}
\label{SOlevelranki}
SO(N)_{K}\quad\longleftrightarrow\quad SO(K)_{-N} ~.
\end{equation} 

Most level-rank dualities in \eqref{SUlevelranki} and \eqref{SOlevelranki} are valid only when the theories involved are spin-TQFTs \cite{Hsin:2016blu,Aharony:2016jvv}.\footnote{ In some cases a duality of spin TQFTs can be promoted to a related duality of non-spin TQFTs. }  If the theory is a spin-TQFT as it stands, no change is needed.  But if it is not, we should make it into a spin-TQFT.  In the unitary theories \eqref{SUlevelranki} this amounts to tensoring with an almost trivial TQFT -- $\{1,\Psi\}$, which can be described by $U(L)_1$ with any $L$.  We can think of it as a theory with two lines, a trivial one and a complex fermion $\Psi$.  (The fermion has to be complex and it should be charged so that the spin/charge relation is satisfied \cite{Seiberg:2016rsg}.) In the orthogonal theories \eqref{SOlevelranki} the analogous almost trivial theory is $\{1,\psi\}$ with real $\psi$, which can be described by $SO(L)_1$ with any $L$. 

It is important to couple the TQFTs in \eqref{SUlevelranki},\eqref{SOlevelranki} to background fields. Moreover, these background fields need specific counterterms for the duality to be valid \cite{Hsin:2016blu,Aharony:2016jvv, Benini:2017dus}.  One such background field is the metric and the necessary counterterm is a gravitational Chern-Simons term.  It was discussed in these papers and therefore, for simplicity, we will suppress it in most of the discussion below.  Another important background is a $U(1)$ field coupled to \eqref{SUlevelranki}.  A careful analysis of this field and its counterterms led to an easy derivation of any two of the dualities in \eqref{SUlevelranki} by assuming the third one \cite{Hsin:2016blu}.

One of the goals of this work is to couple the orthogonal dualities \eqref{SOlevelranki} to background fields $B^{\mathcal{C}}$ and $B^{\mathcal{M}}$ and to determine the necessary counterterms to make the dualities valid.  We will find
\begin{equation}
\label{lrcti}
SO(N)_{K}[B^{\mathcal{C}},B^{\mathcal{M}}]+(K-1)f[B^{\mathcal{C}}]+(N-1)f[B^{\mathcal{M}}]+f[B^{\mathcal{C}}+B^{\mathcal{M}}]\longleftrightarrow
SO(K)_{-N}[B^{\mathcal{M}},B^{\mathcal{C}}]~.
\end{equation}
Here $SO(N)_{K}[B^{\mathcal{C}},B^{\mathcal{M}}]$ denotes the action for $SO(N)_{K}$ coupled to the two background $\mathbb{Z}_2$ gauge fields $B^{\mathcal{C}}$, $B^{\mathcal{M}}$.\footnote{There are two different conventions for $\mathbb{Z}_{n}$ gauge fields.  The first is where they are viewed as $\mathbb{Z}_{n}$ connections with periods $0, \cdots n-1$.  The second is where they are viewed as constrained $U(1)$ gauge fields with periods $0, 2\pi/n, \cdots,2\pi(n-1)/n $.  In general, we use the first convention unless otherwise explicitly indicated.}  The coupling to $B^{\mathcal{C}},$ which is the background gauge field for the charge conjugation symmetry, means that the dynamical gauge field is actually an $O(N)$ gauge field constrained to satisfy $w_1=B^\mathcal{C}$.  (See more details below.)  $f[B]$ is a specific counterterm for a $\mathbb{Z}_2$ gauge field $B$, which is related to the $\eta$ invariant of a massive fermion coupled to $B$.  Its coefficient is an integer modulo $8$.  We will discuss $f[B]$ in detail below.  Note that the background fields $B^{\mathcal{C}}$, $B^{\mathcal{M}}$ are exchanged under the duality.  This reflects the fact that the duality exchanges the global symmetries $\mathcal{C}\longleftrightarrow  \mathcal{M}$ \cite{Aharony:2016jvv}.

The result \eqref{lrcti} makes the duality relation more complete.  It also enables us to derive many other dualities by gauging the $\mathbb{Z}_2 \times \mathbb{Z}_2$ symmetry (or a subgroup of it); i.e.\ by making  $B^{\mathcal{C}}$, or $B^{\mathcal{M}}$, or $B^{\mathcal{C}}+ B^{\mathcal{M}}$, or both of them dynamical.  For example, we will derive
\footnote{It is straightforward to use our techniques to find many additional dualities, for instance, involving $Pin^\pm(N)$ gauge groups or $O(N)$ theories with other levels.  The resulting dual theories are more complicated and we will not discuss them here.
}
\begin{eqnarray}
\label{lrgaugedi}
O(N)^{0}_{K,K} &\quad\longleftrightarrow&\quad Spin(K)_{-N}~, \\
O(N)^{1}_{K,K-1+L} &\quad\longleftrightarrow&\quad O(K)^{1}_{-N,-N+1+L}~. \nonumber
\end{eqnarray}
where the additional superscript and subscript of the orthogonal groups are specific terms in the Lagrangian, which we will discuss in detail below.  (The integer $L\sim L+8$ is arbitrary.)\footnote{Note as a particular consequence that the theories $O(N)^1_{N,N-1}$ and $O(N)^1_{N,N+3}$ are time-reversal invariant quantum mechanically (the time-reversal symmetry flips the signs of all levels $r,K,L$ in $O(N)^r_{K,L}$): 
\begin{equation}
O(N)^{1}_{N,N-1}\leftrightarrow O(N)^{1}_{-N,-N+1}~, \hspace{.5in}O(N)^{1}_{N,N+3}\leftrightarrow O(N)^{1}_{-N,-N-3}~.
\end{equation}
The special case  $O(2)_{2,1}$ is equivalent to the T-Pfaffian spin-TQFT \cite{Bonderson:2013pla,Chen:2013jha}. See appendix \ref{sec:OtwoPfaffian} for details.}

An important special case that will be used repeatedly throughout the following is $N=1$ which yields\footnote{For $O(N)$ with $N\leq 2$ the level encoded by the superscript does not exist and hence we drop it from our notation throughout \cite{Chen:2011pg,Wen:2014zga}.   }
\begin{equation}\label{SpinZ2}
(\mathbb{Z}_{2})_{K} \quad\longleftrightarrow\quad Spin(K)_{-1}~.
\end{equation}
The $\mathbb{Z}_{2}$ level $K$ appearing above is an integer defined modulo eight.  (See appendix \ref{z2appendix} for a detailed discussion.)  In particular, the theory $(\mathbb{Z}_{2})_{1}$ is the (time-reversal of) the Ising TQFT. 

Our conventions are such that $SO(2)_K \cong U(1)_K$ and $Spin(2)_K\cong U(1)_{4K}$.  Therefore, we find from \eqref{SOlevelranki} and \eqref{lrgaugedi} the interesting special cases
\begin{eqnarray}
\label{speciali}
SO(2)_K \cong U(1)_K &\quad\longleftrightarrow&\quad SO(K)_{-2}~, \nonumber \\
Spin(2)_K \cong U(1)_{4K} &\quad\longleftrightarrow&\quad O(K)^0_{-2,-2}~, \\
O(2)_{K,L}  &\quad\longleftrightarrow&\quad O(K)^1_{-2,L-K}~, \nonumber
\end{eqnarray}
which simplify for $K=2$ to
\begin{eqnarray}
\label{specialinew}
U(1)_2 &\quad\longleftrightarrow&\quad U(1)_{-2}~, \nonumber\\
U(1)_{8} &\quad\longleftrightarrow&\quad O(2)_{-2,-2}~,\\
O(2)_{2,L}  &\quad\longleftrightarrow&\quad O(2)_{-2,L-2}~. \nonumber 
\end{eqnarray}

Armed with this understanding of the background fields and their counterterms we can reexamine the CSM dualities of \cite{Metlitski:2016dht,Aharony:2016jvv} between bosons and fermions in the vector representation.  As in the discussion of dualities of TQFTs above, we couple them to the background fields $B^{\mathcal{C}}$ and $B^{\mathcal{M}}$ (which means, e.g.\ that the gauge fields are constrained $O(N)$ fields).  

It is important to clarify the treatment of the fermions.  They are coupled to $O(N)$ gauge fields and integrating over them leads to a phase involving the $\eta$-invariant (see below).  When the fermions are given positive or negative masses this phase becomes a local functional of the gauge fields, but when the fermions are massless the phase of the functional integral is typically non-local.  As is customary, we label the massless theory by the  ``effective level," which is the average of the integral levels for positive and negative fermion masses.  This effective level can be fractional.  

Taking into account the above discussion, we fix conventions such that if no bare counterterm is present in the Lagrangian, the theory of $N_f$ vector fermions is denoted as
\begin{eqnarray}\label{femriond}
SO(N)_{N_{f}/2}[B^{\mathcal{C}}, B^{\mathcal{M}}]
+(N_{f}/2)f[B^{\cal C}]
~\text{with}~N_{f}~ \psi ~. 
\end{eqnarray}
Note, this does not mean that we have a  term with a fractional coefficient in the Lagrangian.
Then we can add to this theory additional, properly-quantized, bare Chern-Simons terms for the dynamical and classical gauge fields.  

Using these conventions we determine the necessary counterterms in the $SO(N)$ boson-fermion duality:
\begin{align} 
&SO(N)_{-K+N_{f}/2}[B^{\mathcal{C}}, B^{\mathcal{M}}]
+(N_{f}/2)f[B^{\cal C}]
~\text{with}~N_{f}~ \psi  \label{CSMi}
\longleftrightarrow \\
&SO(K)_{N}[B^{\mathcal{M}}, B^{\mathcal{C}}]+(N-1)f[B^{\mathcal{M}}]+(K-1) f[B^{\mathcal{C}}]+ f[B^{\mathcal{M}}+B^{\mathcal{C}}] ~\text{with}~N_{f} ~\phi ~.\nonumber
\end{align}
Here the scalars have a $\phi^4$ interaction and the duality is valid only in the infrared.    

\begin{figure}
  \centering
  \subfloat{\includegraphics[width=\textwidth]{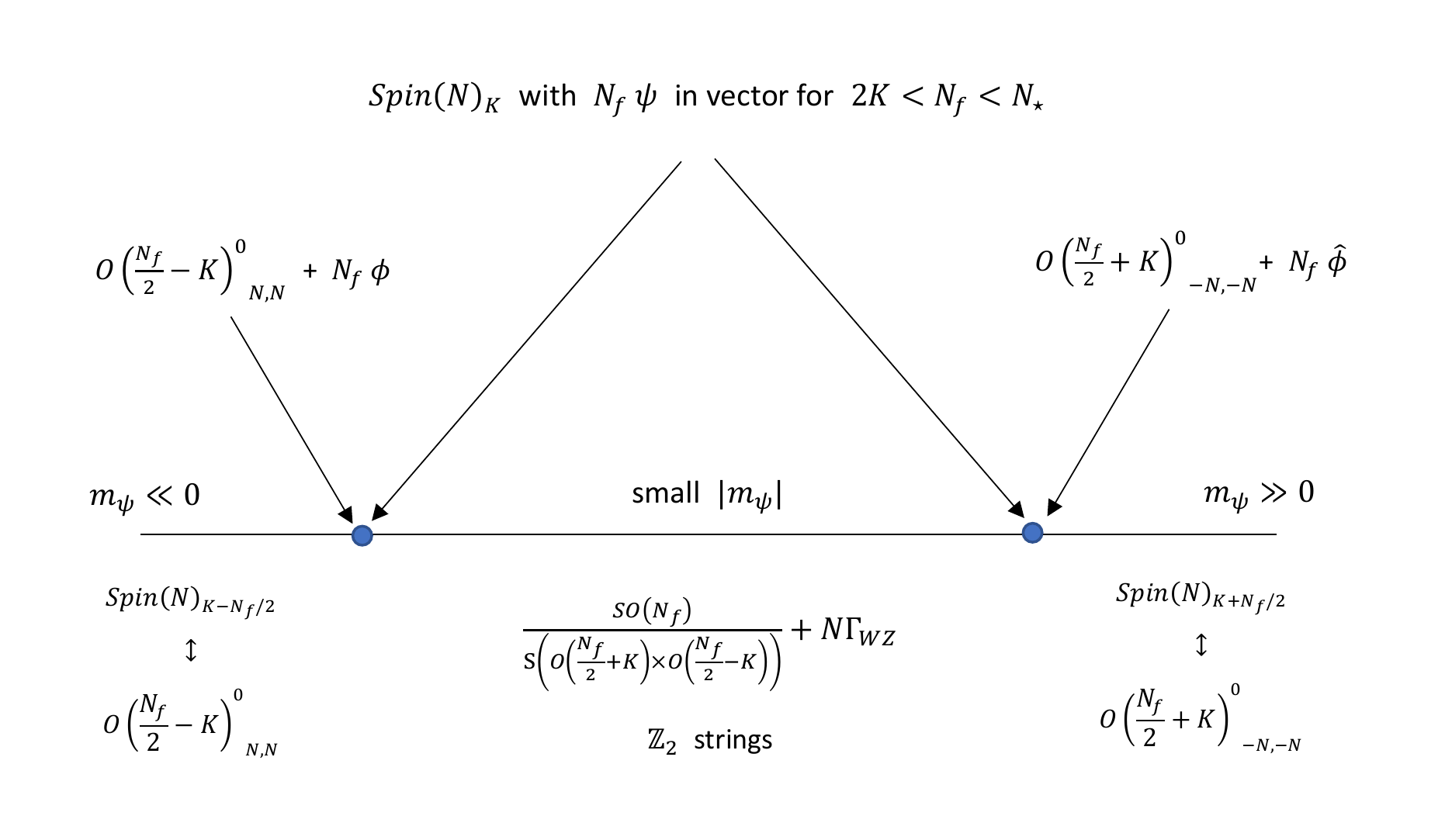}}        
  \caption{The phase diagram of $Spin(N)$ gauge theory coupled to fermions in the vector representation.  For large $|m_{\psi}|$ the infrared is a topological field theory, which is visible semiclassically.  For small $|m_{\psi}|$ the theory spontaneously breaks the flavor symmetry and is described by a sigma model with Grassmannian target. There is a Wess-Zumino term, and stable strings, signaling confinement. The transition from the semiclassical phase to the quantum phase is weakly coupled in the dual bosonic variables ($\phi$ or $\hat{\phi}$).  This diagram completes the discussion in \cite{Komargodski:2017keh} by specifying the discrete $\theta$-parameters needed in the various gauge theories, and provides a non-trivial consistency check on that proposal. }
  \label{figSpinphasevintro}
\end{figure}

In addition to making this duality more precise, we can now make the background fields dynamical; i.e.\ gauge them, and find 
new dualities. Some examples are
\begin{eqnarray}
\label{CSMNi}
O(N)_{K,K}^{0}~\text{with}~N_{f} ~\phi &\quad\longleftrightarrow \quad  &Spin(K)_{-N+N_{f}/2}~\text{with}~N_{f}~ \psi~, \nonumber \\
Spin(N)_{K}~\text{with}~N_{f} ~\phi &\quad\longleftrightarrow \quad&O(K)^{0}_{-N+N_{f}/2, -N+N_{f}/2}~\text{with}~N_{f}~ \psi~, \\
O(N)^{1}_{K,K-1+L}~\text{with}~N_{f} ~\phi &\quad\longleftrightarrow \quad&O(K)^{1}_{-N+N_{f}/2, -N+N_{f}/2+1+L}~\text{with}~N_{f}~ \psi~. \nonumber
\end{eqnarray}
In the above, $N_f$ is required to satisfy the following constraint for the duality to describe a transition point \cite{Aharony:2016jvv}: $N_f\leq N-2$ if $K=1$, $N_f\leq N-1$ if $K=2$, and $N_f\leq N$ if $K>2$.
For other values of $N_f$ below some theory-dependent number $N_\star$ the dualities are conjectured to be valid near different phase transition points with a symmetry-breaking phase in between \cite{Komargodski:2017keh} (see figure \ref{figSpinphasevintro}).

Some interesting special cases of these dualities are highlighted in section \ref{specialcases} below.  For instance with  $K=3$ and $N_{f}=1$ we can use $Spin(3)\cong SU(2)$ to find 
\begin{equation}
\label{SUtwosi}
O(N)_{3,3}^{0}~\text{with\ vector} ~\phi \quad\longleftrightarrow \quad  SU(2)_{-N+1/2}~\text{with\ adjoint}~ \psi~,
\end{equation}
which was used in \cite{Gomis:2017ixy}, with particular interest in $N=1$ and $N=2$.

Essential to our discussion of the dualities \eqref{CSMNi} is the fact that the discrete $\theta$-parameters can be changed by integrating out massive fermions.  For a single Majorana fermion $\lambda$ coupled to a $\mathbb{Z}_{2}$ gauge field, the level $L$ for the gauge field is shifted by one as we transition from negative to positive mass.  This means that we can describe the massless theory as in the discussion above \eqref{femriond} by saying that the effective level is one-half as shown below. 
\begin{equation}
\label{Z2shiftintro}
\begin{tabular}{ccccccccc}
$m_{\lambda}<0$ &&&& $m_{\lambda}=0$ &&&& $m_{\lambda}>0$ \\
\\
$(\mathbb{Z}_{2})_{L}$&&&&$(\mathbb{Z}_{2})_{L+1/2}$&&&&$(\mathbb{Z}_{2})_{L+1}$
\end{tabular}
\end{equation}
This is the notation adopted in \eqref{CSMNi} for expressing the second subscript level of $O(N)^{r}_{K,L}$, where we view the first Stiefel-Whitney class $w_{1}$ as a $\mathbb{Z}_{2}$ gauge field.

Next we will move to an analysis of the suggested phase diagram of orthogonal gauge groups with fermions in symmetric and anti-symmetric tensor representations \cite{Gomis:2017ixy}.  We add to the discussion in \cite{Gomis:2017ixy} the necessary background fields and their counterterms in each phase. This leads to highly non-trivial tests of the proposal.  Again a crucial role is played by the shifts in the counterterms generated by integrating out massive fermions.  In addition to the shifts described by \eqref{Z2shiftintro}, we also find that the $\mathcal{C}$ charge of monopole operators in the $SO(N)$ theory changes when we transition a two-index tensor fermion from negative to positive mass.  This also means that in the $O(N)$ theory, the superscript level (controlling the discrete theta term $\exp(i\pi \int_{X}w_{1}\cup w_{2})$) jumps by one across such a transition.

\begin{figure}
  \centering
  \subfloat{\includegraphics[width=\textwidth]{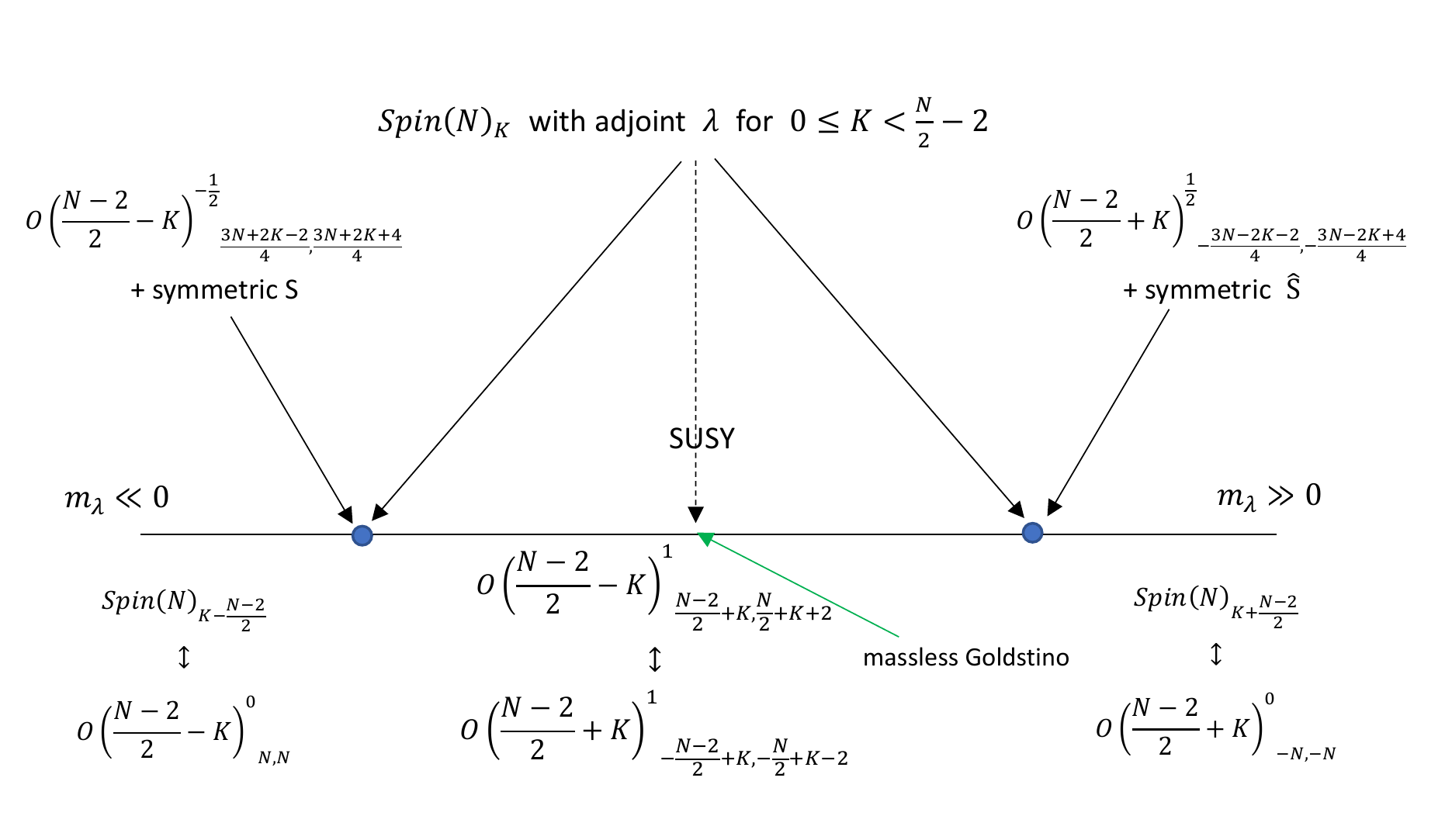}}        
  \caption{The phase diagram of $Spin(N)$ gauge theory coupled to an adjoint fermion.  The infrared TQFTs, together with relevant level-rank duals are shown along the bottom.  The blue dots indicate transitions from the semiclassical phase to a quantum phase. Each transition is weakly coupled in a dual theory, which covers part of the phase diagram.  The dual theory has symmetric tensor fermions ($S$ and $\hat{S}$). Across these transitions the superscript level of the TQFT jumps, i.e.\ $O(N)^{0}$ and $O(N)^{1}$ are exchanged. 
The fractional levels represent the effective contributions from the massless fermions.
At a special value of the mass in the quantum phase, the ultraviolet theory is supersymmetric.  This symmetry is spontaneously broken leading to a massless Goldstino \cite{Witten:1999ds}.  This phase diagram is obtained from that of the $SO(N)$ theory proposed in \cite{Gomis:2017ixy}, by tracking the global symmetries and counterterms and then gauging. }
  \label{figSpinphaseintro}
\end{figure}

As in all our other cases, once these counterterms are set we can easily turn the background fields in the $SO(N)$ phase diagram into dynamical fields and make similar predictions for $Spin(N)$ and $O(N)$ gauge theories.  It is important that these are not logically independent proposals.  They follow from the suggested phase diagram of $SO(N)$ theories.\footnote{We can also add a pair of adjoint gauginos to the Chern-Simons theories to make them $\mathcal{N}=2$ supersymmetric.  After taking into account the appropriate level shifts we find the following $\mathcal{N}=2$ dualities:
\begin{eqnarray}
SU(N)_{K+N}&\leftrightarrow& U(K)_{-K-N,-N}~, \nonumber\\
U(N)_{K+N, K+N}&\leftrightarrow& U(K)_{-K-N, -K-N}~,\nonumber\\
U(N)_{K+N, K-N}&\leftrightarrow& U(K)_{-K-N, K-N}~,\nonumber\\
Sp(N)_{K+N+1}&\leftrightarrow& Sp(K)_{-N-K-1}~,\\
SO(N)_{K+N-2}&\leftrightarrow&SO(K)_{-K-N+2}~,\nonumber\\
Spin(N)_{K+N-2}&\leftrightarrow&O(K)^{1}_{-K-N+2,-K-N+1}~,\nonumber\\
O(N)^{0}_{K+N-2,K+N-2+L}&\leftrightarrow&O(K)^{0}_{-K-N+2,-K-N+2+L}~.\nonumber
\end{eqnarray}  These agree with existing results \cite{Giveon:2008zn, Kapustin:2011gh, Aharony:2013dha, Aharony:2013kma}.  (The second subscript level of the orthogonal group $O(N)$ was previously ignored.)
 }  Therefore, in addition to being interesting in their own right, their consistency also gives further evidence to the original suggestion.  An example phase diagram that we find is shown in figure \ref{figSpinphaseintro}.

In section \ref{reviewso} we review some facts about Chern-Simons theory with Lie algebra $\frak{so}(N)$.
In section \ref{LRDsec} and appendix \ref{derlrdualodd} we derive the level-rank dualities for $SO$ Chern-Simons theories coupled to the gauge fields for the $\mathbb{Z}_2$ symmetries.
In section \ref{matter} we discuss Chern-Simons matter dualities and the phase diagram of QCD with tensor fermion, and conjecture new boson-boson and fermion-fermion dualities.

There are several appendices.
In appendix \ref{dynkin} we summarize some facts about the representation theory of $\mathfrak{so}(N)$.
In appendix \ref{z2appendix} we discuss the spin topological $\mathbb{Z}_2$ gauge theories and their classification, as well as the generalization to other unitary symmetry groups.
In appendix \ref{gauging} we give more detail about the relations between $Pin^-(N),O(N)^1$ and $SO(N)$ gauge theories.
In appendix \ref{CSOddsec} we derive the relation between the $O(N)$ and $SO(N)$ Chern-Simons actions for odd $N$.
In appendix \ref{so2so4sec} we present explicit examples of extended and orbifold chiral algebras.
In appendix \ref{ProjectiveReps} we give examples of Wilson lines in Chern-Simons theory that transform in projective representations of the global symmetry. In appendix \ref{sec:OtwoPfaffian} we discuss the relationship between $O(2)_{2,L}$ and Pfaffian theories.
In appendix \ref{dualityoneform} we present dualities of Chern-Simons theory related to the one-form symmetry.

\section{Chern-Simons Theories with Lie Algebra $\frak{so}(N)$}
\label{reviewso}

\subsection{Groups, Bundles, and Lagrangians}
\label{gaugegrps}

In this section we review aspects of these Chern-Simons theories.  A discussion of the relevant group theory may be found in \cite{lawson1989spin,kirby_taylor_1991,Berg2001,Varlamov2004,Moore2013pin}, while aspects of their bundles are described in \cite{lawson1989spin,kirby_taylor_1991}. 

\subsubsection{Gauge Groups}
\label{sec:gg}

The various global forms of the gauge group can be understood starting from $SO(N),$ the smallest group of interest, and performing extensions by elements of order two.\footnote{For even $N$ one may also consider the group $SO(N)/\mathbb{Z}_{2}$, which we do not discuss here.}  One possible extension is to include reflections, leading to the orthogonal group $O(N)$.  Another extension allows a $2\pi$ rotation to act non-trivially thus permitting spinor representations.  This leads to the group $Spin(N)$.  

The two extensions may be combined leading to the two groups $Pin^{\pm}(N)$.  Both groups are simply connected and contain as their identity component the subgroup $Spin(N)$.  The difference between them is whether the reflection $r$ along a single axis squares to the identity or to $2\pi$ rotation $s$:
\begin{equation}
Pin^{+}(N):~~  r^{2}=1~, \hspace{.5in}Pin(N)^{-}:~~ r^{2}=s~.
\end{equation}
The groups $Pin^{\pm}(N)$ may thus be presented explicitly as semidirect products of $Spin(N)$ with cyclic groups generated by $r$
\begin{equation}
\label{pinsemid}
Pin^{+}(N)\cong Spin(N)\rtimes \mathbb{Z}_{2}~, \hspace{.5in}Pin^{-}(N)\cong \frac{Spin(N)\rtimes \mathbb{Z}_{4}}{\mathbb{Z}_{2}}~.
\end{equation}
A summary of these extensions is illustrated in the diagram below, where each row and column is exact.  
\begin{equation}
\xymatrix{
 \mathbb{Z}_{2}^{s} \ar[d]& \mathbb{Z}_{2}^{s} \ar[d]\\
 Spin(N)\ar[r] \ar[d]&Pin^{\pm}(N) \ar[d] \ar[r] & \mathbb{Z}_{2}^{r} \\
 SO(N) \ar[r]&O(N)\ar[r] & \mathbb{Z}_{2}^{r}
}
\end{equation}

When $N$ is odd, a simultaneous reflection along all axes commutes with $SO(N)$ and we can further simplify $O(N)\cong SO(N)\times \mathbb{Z}_{2}$.  The analogous relation for the groups $Pin^{\pm}(N)$ is \cite{Varlamov2004}
\begin{equation}
\label{pind}
Pin^{\pm}(N)\cong \begin{cases}  Spin(N)\times \mathbb{Z}_{2} & N=\pm1 ~(\text{mod}~4) \\ (Spin(N)\times \mathbb{Z}_{4})/\mathbb{Z}_{2} & N=\mp 1 ~(\text{mod}~4)\end{cases}~,
\end{equation}
where the signs are correlated in each line above.  Note that in contrast to \eqref{pinsemid}, the cyclic factors in the products above are generated by reflection along all axes.  For $N$ even, there are no analogous simplifications in the groups.  

For even $N,$ the reflection that extends $SO(N)$ to $O(N)$ may be understood as an outer automorphism of $SO(N)$, and may be seen from the symmetry of the Dynkin diagram.  We refer to this $\mathbb{Z}_{2}$ symmetry as charge conjugation $\mathcal{C}$.  It acts on the representations of $SO(N)$ by exchanging the Dynkin labels associated with the two permuted nodes.\footnote{In the Lie algebra $\frak{so}(N)$ these are the Dynkin labels associated with the two distinct spinor representations.}  Later, we will see that this outer automorphism gives rise to a global symmetry of the related quantum field theories.

\subsubsection{Bundles}

Next let us discuss the possible topological types of gauge bundles that exist for the various gauge groups defined above.  The topology of the bundles may be parameterized in terms of the Stiefel-Whitney characteristic classes $w_{i}\in H^{i}(X, \mathbb{Z}_{2})$, where $X$ is the spacetime three-manifold and $i=1,2$.  The allowed classes of bundles for each possible gauge group are indicated in table \ref{tab:bun} below \cite{lawson1989spin,kirby_taylor_1991}.
\begin{table}[h!]
\begin{center}
\begin{tabular}{ |c|c|c| } 
Group & $w_{1}$ & $w_{2}$ \\ 
 \hline
$SO(N)$ & 0 & unrestricted \\ 
$O(N)$ & unrestricted & unrestricted \\ 
$Spin(N)$ & 0 & 0 \\ 
$Pin^{+}(N)$ & unrestricted & 0 \\ 
$Pin^{-}(N)$ & unrestricted & $w_{1}\cup w_{1}$ \\ 
 \hline
\end{tabular}
\end{center}
\caption{Allowed bundle topology for various gauge groups.}
\label{tab:bun}
\end{table}
Note that the group $O(N)$ has the largest set of possible gauge bundles, while for other groups the topology of the bundles are restricted.
For $O(N)$ bundles, the classes $w_{i}$ may be defined as follows:
\begin{itemize}
\item $w_{1}$ is a $\mathbb{Z}_{2}$ gauge field for the subgroup of $O(N)$ that reflects a single axis.  It is the obstruction to restricting the structure group of the bundle to $SO(N)$.
\item $w_{2}$ is the obstruction to lifting the structure group of the $O(N)$ bundle to a $Pin^{+}(N)$ bundle.
\end{itemize}

When $N$ is odd, the product structure $O(N)\cong SO(N)\times\mathbb{Z}_{2}$ implies that every $O(N)$ bundle defines an $SO(N)$ bundle.  The Stiefel-Whitney class $w_{2}(SO(N))$ is the obstruction to lifting this $SO(N)$ bundle to a $Spin(N)$ bundle.\footnote{Here and below we denote by $w_{i}(G)$ is the $i$-th Stiefel-Whitney class of a principle bundle with structure group $G$.}  Using the formula \eqref{pind} one may determine that
\begin{equation}\label{eqn:magfluxodd}
w_{2}(O(N))=\begin{cases} w_{2}(SO(N)) & N=+1~(\text{mod}~4)~,\\
 w_{2}(SO(N))+w_{1}\cup w_{1} & N=-1~(\text{mod}~4)~,
\end{cases}
\end{equation}
where we have used the fact that $w_{1}\cup w_{1}$ is the obstruction to lifting $w_{1}$ to a $\mathbb{Z}_{4}$ cohomology class.\footnote{A simple way to see the necessity of this condition is to note that for any degree one cohomology classes $x,y$ we have $x\cup y=-y\cup x.$  Thus, if $x=y$ this vanishes unless the coefficients are $\mathbb{Z}_{2}$.  A more detailed discussion may be found in \cite{steenrod1962cohomology,Hatcher:2001}.}  When $N$ is even there is no analogous formula since in that case an $O(N)$ bundle does not in general define an $SO(N)$ bundle.  For uniformity throughout the following we use the Stiefel-Whitney classes of $O(N)$ bundles unless otherwise explicitly indicated.

\subsubsection{Lagrangians}
\label{sec:lagrangians}

We now turn to a discussion of the possible Lagrangians for Chern-Simons theories based on the gauge groups discussed above.

In each possible theory there is a continuous field variable $A$ that is a connection valued in the Lie algebra $\frak{so}(N).$  We may include in the action a Chern-Simons term $CS(A)$ with a level $K$
\begin{equation}
K \cdot CS(A)=\frac{K}{8\pi}\int _{X}\mathrm{Tr}\left(A \wedge dA +\frac{2}{3}A\wedge A\wedge A\right)~, \label{spincsinv}
\end{equation}
where in the above the trace is defined in the vector representation.  Our normalization is such that the level must be quantized in integral units, $K\in \mathbb{Z}.$  If the level $K$ is odd, then the action \eqref{spincsinv} depends on a choice of spin structure on $X$.  If $K$ is even, no spin structure is required.  For any gauge group $G,$ we denote the Chern-Simons theory with level $K$ by $G_{K}.$ 

In Chern-Simons  gauge theory we perform a path integral over the connection $A$ and the Chern-Simons action $\exp(iKCS(A))$ tells us the weight assigned to each field configuration.  In addition we must also sum over the possible topological types of bundles allowed for the gauge groups, and there are choices for how to weight the various bundles in the sum.  The additional coupling constants that parameterize the possible weights are sometimes referred to as discrete $\theta$-parameters.  As we describe below, they may also be thought of as Chern-Simons interactions for discrete groups.

It is clearest to present the discussion starting from the gauge group $O(N)$ which admits bundles with all possible Stiefel-Whitney classes.  There are two distinct discrete $\theta$-parameters to specify.
\begin{itemize}
\item There is a possible weight $\exp(i\pi\int_{X} w_{1} \cup w_{2}).$   Since the characteristic classes are $\mathbb{Z}_{2}$-valued this evaluates to $\pm1$ for each possible bundle.\footnote{In the special case of $O(2)$ this coupling is trivial \cite{Wen:2014zga}.  We therefore omit the superscript in this case.}  We may introduce a $\mathbb{Z}_{2}$-valued level $p$ that specifies whether this interaction is present ($p=1$) or absent ($p=0$).\footnote{In \cite{Aharony:2013kma} our $O(N)^{0}$ was denoted by $O(N)_{+}$ and our $O(N)^{1}$ was called $O(N)_{-}$. }  We indicate its value in a superscript $O(N)^{p}$.  

\item Another possible coupling depends only on the characteristic class $w_{1}$.  Viewing $w_{1}$ as a $\mathbb{Z}_{2}$ gauge field, the possible additional couplings arise from local effective actions depending on this gauge field.  Since we are interested in spin TQFTs we permit these actions to depend on the spin structure of the underlying three-manifold $X$.  Such local actions have been completely determined \cite{GuLevin,Kapustin:2014dxa,Wang:2016lix,Kapustin:2017jrc}, and admit a $\mathbb{Z}_{8}$ classification.  

We indicate the minimal local action by $f[B]$ where $B$ is any $\mathbb{Z}_{2}$ gauge field.  The properties of these actions are discussed in detail in appendix \ref{z2appendix}.\footnote{The special case $4f[B]$ has an elementary action $\pi \int_{X}B \cup B\cup B$, and represents a bosonic Dijkgraaf-Witten theory \cite{Dijkgraaf:1989pz} for $\mathbb{Z}_2$ gauge group classified by $H^3(B\mathbb{Z}_2,U(1))=\mathbb{Z}_2$.}  The relationship of these $\mathbb{Z}_{2}$ gauge theories to the fermion path integrals is crucial in our discussion of Chern-Simons matter duality in section \ref{matter}.  

In general, we indicate the additional level by a subscript valued mod $8$ (e.g. $(\mathbb{Z}_{2})_{L}$).  For Chern-Simons theory with gauge group $G$ we indicate this $\mathbb{Z}_{8}$-valued level by a second subscript.

\end{itemize}  

In summary, our complete list for possible quantum $O(N)$ Chern-Simons theories and their coupling constants is
\begin{equation}
O(N)_{K,L}^{p}~, \hspace{.5in}K \in \mathbb{Z}~,\hspace{.2in}L\in \mathbb{Z}_{8}~, \hspace{.2in}p \in \mathbb{Z}_{2}~.
\end{equation}
This classification of discrete levels may be repeated for any gauge group $G$, and in particular the $\mathbb{Z}_{8}$-valued level exists for any group that admits non-vanishing first Stiefel-Whitney class $w_{1}$.  Thus for instance, the groups $Pin^{\pm}(N)$ may be used to define Chern-Simons:
\begin{equation}
Pin^{\pm}(N)_{K, L}~, \hspace{.5in}K \in \mathbb{Z}~,\hspace{.2in}L\in \mathbb{Z}_{8}~.
\end{equation}
One of our main results will be explicit level-rank dualities for these gauge theories, including the required maps on the discrete $\theta$-parameters. 

It is important to stress that the gauge theories we describe here are in general spin topological field theories.  That is, they depend explicitly on a choice of spin structure $S$ on the three-manifold $X$.  In particular, all such theories contain a transparent line $\psi$ of spin $1/2$.  

When $K$ is even and $L=0~(\text{mod}~4)$ the theories $O(N)_{K,L}^{p}$ may in addition be defined without the choice of spin structure.  However, both the level-rank dualities and Chern-Simons matter dualities of interest to us only make sense as spin theories.  For instance, the latter involve dynamical fermions where the spin structure is required.  We can promote $O(N)_{K,L}^{p}$ at bosonic values of the levels to a spin theory by tensoring with $\{1, \psi\}$: an almost trivial theory containing two transparent lines, the identity and the spinor.  In particular, the existence of the line $\psi$ restores the dependence on spin structure.  In the following, we use the notation $O(N)_{K,L}^{p}$ to denote the spin version of the theory (i.e.\ tensoring with $\{1, \psi\}$ implicit) unless explicitly indicated.  We adopt the same convention for other global forms of the gauge group.

\subsection{Ordinary Global Symmetries and Counterterms}
\label{ordinarysym}

Let us describe the global symmetries of these gauge theories.  There may be ordinary (zero-from) global symmetries, as well as one-form global symmetries and we describe them in turn.

The ordinary global symmetries may be understood by starting with the smallest group $SO(N),$ which has the largest zero-form symmetry.  The symmetries of $SO(N)_{K}$ depend on the parity of $N$ and $K$.  

If $N$ is even, the Lie algebra $\frak{so}(N)$ has an outer automorphism $\mathcal{C}$ discussed in section \ref{sec:gg}.  This defines a charge conjugation symmetry $\mathcal{C}$ of $SO(N)_{K},$ which acts non-trivially on the lines in the topological field theory.\footnote{See Appendix \ref{dynkin} for a discussion of the representation theory of $\frak{so}(N).$}  Equivalently, one may view this as acting directly on the $SO(N)$ connection.  Expanding in a standard Lie algebra basis of antisymmetric matrices the transformation is 
\begin{equation}\label{cdef}
\mathcal{C}(A^{[ij]})=\begin{cases} \phantom{-}A^{[ij]} & i,j \neq 1~, \\  -A^{[ij]}  & i=1~, \text{or}~ j=1~.   \end{cases}
\end{equation}

If $N$ is odd, the transformation \eqref{cdef} can be achieved by conjugation by an element of $SO(N)$ and hence does not permute representations.  Therefore, for odd $N$ charge conjugation is not a true symmetry of the topological field theory $SO(N)_{K}.$ Nevertheless, when $N$ is odd we still find it useful to discuss a symmetry $\mathcal{C}$ which we will identify with $(-1)^{F}$ where $F$ is the fermion number.  As we will see, this identification is natural from the point of view of level-rank duality.  

Similarly, for $K$ even we may define a magnetic global symmetry $\mathcal{M}$ of $SO(N)_{K}$  that permutes the lines.    Like charge conjugation, the magnetic symmetry is a $\mathbb{Z}_{2}$ global symmetry.  To measure the magnetic charge, one integrates the  second Stiefel-Whitney class $w_{2}$ of the gauge bundle over a two-cycle.  

To understand how $\mathcal{M}$ acts on lines in the TQFT, it is convenient to view the theory $SO(N)_{K}$ as $Spin(N)_{K}/\mathbb{Z}_{2}$ where the quotient means that we gauge a $\mathbb{Z}_{2}$ one-form global symmetry \cite{Kapustin:2014gua,Gaiotto:2014kfa}.   The generating line of this one-form global symmetry gives rise in $SO(N)_{K}$ to a transparent line transforming in the $K$-th symmetric traceless power of the vector representation (i.e.\ Dynkin indices $(K,0,\cdots,0)$), which has spin $K/2$.  In the case of $K$ even this transparent line is a boson, and there are pairs of line defects that are related by fusion with the transparent line.  Those lines of $Spin(N)_{K}$ that are fixed points under fusion with the one-form symmetry generator are doubled in the spectrum of $SO(N)_{K}$. The two lines in a given doublet are exchanged under the action of the symmetry $\mathcal{M}$ \cite{Moore:1989yh}.  An equivalent point of view on the magnetic symmetry $\mathcal{M}$ via chiral algebras is discussed in section \ref{chiral} below.

When $K$ is odd one may similarly define a conserved charge $\mathcal{M}$ measured by $w_{2}.$  However the associated symmetry acts trivially on lines since pairs related by fusion with the transparent line may be distinguished by their spin.  The meaning of this magnetic symmetry is the following.  For odd level the theory $SO(N)_{K}$ depends on the spin structure $S$.  We can probe this dependence by shifting $S$ by a $\mathbb{Z}_{2}$ gauge field $B$.  Under this transformation the minimally quantized Chern-Simons action responds as \cite{Jenquin2005CS}
\begin{equation}\label{eqn:shiftspinstruc}
 CS(A,S+B)=CS(A,S)+ \pi \int_{X}w_{2}\cup B~.
\end{equation}
Thus, for odd level, $B$ may be interpreted as a background gauge field for $\mathcal{M}$.  More physically, at odd level the transparent line is a spinor and the magnetic charge is identified with fermion number mod 2.

We summarize the identifications between the charge conjugation and magnetic symmetry in table \ref{0formsym}.
\begin{table}[h!]
\setstretch{1.5}
\begin{equation}
\begin{tabular}{ |c|c|c| } 
 & $N=0~(\text{mod}~2)$ & $N=1~(\text{mod}~2)$\\ 
 \hline
$K=0~(\text{mod}~2)$ & $\mathcal{C}, ~\mathcal{M},~ (-1)^{F}$  & $\mathcal{M},~ \mathcal{C}=(-1)^{F}$ \\
\hline
$K=1~(\text{mod}~2)$ & $\mathcal{C},~ \mathcal{M}=(-1)^{F}$  &$\mathcal{C}= \mathcal{M}=(-1)^{F}$     \\
\hline
\end{tabular}
\nonumber
\end{equation}
\caption{Identifications between the zero-form global symmetries $\mathcal{C}, \mathcal{M}, (-1)^{F}$ of $SO(N)_{K}$.}
\label{0formsym}
\end{table}

Depending on $N$ and $K$ there may be additional ordinary global symmetries.  For instance, if $N$ and $K$ are even and $NK$ is an odd multiple of $8$ there is a $\mathbb{Z}_{2}$ quantum zero-form symmetry that permutes the anyons \cite{Aharony:2016jvv} defined by fusing the anyons charged under the $\mathbb{Z}_2$ one-form symmetry with the generator of the one-form symmetry.  In the following we focus on the symmetries $\mathcal{C}$ and $\mathcal{M}.$

In the theory  $SO(N)_{K}$ with even $N$ and $K$ the symmetries $\mathcal{C}$ and $\mathcal{M}$ both square to one and commute,\footnote{Interestingly, the symmetries are represented projectively on some lines.  See Appendix \ref{ProjectiveReps} for details.} and hence the zero-form symmetry is $\mathbb{Z}_{2} \times \mathbb{Z}_{2}$.  We may couple our system to background gauge fields $B^{\mathcal{C}}$ and $B^{\mathcal{M}}$ for these $\mathbb{Z}_{2}$ symmetries.  Additionally we may add to the action local counterterms for these gauge fields.  The possible local counterterms have been determined and admit a $\mathbb{Z}_{8}\times \mathbb{Z}_{8}\times \mathbb{Z}_{4}$ classification \cite{Wang:2016lix}.  Each $\mathbb{Z}_{8}$ factor is the level for the individual gauge fields $B^{\mathcal{C}}$ and $B^{\mathcal{M}}$ (i.e.\ if it is gauged one obtains $(\mathbb{Z}_{2})_{L}$),  while the $\mathbb{Z}_{4}$ controls a counterterm that depends on both backgrounds.   We express the counterterm action as
\begin{equation}
\label{ctgendef}
S_{\text{counterterm}}=xf[B^{\mathcal{C}}]+yf[B^{\mathcal{M}}]+zf[B^{\mathcal{C}}+B^{\mathcal{M}}]~,
\end{equation}
where the parameters enjoy the identification
\begin{equation}
x\sim x+8~, \hspace{.5in}y\sim y+8~, \hspace{.5in} (x,y,z)\sim(x+4,y+4,z+4)~.
\end{equation}
More details on these counterterns are presented in appendix \ref{z2appendix}.  

\begin{figure}
\begin{equation}
\nonumber
\xymatrix{
&&&&&&Spin(N)\ar[dr]^{\mathcal{C}}\\
Pin^{-}(N)&&&&& \ar[lllll]_{~\mathcal{C}\&\mathcal{M}+(x,y,z)=(-2,-2,\,2)}SO(N) \ar[r]^{\mathcal{CM}} \ar[ur]^{\mathcal{M}}\ar[dr]_{\mathcal{C}}& O(N)^{1}\ar[r]^{\mathcal{M}} & Pin^{+}(N)\\
&&&&&&  O(N)^{0}\ar[ur]_{\mathcal{M}}
}
\end{equation}
\caption{A map of possible gauge groups obtained by starting with $SO(N)$ and gauging symmetries ($\mathbb{Z}_{2}$ levels suppressed).  In  Appendix \ref{gauging} we discuss some details of this map.}
\label{gaugemap}
\end{figure}
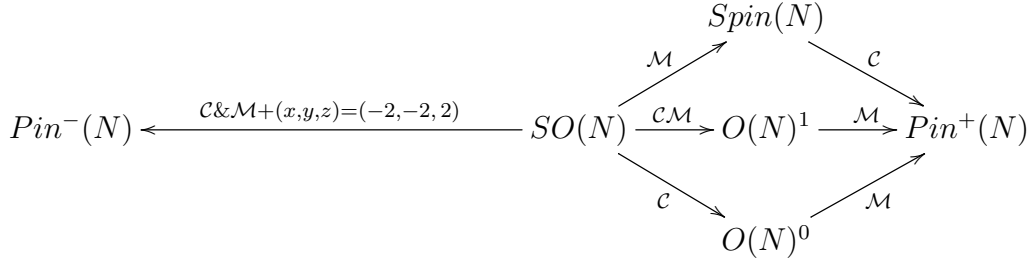

Starting from the case of $SO(N)_{K}$ we can gauge any of the ordinary global symmetries with specified values of the counterterms.  In this way we can produce all the gauge groups discussed in section \ref{gaugegrps} together with specified discrete $\theta$-parameters, as well as other models.  The zero-form symmetries of the resulting theories are the quotient of $\mathbb{Z}_{2}\times \mathbb{Z}_{2}$ by the subgroup that is made dynamical.   We illustrate this in figure \ref{gaugemap}.

As an illustrative example of this process, consider starting with $SO(N)_{K}$ and gauging the magnetic symmetry $\mathcal{M}$.  This inserts in the action a term $\pi \int_{X} w_{2}\cup B^{\mathcal{M}}$, with $B^{\mathcal{M}}$ now a dynamical $\mathbb{Z}_{2}$ gauge field.  The resulting sum over $B^{\mathcal{M}}$ constrains $w_{2}$ to vanish and implies that the gauge group is $Spin(N)$.  If we include the counterterm $Lf[B^{\mathcal{M}}]$ then we construct the theory $Spin(N)_{K,L}$ defined in \eqref{spindiscretetheta}.  This corresponds to the choice $y+z=L$ in \eqref{ctgendef}.

Analogously, starting from $SO(N)_{K}$ and gauging either $\mathcal{C}$ or the product $\mathcal{CM}$ with fixed counterterms produces $O(N)^{0}_{K,L}$ and $O(N)^{1}_{K,L}$ respectively.

In the special case where $N$ and $K$ are both odd the identification of global symmetries discussed around table \ref{0formsym} implies that $\mathcal{M}$=$\mathcal{C}$ and hence gauging them should produce equivalent field theories.  To make this precise, recall that for odd $N$ we have $O(N)\cong SO(N)\times \mathbb{Z}_{2}$.  The relationship between the $O(N)$ Chern-Simons action (with our choice of charge conjugation \eqref{cdef}) and that of the factorized $SO(N)\times \mathbb{Z}_{2}$ variables is derived in appendix \ref{CSOddsec} 
\begin{equation}
\label{CS(O(odd))}
CS(O(N))=CS(SO(N))+\pi \int_X w_{2}(SO(N))\cup w_{1}+(N-1)f[w_{1}]~.
\end{equation}
Therefore, for $N$ and $K$ both odd, we have the following equivalence of spin Chern-Simons theories\footnote{
The equivalence only holds as spin TQFTs.  In particular for $Spin(N)_K$ we must promote it to a spin theory by tensoring with $\{1,\psi\}$.
}
\begin{equation}
O(N)^{0}_{K,K-NK}\quad\longleftrightarrow\quad Spin(N)_{K}~. \label{oddduality}
\end{equation}
No analogous relationship holds for even $N$ or $K$.

\subsection{One-form Global Symmetries}

Now let us turn to the one-form global symmetry of these models.  A starting point to understanding these symmetries is via the center of the gauge groups shown in table \ref{center} \cite{lawson1989spin,Berg2001,Varlamov2004,Moore2013pin}. 

\begin{table}[h!]
\setstretch{1.5}
\begin{equation}
\begin{tabular}{ |c|c|c|c|c| } 
Group & $N=0~(\text{mod}~4)$ & $N=1~(\text{mod}~4)$& $N=2~(\text{mod}~4)$ & $N=3~(\text{mod}~4)$\\ 
 \hline
$SO(N)$ & $\mathbb{Z}_{2}$  & 1 & $\mathbb{Z}_{2}$ & 1\\
\hline
$O(N)$ & $\mathbb{Z}_{2}$  &$\mathbb{Z}_{2}$   & $\mathbb{Z}_{2}$ &$\mathbb{Z}_{2}$  \\
\hline
$Spin(N)$ & $\mathbb{Z}_{2}\times \mathbb{Z}_{2}$ &  $\mathbb{Z}_{2}$& $\mathbb{Z}_{4}$&$\mathbb{Z}_{2}$ \\
\hline
$Pin^{+}(N)$ &$\mathbb{Z}_{2}$ & $\mathbb{Z}_{2}\times \mathbb{Z}_{2}$ & $\mathbb{Z}_{2}$&$\mathbb{Z}_{4}$ \\
\hline
$Pin^{-}(N)$ & $\mathbb{Z}_{2}$ &$\mathbb{Z}_{4}$  & $\mathbb{Z}_{2}$&$\mathbb{Z}_{2}\times \mathbb{Z}_{2}$ \\
\hline
\end{tabular}
\nonumber
\end{equation}
\caption{Centers of the Gauge Groups ($N>2$).}
\label{center}
\end{table}

In general, each element of the center of the gauge group may give rise to a one-form global symmetry.  For each Wilson line in a representation $\mathbf{R}$ of the gauge group, the  charge of the line is the phase defined by the action of the associated element of the center in the representation $\mathbf{R}$.  In order for this definition to truly give rise to a global symmetry, it is necessary that there exists a line in the spectrum of the theory that can measure this phase by linking.  This is the case in all examples in table \ref{center} except  for $SO(N)_{K}$ and $O(N)_{K}$.  The line generating the one-form symmetry associated to the center of the gauge group has Dynkin indices $(0,\cdots,0,K)$.  If $K$ is odd this representation is a spinorial and excluded from the spectrum.  If $K$ is even it generates a $\mathbb{Z}_{2}$ global symmetry.

For $SO(N)_{K},$ and $O(N)_{K}$ the center symmetry at even level exhausts the one-form global symmetry.  However for other gauge groups there may exist additional quantum one-form symmetries that are not manifest at the level of the classical action.  Let us first consider the case of vanishing $\mathbb{Z}_{2}$ discrete $\theta$-parameter and subsequently generalize. A practical way to understand the additional symmetries is as bonus symmetries that arise from gauging ordinary global symmetries.  

In general, if we begin with any three-dimensional quantum field theory and gauge a $\mathbb{Z}_{r}$  global symmetry, the resulting theory has an emergent $\mathbb{Z}_{r}$ one-form global symmetry whose generating line is the Wilson line of the $\mathbb{Z}_{r}$ gauge theory.  The inverse process also exists.  Namely, gauging the emergent $\mathbb{Z}_{r}$ one-form global symmetry in the resulting theory produces an emergent zero-form global symmetry generated by the Wilson surface of the $\mathbb{Z}_{r}$ two-form gauge theory, and the result is the original theory.\footnote{
We can also gauge the $\mathbb{Z}_r$ zero-form symmetry with different counterterms for the $\mathbb{Z}_r$ gauge field to produce different theories, and each of them has the dual $\mathbb{Z}_r$ one-form symmetry. Gauging this one-form symmetry takes us back to the original theory.
}

We can apply this logic to the gauge groups appearing in figure \ref{gaugemap}, starting from the one-form symmetry of $SO(N)_{K}$.  For instance if $N$ and $K$ are even we deduce that $O(N)^{0}_{K,0},O(N)^{1}_{K,0}$ and $Pin^{\pm}(N)_{K,0}$ must have quantum $\mathbb{Z}_{2}$ one-form global symmetries.  The exact form of the resulting one-form global symmetry group is either $\mathbb{Z}_{2}\times \mathbb{Z}_{2}$ or $\mathbb{Z}_{4}$ depending on the level.  Some useful explicit examples are ($N$ even):
\begin{eqnarray}
\label{Opm1form}
\text{one-form symmetry}\left(O(N)^{0}_{K,0}\right)& = &\begin{cases} \mathbb{Z}_{2}\times \mathbb{Z}_{2} & K=0~ (\text{mod}~4) \\ \mathbb{Z}_{4} &  K=2~ (\text{mod}~4)  ~,\end{cases}\\
\text{one-form symmetry}\left(O(N)^{1}_{K,0}\right)& = &\begin{cases} \mathbb{Z}_{4} & N+K=0~ (\text{mod}~4) \\  \mathbb{Z}_{2}\times \mathbb{Z}_{2} &  N+K=2~ (\text{mod}~4) ~. \end{cases} \nonumber
\end{eqnarray}

We can also easily add the $\mathbb{Z}_{2}$ discrete $\theta$-parameter to the discussion.  Instead of viewing the $\theta$-parameter as an added $Lf(w_1)$ in the Lagrangian, as we did above, we use \eqref{Opm1form}.   The theory $(\mathbb{Z}_{2})_{L}$ has one-form global symmetry $\mathbb{Z}_{2}$ if $L$ is odd, $\mathbb{Z}_{2}\times \mathbb{Z}_{2}$ if $L=0$ mod 4, and $\mathbb{Z}_4$ if $L=2$ mod 4.  Then if $G$ is any gauge group discussed in section \ref{gaugegrps}, we can express the models with discrete $\theta$-parameter as
\begin{equation}\label{eqn:GKL}
G_{K,L}=\frac{G_{K,0}\times (\mathbb{Z}_{2})_{L}}{\mathbb{Z}_{2}}~.
\end{equation}
In this expression we deviate from our standard notation of labeling the TQFT by the gauge group of the Chern-Simons theory.  Instead, here the quotient means that we gauge a $\mathbb{Z}_{2}$ one-form global symmetry of the TQFT in the numerator.  This $\mathbb{Z}_2$ one-form symmetry is generated by a product of a Wilson line in $G_{K,0}$ and the electric Wilson line in the $(\mathbb{Z}_{2})_L$ factor.
The remaining one-form symmetry after gauging is spanned by the subset of Abelian anyons in $G_{K,0}\times (\mathbb{Z}_{2})_{L}$ that are uncharged under the gauged $\mathbb{Z}_2$, with the identification of lines that differ by fusion with the $\mathbb{Z}_{2}$ generator.\footnote{
In the case where the generating line of $\mathbb{Z}_2$ has half-integral spin, we do not identify lines that differ by fusion with the one-form symmetry generator since they have different spin and can be distinguished.  However, the difference between any two such lines is the transparent spin one-half line.  Since the transparent line is local with respect to all lines in the theory, it does not generate a non-trivial one-form global symmetry and thus does not affect the resulting one-form symmetry in the quotient.
Further discussion of the spectrum of lines after gauging a one-form symmetry is presented in section \ref{chiral}.
}

In any theory with one-form global symmetry we may activate background two-form gauge fields.  For instance, consider any gauge group that admits bundles with non-trivial $w_{1}.$   If we insert a line defect, then the integral of $w_{1}$ around a cycle linking the line measures a one-form charge.  We may refine our observables by coupling this global symmetry to background fields by adding to the action $\pi \int_{X}w_{1}\cup B_2$ where $B_2$ is a $\mathbb{Z}_{2}$ two-form gauge field.

\subsection{'t Hooft Anomalies of the Global Symmetries}

The  global symmetries described in the previous sections participate in 't Hooft anomalies.  We present the anomalies of $SO(N)_{K}$ with $N$ and $K$ even; those of other models may be deduced by gauging.  

As described in section \ref{ordinarysym}, there is a $\mathbb{Z}_{2}\times \mathbb{Z}_{2}$ global symmetry generated by $\mathcal{M}$ and $\mathcal{C}$.  These zero-form symmetries do not have any anomalies amongst themselves.  Indeed, they may clearly be gauged.

Next consider the possible anomalies involving only one-form symmetries.   The generating line of the $\mathbb{Z}_{2}$ one-form global symmetry has spin $NK/16$.  Since we are considering spin theories, this symmetry may be gauged if the spin is half-integral.  Thus in general there is an anomaly, which may be represented by a classical action on a four-manifold $Y$ with boundary $X$.  This action is written in terms of the background two-form gauge field $B_2$ as \cite{Kapustin:2014gua}
\begin{equation}\label{eqn:oneformself}
\frac{\pi NK}{4}\int_{Y}\frac{ \mathcal{P}(B_2)}{2}~,
\end{equation}
where $\mathcal{P}: H^{2}(Y,\mathbb{Z}_{2})\rightarrow H^{4}(Y,\mathbb{Z}_{4})$ is the Pontryagin square, and we have used the fact that on a spin four-manifold $Y$, $\mathcal{P}(B_2)$ is divisible by two.

Finally, let us describe the mixed anomalies between the zero-form and one-form global symmetries.  These are encoded in the following four-dimensional action:
\begin{equation}
\label{onezeroanom}
\pi \int_{Y} B_2 \cup\left[\frac{K}{2}B^{\mathcal{C}}\cup B^{\mathcal{C}}+\frac{N}{2}B^{\mathcal{M}}\cup B^{\mathcal{M}} +B^{\mathcal{C}}\cup B^{\mathcal{M}}\right]~.
\end{equation}
To derive the formula \eqref{onezeroanom}, we must consider the one-form symmetry after gauging any of the ordinary global symmetries.  For instance, consider gauging $\mathcal{M}$ with trivial background $B_{2}$ and $B^{\mathcal{C}}$.  Since the anomaly action \eqref{onezeroanom} vanishes, there is no obstruction to this process.  Now suppose that we turn on a background two-form $B_{2}$.  In this case \eqref{onezeroanom} is generally non-zero if $N=2~(\text{mod}~4)$. This means that the allowed class of background $B_{2}$ is restricted.

To deduce the allowed backgrounds, let $\beta: H^{j}(Y,\mathbb{Z}_{2})\rightarrow H^{j+1}(Y,\mathbb{Z}_{2})$ be the Bockstein homomorphism.  Then using the fact that $\beta$ obeys a Leibniz rule we can reexpress the anomaly involving $B_{2}$ and $B^{\mathcal{M}}$ as
\begin{equation}
\label{onezeroanom1}
\frac{\pi N}{2} \int_{Y} B_2 \cup B^{\mathcal{M}}\cup B^{\mathcal{M}}=\frac{\pi N}{2} \int_{Y} B_2 \cup \beta( B^{\mathcal{M}})=\frac{\pi N}{2} \int_{Y} \beta(B_2) \cup B^{\mathcal{M}}~.
\end{equation}
Thus the allowed class of two-form backgrounds are those where $\beta(B_{2})$ vanishes.  The Bockstein map is the obstruction to lifting the class $B_{2}$ from a $\mathbb{Z}_{2}$ gauge field to a $\mathbb{Z}_{4}$ gauge field.   Therefore, we deduce that after gauging $B^{\mathcal{M}}$ our topological field theory can naturally couple to a $\mathbb{Z}_{4}$ two-form gauge field.

We will now show that this $\mathbb{Z}_{4}$ is the one-form global symmetry after gauging and that the $\mathbb{Z}_{4}$ two-form gauge field is a background field for this symmetry.  When we promote $B^{\mathcal{M}}$ to a dynamical field there is an emergent $\mathbb{Z}_{2}$ one-form global symmetry.  This emergent symmetry couples to a background field $B_2'$  by modifying the action to include a term $\pi\int_X B_2'\cup B^{\cal M}.$  Let us consider the behavior of the action, including the coupling to $B_{2}'$, under gauge transformation of $B^{\cal M}$. Since $B^{\cal M}$ is now a dynamical field the total action must be invariant under any such transformation, i.e.\ the anomalous transformation of the action under $B^{\cal M}$ gauge transformations must cancel.  This is achieved by imposing the following relation (the symbol $\delta$ is the coboundary in cohomology) 
\begin{equation}
\delta B_{2}'=\beta(B_{2})~. \label{extensionform}
\end{equation}

Note that in theories without a mixed anomaly \eqref{onezeroanom1}, we require that $B_{2}'$ is closed.  The modification \eqref{extensionform} cancels the anomalous variation of the action and ensures that $B^{\mathcal{M}}$ can be consistently gauged.
Observe also that the left-hand-side of \eqref{extensionform} is trivial in cohomology ensuring that $B_{2}$ may be extended to a class with $\mathbb{Z}_{4}$ coefficients.  Let $\tilde{B}_{2}$ be any $\mathbb{Z}_{4}$ cochain extending $B_{2}$.  We construct a $\mathbb{Z}_{4}$ cocycle by the combination $\tilde{B}_{2}+2B_{2}',$ which is closed by \eqref{extensionform}.  In particular, the $\mathbb{Z}_{2}$  emergent one-form symmetry that couples to $B_{2}'$ is the $\mathbb{Z}_{2}$ subgroup of this $\mathbb{Z}_{4}$.  Therefore we conclude that the one-form global symmetry after gauging is extended to $\mathbb{Z}_{4}.$\footnote{A similar analysis shows that if we gauge the one-form symmetry, the mixed anomaly \eqref{onezeroanom1} forces the magnetic symmetry of $SO(N)/\mathbb{Z}_{2}$ to be extended to $\mathbb{Z}_4$.}

This matches with the expected one-form global symmetry.  Indeed when the anomaly is non-trivial i.e.\ $N=2~(\text{mod}~4),$ the one-form symmetry of $Spin(N)_{K}$ is $\mathbb{Z}_{4}$  (see table \ref{center}). 

Similarly, one can deduce the other terms in \eqref{onezeroanom} by gauging either $\mathcal{C}$ or $\mathcal{CM}$ and using the one-form symmetries of $O(N)^{0}_{K,0},O(N)^{1}_{K,0}$ given in \eqref{Opm1form}.
This interplay between mixed `t Hooft anomalies and extensions of the global symmetry is analogous to the discussion in \cite{Gaiotto:2017yup} involving CP symmetry and the one-form symmetry.

\subsection{Chiral Algebras}
\label{chiral}

The properties of the Chern-Simons theories described in previous sections are also encoded in their associated chiral algebras.  We discuss examples in section \ref{sec:examples} and appendix \ref{so2so4sec}.

The most familiar case is that of the simply connected group associated to the algebra $\frak{so}(N)$, i.e.\  $Spin(N)_{K}$.  In this case the associated chiral algebra is the (enveloping algebra of the) Kac-Moody algebra of currents.

To obtain the chiral algebra of $SO(N)_{K}$ from that of $Spin(N)_{K}$ we must extend the chiral algebra \cite{Moore:1988ss}.  This is the chiral algebra description of gauging a one-form global symmetry in the associated Chern-Simons theory \cite{Moore:1989yh}.  Let $a$ denote the generator of the one-form global symmetry, and assume that $a$ has integer spin.  In the Chern-Simons theory gauging $a$ has the following effects on the spectrum of lines \cite{Moore:1988ss,Moore:1989yh,Bais:2008ni}:
\begin{itemize}
\item We exclude from the spectrum all lines that carry non-trivial charge under $a$.
\item We identify lines $b$ and $a\cdot b$ that differ by fusion with $a$.
\item Lines $b$ that are stabilized under fusion (i.e.\ $b=a\cdot b$) are taken as several distinct lines in the spectrum of the gauged theory. 
\end{itemize}
The associated operation on chiral algebras is extension.  The chiral algebra is enlarged to include the current associated to the line $a$, forcing all allowed modules to be local with respect to these additional currents.  The modules of the extended chiral algebra are thus enlarged to include the action of modes of $a$.  Finally, those modules of the original theory, which are stabilized under the action of the $a$ modes may now be assigned a phase under the $a$ action and hence give rise to several distinct modules of the extended chiral algebra.

If $a$ has half-integral spin, we still exclude all lines that carry non-trivial charge under $a$, but we do not identify $b$ with $a\cdot b$ since their spins differ by a half integer, and hence they are always distinct.

We may apply this general procedure to obtain the chiral algebra of $SO(N)_{K}$.  The extending representation is the transparent line  of spin $K/2$ in the $K$-th symmetric traceless power of the vector representation (Dynkin labels $(K,0,\cdots,0)$).  The magnetic symmetry $\mathcal{M}$ acts to exchange the lines that are doubled under this extension. 

To obtain the chiral algebras for other gauge groups we can apply the chiral algebra analogue of gauging a zero-form global symmetry.  This is an orbifold \cite{Moore:1989yh}.  The zero-form symmetries $\mathcal{C}$ and $\mathcal{M}$ act on the $SO(N)_{K}$ chiral algebra as automorphisms. 
In particular, for even $K$ the magnetic symmetry ${\cal M}$ permutes the two distinct modules in each pair $b_\pm$ satisfying $b\cdot a=b$.
By suitably introducing twisted sectors \cite{Dijkgraaf:1989hb} for either $\mathcal{C}$ or $\mathcal{CM}$ we obtain the chiral algebra of $O(N)^{0}_{K,0}$ and $O(N)^{1}_{K,0}$.
The chiral algebras of $O(N)^{0}_{K,L}$ and $O(N)^{1}_{K,L}$ can then be constructed as in \eqref{eqn:GKL}.

\section{Level-Rank Duality}
\label{LRDsec}
In this section we derive level-rank dualities for Chern-Simons theories with gauge group based on the Lie algebra $\frak{so}(N)$.  

One way to phrase our result is to consider $SO(N)_{K}$ coupled to background $\mathbb{Z}_{2}\times \mathbb{Z}_{2}$ gauge fields $B^{\mathcal{C}}$ and $B^{\mathcal{M}}$ for the global symmetry $\mathcal{C}$ and $\mathcal{M}$.  The level-rank duality then states how the correlation functions in the presence of background fields are related including the required map on counterterms for these gauge fields. Let us denote by $SO(N)_{K}[B^{\mathcal{C}},B^{\mathcal{M}}]$, the $SO(N)_{K}$ topological field theory coupled to background fields.  Then, our explicit duality is
\begin{equation}
\label{lrct}
SO(N)_{K}[B^{\mathcal{C}},B^{\mathcal{M}}]\quad\longleftrightarrow\quad
SO(K)_{-N}[B^{\mathcal{M}},B^{\mathcal{C}}]-(K-1)f[B^{\mathcal{C}}]-(N-1)f[B^{\mathcal{M}}]-f[B^{\mathcal{C}}+B^{\mathcal{M}}]~.
\end{equation}
A special case of this result, with $K=N=2$ so that both theories are Abelian, was discussed in \cite{Seiberg:2016rsg}.

Note crucially that under level-rank duality the symmetries $\mathcal{C}$ and $\mathcal{M}$ (and hence also their background fields) are exchanged \cite{Aharony:2016jvv}.  We explain this feature in more detail in section \ref{sec:embed} below.  

Starting from this result we may add any desired counterterms and gauge the global symmetries to obtain a host of other level-rank dualities.  Several examples are
\begin{eqnarray}
\label{lrgauged}
O(N)^{0}_{K,K} &\quad\longleftrightarrow&\quad Spin(K)_{-N}~, \\
O(N)^{1}_{K,K-1+L} &\quad\longleftrightarrow&\quad O(K)^{1}_{-N,-N+1+L}~. \nonumber 
\end{eqnarray}
Note that these are dualities among spin TQFTs, thus keeping with our earlier convention, we add the transparent spin $1/2$ line $\psi$ to the spectrum if required.  Without this addition the dualities are in general false.  Additionally, we observe that the second duality in \eqref{lrgauged} holds for arbitrary $L$.  This is possible since the symmetry $\mathcal{CM}$ is mapped to itself under level-rank duality.

We claim that the dualities \eqref{lrct} and \eqref{lrgauged} hold for all $N$ and $K.$ In order to make sense of this for either $N$ or $K$ odd, where some of the global symmetries do not act on $SO(N)_{K}$ we use our general discussion in section \ref{ordinarysym} to define $\mathcal{C}$ and $\mathcal{M}$.  This implicitly defines our conventions for $O(N)^{0}_{K,L}$ and $O(N)^{1}_{K,L}$ for odd $N$ (where the group is a product $O(N)=SO(N)\times \mathbb{Z}_{2}$) by starting from $SO(N)_{K}$ with counterterms and gauging.

In the remainder of this section we prove the dualities \eqref{lrgauged}.  This also establishes the level-rank duality \eqref{lrct} of $SO(N)_{K}$ coupled to background fields.

\subsection{Conformal Embeddings and Non-Spin Dualities}
\label{sec:embed}

Our strategy for deriving the result \eqref{lrgauged} will be to bootstrap our way up starting from the level-rank dualities that arise for non-spin theories, and then subsequently generalize to the spin theories of interest.  Thus in this section, unlike others, we discuss non-spin TQFTs and their associated chiral algebras (i.e. we do not tensor with the $\{1, \psi\}$ sector). 

The basic tool is the conformal embedding of non-spin chiral algebras.  Consider $NK$ real two-dimensional fermions $\zeta^{\alpha}_{a}$ where $\alpha=1,\cdots, N$ and $a=1,\cdots K$.  These generate the current algebra $Spin(NK)_{1}$.  By forming singlets out of the $a$ or $\alpha$ indices, we can generate various currents.  The first are
\begin{equation}
J^{[\alpha\beta]}=\delta^{ab}\zeta^{\alpha}_{a}\zeta^{\beta}_{b}~,\hspace{.5in}J_{[ab]}=\delta_{\alpha\beta}\zeta^{\alpha}_{a}\zeta^{\beta}_{b}~, \label{prodcur}
\end{equation}
which generate Kac-Moody algebras for $Spin(N)_{K},$ and $Spin(K)_{N}$ respectively.  Other objects of interest are: 
\begin{equation}
\Lambda^{(\alpha_{1}\alpha_{2}\cdots \alpha_{K})}=\varepsilon^{a_{1}a_{2}\cdots a_{K}}\zeta^{\alpha_{1}}_{a_{1}}\zeta^{\alpha_{2}}_{a_{2}}\cdots \zeta^{\alpha_{K}}_{a_{K}}~,\hspace{.5in}\Lambda_{(a_{1}a_{2}\cdots a_{N})}=\varepsilon_{\alpha_{1}\alpha_{2}\cdots \alpha_{N}}\zeta^{\alpha_{1}}_{a_{1}}\zeta^{\alpha_{2}}_{a_{2}}\cdots \zeta^{\alpha_{N}}_{a_{N}}~.
\end{equation}
If $K$ is even, $\Lambda^{(\alpha_{1}\alpha_{2}\cdots \alpha_{K})}$ has integer dimension and can be used to extend the chiral algebra from $Spin(N)_{K}$ to $SO(N)_{K}.$  Similarly if $N$ is even $\Lambda_{(a_{1}a_{2}\cdots a_{N})}$ can be used to extend the chiral algebra to $SO(K)_{N}.$ (See section \ref{chiral}.)

Thus we deduce the following conformal embeddings of chiral algebras (see also \cite{Hasegawa:1989741,Verstegen:1990at}):
\begin{eqnarray}
SO(N)_{K}\times SO(K)_{N}&\subset& Spin(NK)_{1}~~~~~N~\text{even},K~ \text{even}~, \nonumber \\ 
Spin(N)_{K}\times SO(K)_{N}&\subset& Spin(NK)_{1}~~~~~N~\text{even},K~ \text{odd}~, \label{confembedding}\\ 
Spin(N)_{K}\times Spin(K)_{N}&\subset& Spin(NK)_{1}~~~~~N~\text{odd},K~ \text{odd}~, \nonumber
\end{eqnarray}
Note that in the above embeddings, the subalgebras on the left are not in general maximal, i.e.\ depending on $N$ and $K,$ there may be strictly larger chiral algebras that embed inside $Spin(NK)_{1}$.  In particular, the centers on the left and right of \eqref{confembedding} need not agree.  

A related point is that in general there are modules of the subalgebras that do not occur in the decomposition of any module of $Spin(NK)_{1}$.  One might attempt to remedy this by quotienting the groups on the left by elements of their center, but in general there is no quotient that imposes the required selection rules.\footnote{For instance, consider a primary of spins $(1,1,1,1)$ for the group $Spin(4)_{4}\times Spin(4)_{4}\cong SU(2)_{4}\times SU(2)_{4}\times SU(2)_{4}\times SU(2)_{4} \subset Spin(16)_{1}.$  There is no quotient that can remove this representation, and it never occurs in a module of $Spin(16)_{1}$. }  A consequence of this is that in general there is no map between the simple currents on the left and right of the conformal embeddings \eqref{confembedding}.  Indeed, since there are more modules of the subalgebra, the simple currents of the subalgebra need not act faithfully on the subset of modules that arise via decomposition of modules of the larger algebra.  Similarly, the simple currents of the larger algebra may fail to be local with respect to modules of the subalgebra that do not arise from the decomposition.\footnote{This corrects some misleading comments about the centers of conformal embeddings in \cite{Aharony:2016jvv}.}

Although the centers in the conformal embeddings \eqref{confembedding} are not in general matched, it is crucial for the derivation of level-rank duality that each factor of the subalgebra acts faithfully.  Thus, every module of one of the factors occurs in some decomposition of a module of $Spin(NK)_{1}.$ 

For our application, we aim to gauge some of the zero-form symmetries in \eqref{confembedding}.  Thus, we must understand their action in more detail.  For concreteness, we focus on the case $N$ and $K$ both even where $\mathcal{C}$ and $\mathcal{M}$ act as true symmetries of the topological field theory that permute anyons.  The required generalizations of level-rank duality for odd $N$ or $K$ are discussed in Appendix \ref{derlrdualodd}.

Let us explain why $\mathcal{C}$ and $\mathcal{M}$ are exchanged via level-rank duality.  We will do this by direct inspection of the representations of the chiral algebras.

The representations of the group $SO(N)$ are labelled by Young tableau, $\mathcal{Y}$, built from the vector representation.  Such a tableau is specified by a tuple $(l_{1}, \cdots, l_{N/2})$ of row lengths (see Appendix \ref{dynkin} for conventions).  The $l_{i}$ for $i<N/2$ are non-negative integers.  The final element $l_{N/2}$ is an integer of either sign accounting for the two chiralities of spinors.\footnote{When drawing such a tableau one typically takes the final row length to be $l_{N/2}$ and adds an extra sign label.}

When we discuss modules of the chiral algebra $SO(N)_{K}$ we must take into account the fact that the chiral algebra of $SO(N)_{K}$ is larger than that of $Spin(N)_{K}$ (see section \ref{chiral}).  The extending representation is a current in the $K$-th symmetric power of the vector representation, which is described by a tableau with $(l_{1}, l_{2}, \cdots, l_{N/2})=(K,0,\cdots,0).$  This means that as modules of $SO(N)_{K}$ there is the following spectral flow identification \cite{Moore:1988ss, Moore:1989yh}:
\begin{equation}
(l_{1}, l_{2}, \cdots, l_{N/2-1},l_{N/2}) \sim (K-l_{1}, l_{2}, \cdots, l_{N/2-1}, -l_{N/2})~. \label{spectralflow}
\end{equation}
Using this identification we may restrict to those primaries with $l_{1}\leq K/2$.  

We can also explicitly see the primaries of representations that are doubled due to the extension.  A representation is fixed under spectral flow if $l_{1}=K/2$ and $l_{N/2}=0$.  Each such representation gives rise to two modules of the chiral algebra $SO(N)_{K}$.  They are permuted by the magnetic symmetry $\mathcal{M}$.  The charge conjugation symmetry $\mathcal{C}$ is also visible, it simply acts on the final label by $l_{N/2}\rightarrow -l_{N/2}$.

Next let us inspect the modules of the chiral algebra $Spin(NK)_{1}$.  Since $NK$ is even, there are four distinct representations: $1, \zeta, \sigma_{\pm}$.  We ask how these modules decompose under the embedded chiral algebra $SO(N)_{K}\times SO(K)_{N}.$ The current  generating $Spin(NK)_{1}$ is a general bilinear in the fermions $\zeta$.  As a representation of $SO(N)\times SO(K)$ it decomposes as:\footnote{Here $\yng(2)$ means the symmetric tensor with the trace removed. }
\begin{equation}
J_{Spin(NK)_{1}}=\zeta^{\alpha}_{a}\zeta^{\beta}_{b}=\left(\young( \alpha \beta)~,\young(a,b)\right) \oplus \left(\young(\alpha,\beta)~,\young(ab)\right) \oplus \left(1~,\young(a,b)\right) \oplus \left(\young(\alpha,\beta)~,1\right)~.
\end{equation}
The last two terms in the direct sum above are simply the currents of the embedded chiral algebra identified in \eqref{prodcur}.  The first two terms are new primaries that appear in the decomposition of the identity under the subalgebra $SO(N)_{K}\times SO(K)_{N}$.  Notice that these primaries  are labelled by pairs of tableaux each with an even number of boxes that are related by transposition.  This remains true for all other primaries obtained from this decomposition.  Indeed, the current algebra primaries of each of the subalgebra factors do not involve derivatives of the fermions $\zeta^{\alpha}_{a}$ and hence are subject to Fermi statistics.  This implies that whenever two upper indices $\alpha,\beta$ are symmetrized, the corresponding lower indices $a,b$ must be anti-symmetrized and vice versa.

Based on this logic, we can readily deduce properties of decompositions of modules of  $Spin(NK)_{1}$ under $SO(N)_{K}\times SO(K)_{N}.$ We have 
\begin{eqnarray}
1_{Spin(NK)_{1}}&\longrightarrow &\bigoplus(\mathcal{Y}_{\text{even}},\mathcal{Y}^{T}_{\text{even}})~, \label{youngdecomp}\\
\zeta_{Spin(NK)_{1}}&\longrightarrow &\bigoplus(\mathcal{Y}_{\text{odd}},\mathcal{Y}^{T}_{\text{odd}})~, \nonumber
\end{eqnarray}
where the subscript on the tableau indicates the parity of the number of boxes.  
Similarly, we can also deduce properties of the decomposition of the modules $\sigma_{+}$ and $\sigma_-=\sigma_+\cdot \zeta$.
In fact, at least one of them contains the simple current of $SO(N)_{K}$ or $SO(K)_{N}$ \cite{Hasegawa:1989741,Verstegen:1990at}, and thus the decompositions can be obtained by fusion with the right-hand-sides of \eqref{youngdecomp}.

The fact that Young tableaux are paired with their transpose in the decomposition of modules of $Spin(NK)_{1}$ clarifies the map of global symmetries $\mathcal{C} \leftrightarrow \mathcal{M}$ under level-rank duality.  The basic point is simply that if $\mathcal{Y}$ is any tableau which is acted on non-trivially by $\mathcal{C}$ then $\mathcal{M}$ acts non-trivially on $\mathcal{Y}^{T}.$  Indeed, $\mathcal{C}$ acts on a tableau if $l_{1}<K/2$ and $l_{N/2} \neq 0.$ When this is so, the transposed Young tableau has $l^{T}_{1}=N/2$ and $l_{K/2}^{T}=0.$  In particular such a representation is stabilized by the spectral flow \eqref{spectralflow} and hence is acted on by $\mathcal{M}$.

We are now equipped to discuss gauging the ordinary global symmetries in the chiral algebra embeddings.  Consider the symmetry that acts with $\mathcal{C}$ on the first factor and $\mathcal{M}$ on the second factor in \eqref{confembedding}.   It is achieved by an inner automorphism of $Spin(NK)_{1},$ which acts on the fermions $\zeta^{\alpha}_{a}$ by giving some of them periodic boundary conditions on the cylinder (compare to the definition \eqref{cdef})\footnote{Recall that when we study the RCFT on a cylinder the representations \eqref{youngdecomp} arise in the NS sector where the fermions $\zeta_a^\alpha$ are antiperiodic.}
\begin{equation}
\text{periodic:}~~\zeta^{1}_{a} ~, \hspace{.5in}\text{antiperiodic:}~~\zeta^{\alpha>1}_{a} ~. \label{twist1}
\end{equation}
Since we have mapped the symmetries between the two sides of the conformal embedding we can gauge to produce new embeddings.  Gauging the symmetry $(\mathcal{C}, \mathcal{M})$ is equivalent to an inner-automorphism orbifold of $Spin(NK)_{1}$ with the twists \eqref{twist1}.  This leads to the conformal embedding:
\begin{equation}
O(N)_{K,0}^{0}\times Spin(K)_{N}\subset Spin(K)_{1}\times Spin(NK-K)_{1}~. \label{Opembed}
\end{equation}

Similarly we can gauge the symmetry $(\mathcal{CM}, \mathcal{CM})$.  The related inner automorphism of $Spin(NK)_{1}$ now acts on the fermions as
\begin{equation}
\text{periodic:}~~\zeta^{1}_{a>1}~, \zeta^{\alpha>1}_{1}~, \hspace{.5in}\text{antiperiodic:}~~\zeta^{\alpha>1}_{a>1}~, \zeta^{1}_{1} ~.
\end{equation}
Therefore we deduce the conformal embedding\footnote{The special case $K=2$ or $N=2$ coincides with (\ref{Opembed}) since $O(2)^0_K\leftrightarrow O(2)^1_K$, $O(N)^1_2\leftrightarrow Spin(N)_2$. This uses the fact that  ${\cal M}$ and ${\cal C}$ are not outer automorphisms of $SO(2)_K$ and $SO(N)_2$ respectively.
}
\begin{equation}
O(N)_{K,0}^{1}\times O(K)^{1}_{N,0}\subset Spin(N+K-2)_{1}\times Spin(NK-K-N+2)_{1}~.\label{Omembed}
\end{equation}

The embeddings \eqref{Opembed}-\eqref{Omembed} allow us to obtain dualities among non-spin Chern-Simons theories.  First we have the following equivalences of chiral algebras\footnote{The case $N=2$ of the first equation was studied in \cite{Ginsparg:1988ui}.}
\begin{eqnarray}
O(N)_{K,0}^{0} &\cong& \frac{Spin(K)_{1}\times Spin(NK-K)_{1}}{ Spin(K)_{N}}~,\label{eqn:equivnonspinalg}\\
O(N)_{K,0}^{1} &\cong&\frac{Spin(N+K-2)_{1}\times Spin(NK-N-K-2)_{1}}{O(K)^{1}_{N,0}}~, \nonumber
\end{eqnarray}
where on the right-hand side above the chiral algebras are GKO cosets.   Here we make use of the fact mentioned above that in the conformal embeddings \eqref{confembedding} each factor of the subalgebra acts faithfully.  Thus every module of the left-hand side of \eqref{eqn:equivnonspinalg} is an allowed representation of the coset.  In particular this means that the simple currents must match between the left and right of \eqref{eqn:equivnonspinalg}, since if they did not, some modules would necessarily be forbidden.

From (\ref{eqn:equivnonspinalg}) we obtain the Chern-Simons dualities
\begin{eqnarray}
O(N)_{K,0}^{0} &\quad\longleftrightarrow&\quad \frac{Spin(K)_{-N}\times Spin(NK-K)_{1}\times Spin(K)_{1}}{ \mathcal{Z}}~, \label{nonspineven} \\
O(N)_{K,0}^{1}&\quad\longleftrightarrow&\quad \frac{O(K)^{1}_{-N,0}\times Spin(N+K-2)_{1}\times Spin(NK-N-K-2)_{1}}{\mathcal{Z'}}~, \nonumber
\end{eqnarray}
where the common one-form symmetries $\mathcal{Z}$ and $\mathcal{Z}'$ are
\begin{equation}
\mathcal{Z}= \begin{cases} \mathbb{Z}_{2}\times \mathbb{Z}_{2} & K=0~(\text{mod}~4)~,\\ \mathbb{Z}_{4}& K=2~(\text{mod}~4)~,\end{cases} \hspace{.5in} \mathcal{Z}'= \begin{cases} \mathbb{Z}_{4} & N+K=0~(\text{mod}~4)~,\\ \mathbb{Z}_{2}\times \mathbb{Z}_{2}& N+K=2~(\text{mod}~4)~.\end{cases}
\end{equation}
In these equations the $\mathbb{Z}_{2}\times \mathbb{Z}_{2}$ quotients use the two generators of the one-form symmetry $\chi$ and $j$ from each factor in the numerator of \eqref{nonspineven}.   In the $\mathbb{Z}_{4}$ case the quotient uses the generator $j$ (with $\chi=j^{2}$) from each factor.\footnote{\label{foot}The Abelian anyon $\chi$ is the line in $Spin(K)_{-N}$ transforming in the $N$-th symmetric power of the vector representation, i.e.\ Dynkin indices $(N,0,\cdots,0)$.  Similarly $j$ transforms as a power of the spinor representation with Dynkin indices $(0, \cdots, 0,N)$.} The consistency of the quotient depends on the spin of these generators being integral.  These spins are the sum of the contributions from each factor in \eqref{nonspineven}.    (In checking this we use the fact that the spins of $\chi$ and $j$ in the first factors are $0$ and $-\frac{NK}{16}$.)   We also require that $\chi$ and $j$ have trivial mutual braiding so that there is no anomaly \cite{Gaiotto:2014kfa}. 

We stress that these dualities, and the associated chiral algebra isomorphisms can be rigorously proven. 

\subsection{Level-Rank Duality for Spin Chern-Simons Theory}

We now promote the non-spin dualities \eqref{nonspineven} to spin dualities and simplify.  Thus in this section we restore our convention that all theories are spin TQFTs.

The first step is to use the duality $Spin(L)_1\leftrightarrow (\mathbb{Z}_2)_{-L}$ discussed in appendix \ref{z2appendix}.  This gives
\begin{eqnarray}
O(N)^0_{K,0} &\longleftrightarrow& 
{Spin(K)_{-N}\times (\mathbb{Z}_2)_{-K} \times (\mathbb{Z}_2)_{-NK+K}\over {\cal Z}} \label{spineven} \\
O(N)^1_{K,0}&\longleftrightarrow&
{O(K)^1_{-N,0}\times (\mathbb{Z}_2)_{-(N+K-2)} \times (\mathbb{Z}_2)_{-NK+N+K-2}  \over {\cal Z}'}~.\nonumber
\end{eqnarray}
When the common one-form symmetry is  $\mathbb{Z}_4$ the quotient is generated by a product of the Wilson line $j$ in the continuous factor and the basic magnetic line in each $\mathbb{Z}_2$ factor.  When the common one-form symmetry is $\mathbb{Z}_2\times\mathbb{Z}_2$, one $\mathbb{Z}_{2}$ is generated by a product of the Wilson line $j$ in the continuous factor and the basic magnetic line in each $\mathbb{Z}_2$ factor.  The quotient by this $\mathbb{Z}_{2}$ may be viewed as a quotient on the gauge group in the numerator of \eqref{spineven}.  However, the second $\mathbb{Z}_{2}$ is generated by a product of Wilson lines, $\chi$ from the continuous group and the electric Wilson line in the two $\mathbb{Z}_2$ groups.  The quotient by this $\mathbb{Z}_{2}$ is not a quotient of the gauge group, but of the TQFT and thus the above deviates from our convention of labeling TQFTs by the Chern-Simons gauge groups.  (See the  discussion around \eqref{eqn:GKL}.)

When $j$ generates a $\mathbb{Z}_2$ one-form symmetry, we have (see Appendix \ref{dualityoneform} for proof)
\begin{eqnarray}
Spin(K)_{-N} &\quad\longleftrightarrow&\quad {Spin(K)_{-N}\times (\mathbb{Z}_2)_{-NK}\over \mathbb{Z}_2}~, \label{oneformztwo}\\
O(K)^1_{-N,0} &\quad\longleftrightarrow&\quad {O(K)^1_{-N,0}\times (\mathbb{Z}_2)_{-NK}\over \mathbb{Z}_2}~, \nonumber
\end{eqnarray}
where the quotient on $(\mathbb{Z}_2)_{-NK}$ uses the basic magnetic line.  When $j$ generates a $\mathbb{Z}_4$ one-form symmetry (with $j^2=\chi$) we instead find
\begin{eqnarray}
Spin(K)_{-N} &\quad\longleftrightarrow&\quad {Spin(K)_{-N}\times (Z_4)_{-2NK}\over \mathbb{Z}_4}~,\label{oneformzfour} \\
O(K)^1_{-N,0} &\quad\longleftrightarrow&\quad {O(K)^1_{-N,0}\times (Z_4)_{-2NK}\over \mathbb{Z}_4}~, \nonumber
\end{eqnarray}
where $(Z_4)_M$ is the Abelian Chern-Simons theory $(M/ 4\pi)udu+(4/2\pi)udv$ with $U(1)$ gauge fields $u,v$, and the quotient on $(Z_4)_{-2NK}$ uses the basic magnetic line  (i.e.\ $\exp(i\oint v)$). (Note that the $Z_{4}$ level is that defined by the Abelian Chern-Simons theory, which differs by a factor of two from our conventions for the $\mathbb{Z}_{2}$ level.  In order to remind us of this difference we change the fonts, e.g.\ $\mathbb{Z}_{2}$ vs.\ $Z_{4}$).

We next use the following Abelian Chern-Simons dualities:
\begin{eqnarray}
(\mathbb{Z}_2)_{4M}\times (\mathbb{Z}_2)_{2L}&\quad\longleftrightarrow&\quad (\mathbb{Z}_2)_{4M-2L}\times (\mathbb{Z}_2)_{2L}\quad {\rm even}\ L~\label{CSuone} \\ 
{(Z_4)_{8M}\times (\mathbb{Z}_2)_{2L}\over \mathbb{Z}_2}&\quad\longleftrightarrow&\quad (\mathbb{Z}_2)_{4M-2L}\times (\mathbb{Z}_2)_{2L}\quad {\rm odd}\ L~,
\end{eqnarray}
where the quotient uses the product of the charge two magnetic line in $(Z_{4})_{8M}$ and the basic Wilson line in $(\mathbb{Z}_2)_{2L}$.  These may be established by changing variables.  To prove the first, write
\begin{equation}
{2M\over 4\pi}ada+{L\over 4\pi}bdb+{2\over 2\pi}adx+{2\over 2\pi}bdy
=
{2M-L\over 4\pi}a'da'+{L\over 4\pi}b'db'+{2\over 2\pi}a'dx'+{2\over 2\pi}b'dy'~, \label{CSuonepf}
\end{equation}
where $a'=a,b'=b+a,x'=x-y-(L/2)b,y'=y$. To prove the second line of \eqref{CSuone}, proceed as
\begin{equation}
{2M-L\over 4\pi}ada+{L\over 4\pi}bdb+{2\over 2\pi}adx+{2\over 2\pi}bdy
=
{8M\over 4\pi}a'da'+{L\over 4\pi}b'db'+{4\over 2\pi}a'dx'+{2\over 2\pi}b'dy'~, \label{CSuoneps}
\end{equation}
where $a=2a',b=b'+2a',x=x'-y'+La',y=y'-La'$.
The $\mathbb{Z}_2$ quotient is generated by the line $\exp{\left(-iL\oint a+2i\oint x\right)}=\exp{\left(2i\oint x'-2i\oint y'\right)}=\exp{\left(2i\oint x'+i\oint b'\right)}$.

Consider the first duality in \eqref{spineven}. We can simplify the right hand side as follows. Suppose $L=-K/2$ is even, we tensor the first duality in \eqref{oneformztwo} with $(\mathbb{Z}_2)_{2L}$ and take the $\mathbb{Z}_2$ quotient generated by the product of  $\chi$ and the basic Wilson line on the left. Using the change of variables \eqref{CSuone} for the the numerator on the right makes the quotient diagonal and we find the right hand side of \eqref{spineven} for the first duality. The left hand side is thus dual to $[Spin(K)_{-N}\times (\mathbb{Z}_2)_{2L}]/\mathbb{Z}_2$.

Similarly for odd $L=-K/2$ we tensor the first duality in \eqref{oneformzfour} with $(\mathbb{Z}_2)_{2L}$ and take the $\mathbb{Z}_2$ quotient generated by the product of  $\chi$ and the basic Wilson line on the left. Using the change of variables \eqref{CSuone} for the numerator on the right makes the quotient diagonal and we find the right hand side of \eqref{spineven} for the first duality. The left hand side is again dual to $[Spin(K)_{-N}\times (\mathbb{Z}_2)_{2L}]/\mathbb{Z}_2$.

Therefore we prove that for all even $N,K$ (rearranging $(\mathbb{Z}_2)_{2L}$),
\begin{equation}\label{spindualspinO}
O(N)^0_{K,K} \quad\longleftrightarrow\quad Spin(K)_{-N}~.
\end{equation}

By repeating the same steps with $L=-(N+K-2)/2$ we similarly establish
\begin{equation}\label{spindualOO}
O(N)^1_{K,K-1} \quad\longleftrightarrow\quad O(K)^1_{-N,-N+1}~.
\end{equation}

From the dualities (\ref{spindualspinO}) and (\ref{spindualOO}) we conclude the level-rank duality map of $SO$ spin Chern-Simons theories coupled to the backgrounds for ${\cal C,M}$ symmetries stated in \eqref{lrct}.

\subsection{Consistency Checks}
\label{sec:examples}

In this section we present a few simple consistency checks of the level-rank dualities stated in equation (\ref{lrgauged}).   In appendix \ref{so2so4sec} we also discuss various low-rank examples of chiral algebras involved in the dualities.

\subsubsection{Conformal Dimensions in $O(N)^0_{K,K}$ and $Spin(K)_{-N}$}

Consider the duality with even $N$. The lowest dimension twisted operator in the chiral algebra $O(N)^0_{K,K}$ is a product of the twist operator of $O(N)^0_K$ and the spinor of $Spin(K)_{-1}\leftrightarrow (\mathbb{Z}_{2})_{K}$. On the other side of duality the lowest dimension twisted operator in $Spin(K)_N$ is the spinor of Dynkin label $(0,\cdots,0,1)$.  We will compute their dimensions and demonstrate that they agree.

The conformal dimension of the twist operator $\sigma$ of $O(N)^0_K$ can be computed as follows.
Denote the Kac-Moody current by $J_{\mu\nu}$, the ${\cal C}$ symmetry changes the sign of $J_{1i}$ with $i=2,\cdots N$.
We evaluate the one-point function of the Sugawara stress tensor in the presence of the twist operator by expanding $J_{1i}$ in half-integral modes:
\begin{eqnarray}
\langle T(z)\rangle_{\rm twisted}&=&\lim_{w\to z}\langle {1\over 2(N+K-2)}\left(\sum_{m,m'\in\mathbb{Z},i,j=2,\cdots N,i<j}J^{m}_{ij}J^{m'}_{ij} z^{-m-1} w^{-m'-1}\right. \nonumber \\
&+&\left.\sum_{m,m'\in \mathbb{Z}+{1\over 2},l=2,\cdots N}J^{m}_{1l}J^{m'}_{1l} z^{-m-1} w^{-m'-1} \right)\rangle-{\rm divergence}\\
&=&{1\over 2(N+K-2)}(N-1)\lim_{w\to z}\left(K{\sqrt{z/w}+\sqrt{w/z}\over 2(z-w)^2}-{K\over (z-w)^2}\right) \nonumber\\
&=&{(N-1)K\over 16(N+K-2)z^2}~, \nonumber
\end{eqnarray}
where we used the central term in the current commutator $Km\delta_{m+m',0}$, thus for integral modes it vanishes after summing over positive and negative integers, while for half-integral modes it gives a non-trivial contribution.
Since $T(z)\sigma(0)|0\rangle\sim {h_\sigma(0)\sigma(0)\over z^2}|0\rangle+\cdots$ this determines the conformal weight of the twist operator $\sigma$ to be ${(N-1)K\over 16(N+K-2)}$. The same result was obtained in \cite{Ganor:2002ya}.

Using the fact that the spinor of $Spin(K)_{-1}$ has spin $-K/16$ we see that
\begin{eqnarray}
h[\sigma_{O(N)^0_{K,K}}] &=& {(N-1)K\over 16(N+K-2)}-{K\over 16} ={-K(K-1)\over 16(K+N-2)}~.
\end{eqnarray}
This agrees with the spin of the spinor representation of $Spin(K)_{-N}$, which can be computed from the Casimir (see e.g. \cite{di1997conformal} and appendix \ref{dynkin}).  Therefore we find the lines have the same spin as expected from the duality.

\subsubsection{Counting the Lines}

As another consistency check, we count the number of lines in each theory and show that they match in the level-rank dualities stated in equation \eqref{lrgauged}. We take even $N,K>2$.

The number of lines in $Spin(N)_K$ with even $N,K$ is the number of non-negative integral solutions for the affine Dynkin labels $\lambda_0+\lambda_1+\lambda_{{N\over 2}-1}+\lambda_{{N\over 2}}+2\sum_{i=2}^{N/2-2}\lambda_i=K$ where $\lambda_0$ is the label for the extended node in the affine Dynkin diagram.
The tensor representations have Dynkin labels $\lambda_{N/2-1},\lambda_{N/2}$ both even or both odd, otherwise the representation is spinorial.

The number of tensor and spinor representations are given by \begin{equation}\label{countSpin}
N_{\rm tensor}=
4\binom{(N+K-2)/2}{(K-2)/2}+\binom{(N+K-4)/2}{K/2},\quad
N_{\rm spinor}=
4\binom{(N+K-2)/2}{(K-2)/2}~.
\end{equation}

As discussed in section \ref{chiral}, the chiral algebra of $SO(N)_K$ is the chiral algebra of $Spin(N)_K$ extended by the representation of Dynkin label $(K,0,\cdots,0)$. The spinor representations are projected out and the tensor representations are identified with each other or doubled. For even $N,K$ the number of lines is
\begin{equation}\label{countSO}
\binom{(N+K-4)/2}{(N-2)/2}+\binom{(N+K-4)/2}{(K-2)/2}
+2\binom{(N+K-4)/2}{N/2}+2\binom{(N+K-4)/2}{K/2}~,
\end{equation}
which has $2\binom{(N+K-4)/2}{K/2}$ doubled representations permuted by the magnetic symmetry $\mathcal{M}$.

The chiral algebra $O(N)^0_{K,0}$ for even $N,K$ can be obtained from $SO(N)_K$ by an orbifold with the symmetry ${\cal C},$ which acts on the Dynkin labels as $\lambda_{N/2-1}\leftrightarrow\lambda_{N/2}$.
The number of untwisted and twisted primaries in the chiral algebra of $O(N)^0_K$ can be derived by \cite{Dijkgraaf:1989hb}:
\begin{equation}\label{countOplus}
N_{\rm untwisted}=
4\binom{(N+K-2)/2}{(N-2)/2}+\binom{(N+K-4)/2}{N/2},\quad
N_{\rm twisted}=
4\binom{(N+K-2)/2}{(N-2)/2}~.
\end{equation}

The chiral algebra $O(N)^1_{K,0}$ for even $N,K$ can be obtained from $SO(N)_K$ by orbifold with the symmetry ${\cal CM}$, which acts as $\lambda_{N/2-1}\leftrightarrow\lambda_{N/2}$, and permutes the two primaries in every doubled representations.
The number of untwisted and twisted primaries in the chiral algebra of $O(N)^1_{K,0}$ can be derived by \cite{Dijkgraaf:1989hb}:
\begin{eqnarray}\label{countOplus}
N_{\rm untwisted}&=&
\binom{(N+K)/2}{K/2}+\binom{(N+K-4)/2}{(N-2)/2}+\binom{(N+K-4)/2}{(K-2)/2}~,\\ \nonumber
N_{\rm twisted}&=&
2\binom{(N+K-4)/2}{(N-2)/2}+2\binom{(N+K-4)/2}{(K-2)/2}~.
\end{eqnarray}

Using the above formulas, one can verify that the number of lines match in the dualities
\begin{eqnarray}
SO(N)_K&\quad\longleftrightarrow&\quad SO(K)_{-N}~, \nonumber\\
O(N)^0_{K,K} &\quad\longleftrightarrow&\quad Spin(K)_{-N}~, \label{dualnumber}\\
O(N)^1_{K,K-1}&\quad\longleftrightarrow&\quad O(K)^1_{-N,-N+1}~. \nonumber
\end{eqnarray}
Note that for even $K'$ and as spin theories, $O(N)^0_{K,K'}$ (or $O(N)^1_{K,K'}$) has the same number of lines as $O(N)^0_{K,0}$ (or $O(N)^1_{K,0}$).  Meanwhile for odd $K',$ the number of untwisted primaries is doubled by the presence of the lines $1,\chi$ of $Spin(K')_{-1}$, but the number of twisted primaries remains the same.  In particular in the last two dualities of \eqref{dualnumber} the number of primaries matches for the untwisted/twisted sectors respectively in the dual chiral algebras.

\section{Chern-Simons Matter Duality}
\label{matter}

In this section we derive new Chern-Simons matter dualities.  We extend the previously conjectured dualities in \cite{Metlitski:2016dht,Aharony:2016jvv} by coupling them to background fields for the global symmetries.  Gauging the global symmetries we find new dualities.  We similarly extend the phase diagrams of \cite{Gomis:2017ixy}.

\subsection{Fermion Path Integrals and Counterterms}
\label{levelshifts}

In order to utilize level-rank dualities in the context of Chern-Simons matter theories, we must understand how the counterterms we have studied in the previous sections may be generated by integrating out massive degrees of freedom.  For a review of many of the elements described below see \cite{Witten:2015aba}.

We first consider a real (Majorana) fermion $\lambda$ coupled to an $SO(N)$ gauge field $A$ and transforming in representation $\mathbf{R}$.  Since $\mathbf{R}$ is real, $\lambda$ may be given an $SO(N)$ invariant mass $m$.  The phase of the fermion partition function depends on the sign of $m$.  We fix conventions such that the phase is\footnote{Note that these conventions imply a choice of scheme for the Fermion path integral.  A different choice of scheme would shift the Chern-Simons level for both positive and negative mass by an integer $k$.  The invariant scheme independent statement is that the difference between positive and negative mass is a level $c_{\mathbf{R}}$ Chern-Simons term.  }
\begin{equation}
Z_{\lambda}[A]|_{m>0}=|Z_{\lambda}|\exp\left(\frac{i\pi}{2}\eta(A)\right)~, \hspace{.5in}Z_{\lambda}[A]|_{m<0}=|Z_{\lambda}|~.
\end{equation}
In particular, in the massless theory we take the phase to be the average between these, that is $\exp\left(\frac{i\pi}{4} \eta(A)\right)$.  In these formulas, $\eta(A)$ is the eta-invariant defined by a regularized sum of eigenvalues of the Dirac operator.\footnote{For instance using zeta function regularization we have: $\eta(A)=\lim_{s\rightarrow 0}\sum_{k}\text{sign}(\alpha_{k})|\alpha_{k}|^{-s},$  where $\alpha_{k}$ are the eigenvalues of the Dirac operator coupled to $A$.}

The APS index theorem \cite{Atiyah:1975jf} relates the eta invariant to the minimally quantized spin Chern-Simons action $CS(A)$ as well as a gravitational Chern-Simons term $ CS_{\text{grav}}$.  
\begin{equation}
\exp\left(\frac{i\pi}{2} \eta(A)\right)= \exp\left(i c_{\mathbf{R}}\int_{X} CS(A)+i \text{dim}(\mathbf{R}) \int_{X}CS_{\text{grav}}\right)~. \label{apsreal}
\end{equation}
In the above $c_{\mathbf{R}}$ is the quadratic Casimir of the representation $\mathbf{R}$, and the gravitational Chern-Simons terms is defined by 
\begin{equation}
\int_{X}CS_{\text{grav}}=\frac{1}{192\pi}\int_{Y} \text{tr}(R\wedge R)~,
\end{equation}
where $Y$ is a four-manifold with boundary $X$ and $R$ is the curvature two-form.  In practice we will use formula \eqref{apsreal} for vectors and two index tensors for which:
\begin{equation}
c_{\mathbf{R}}\left(\hspace{-.17in}\phantom{\int}\yng(1) \right)=1~,\hspace{.5in} c_{\mathbf{R}}\left(\hspace{-.17in}\phantom{\int}\yng(1,1)\right)=N-2~,\hspace{.5in} c_{\mathbf{R}}\left(\hspace{-.17in}\phantom{\int}\yng(2)\right)=N+2~.
\end{equation}

Note that the partition function of the massive fermion (of either sign of $m$) defines a local effective action of the gauge field $A$.  By contrast the massless fermion gives a non-local phase.  In general, when describing the level of a Chern-Simons matter theory, we follow the convention that the theory is labelled by an effective level, which is the average of the level at $m>0$ and $m<0$.

Similar analysis applies to the case of complex fermions.  For instance, a massive complex fermion of charge $q$ coupled to a $U(1)$ gauge field gives a contribution to the effective level of
\begin{equation}
k_{m>0}-k_{m<0}=q^{2}~.\label{Abelianlevelshift}
\end{equation}  
As in the case of real fermions, one typically labels the massless theory by the effective level, which receives a contribution $q^{2}/2$ from such a fermion.

Now let us extend our discussion to a single real Majorana fermion $\lambda$ coupled to a $\mathbb{Z}_{2}$ gauge field $B$.  Again we fix conventions such that for negative mass the phase of the partition function is trivial, whereas for positive mass it is non-trivial and defined by the eta invariant.  In particular, the massless theory has phase $\exp\left(\frac{i\pi}{4} \eta(B)\right).$  The analog of the APS index theorem is \cite{Atiyah:1975jf}
\begin{equation}
\exp\left(\frac{i\pi}{2} \eta(B)\right)= \exp\left(i f[B]+i \int_{X}CS_{\text{grav}}\right)~.  \label{etadiscrete}
\end{equation}
Here $f[B]$ is the basic $\mathbb{Z}_{2}$ action discussed in section \ref{gaugegrps} and Appendix \ref{z2appendix}. We again label theories by an effective level, which includes the bare coupling (an integer mod $8$) as well as a fraction arising from fermions that are odd under the $\mathbb{Z}_{2}$ symmetry.  With these conventions, \eqref{etadiscrete} implies that integrating out a massive fermion shifts the $\mathbb{Z}_{2}$ level by $\text{sign}(m)/2$

One simple way to check \eqref{etadiscrete} is to consider two Majorana fermions, which we combine into a single complex fermion.  If both fermions are odd under the $\mathbb{Z}_{2},$ the effective action for $m>0$ is simply the square of \eqref{etadiscrete}.  Thus, neglecting gravitational counterterms, the phase of the partition function for positive mass is $\exp(2if[B])$.  Since the level is even, we can represent this effective action by Abelian Chern-Simons theory following \eqref{acsdescription} and compare to \eqref{Abelianlevelshift}.  We see that, for positive mass, a charge $q$ complex fermion yields $(\mathbb{Z}_{2})_{2q^{2}}.$  This is the expected answer since, when $q$ is even, $2q^{2}=0$  (mod) 8 and hence trivial for a $\mathbb{Z}_{2}$ gauge field, while if $q$ is odd $2q^{2}=2$ (mod) 8.

In the following applications we will be interested in how the levels for the $\mathbb{Z}_{2}\times \mathbb{Z}_{2}$ global symmetry defined by $\mathcal{C}$ and $\mathcal{M}$ are changed by integrating out massive degrees of freedom.  We define the action of charge conjugation to be compatible with the action \eqref{cdef} on gauge fields.  In particular, $\mathcal{C}$ acting on a field (either scalar or fermion) reflects the first index in a vector representation, and acts trivially on other indices:
\begin{equation}
\mathcal{C}(\rho^{i_{1}i_{2}\cdots i_{\ell}})=(-1)^{\# 1's}\rho^{i_{1}i_{2}\cdots i_{\ell}}~.
\end{equation}
Let $x_{\mathbf{R}}$ indicate the number of components of an $SO(N)$ representation that are charged under $\mathcal{C}$.  For the representations of interest to us:
\begin{equation}
x_{\mathbf{R}}\left(\hspace{-.17in}\phantom{\int}\yng(1) \right)=1~,\hspace{.5in} x_{\mathbf{R}}\left(\hspace{-.17in}\phantom{\int}\yng(1,1)\right)=N-1~,\hspace{.5in} x_{\mathbf{R}}\left(\hspace{-.17in}\phantom{\int}\yng(2)\right)=N-1~.
\end{equation}
We take a parameterization of the counterterms in the presence of massless fields to be 
\begin{equation}
xf[B^{\mathcal{C}}]+yf[B^{\mathcal{M}}]+zf[B^{\mathcal{C}}+B^{\mathcal{M}}]~.
\end{equation}
Upon activating a mass $m$ for a fermion in representation $\mathbf{R}$ these counterterms shift.  Since the magnetic symmetry $\mathcal{M}$ is non-perturbative and all level shifts are one-loop exact, the coefficients $y$ and $z$ above are unmodified.  The total effect is therefore
\begin{equation}
(x,y,z)\longrightarrow \left(x+\frac{\text{sign}(m)x_{\mathbf{R}}}{2},y,z\right)~. \label{discreteshifts}
\end{equation}

To further apply our results to theories with more general tensor matter, it is necessary to investigate the properties of monopole operators in more detail.  In particular we will argue that the charge of a monopole operator $V$ under the charge conjugation symmetry $\mathcal{C}$ may be changed by transition through a point where a fermion becomes massless.

We first consider the theory of $SO(2)_{K+1/2}\cong U(1)_{K+1/2}$ coupled to a single complex fermion $\psi$ of charge one.  Since the Chern-Simons level jumps by one between positive and negative fermion mass, the gauge charge of the bare classical monopole differs by one as well.  Thus, if $V$ represents the gauge invariant monopole operator for negative mass, the operator for positive mass must be dressed by an additional fermion.  Therefore, the monopole operator for positive mass is $V \psi$.  

This analysis carries over straightforwardly to $SO(N)_{K}$ coupled to fermions in the vector representation.  The only essential difference is that in this case, the monopole charge is only conserved mod 2.  Across a vector fermion transition, the $\mathcal{M}$ charge is unchanged, the electric charge is shifted, and the gauge invariant monopole acquires an additional fermion across the transition.  

If we now consider $SO(N)$ Chern-Simons theory coupled to a fermion $\psi^{[i,j]}$ transforming in the adjoint representation we find additional effects.  We examine a GNO monopole embedded in the $SO(2)$ subgroup of $SO(N)$ which rotates the two-plane spanned by the last two indices.  We refer to this operator as $V^{[N,N-1]}.$  Note that this is the minimally charged monopole.  Across a massless transition for $\psi^{[i,j]}$, the electric charge of $V^{[N,N-1]}$ is changed by $N-2$.  Therefore across the transition the gauge invariant monopole operator jumps as
\begin{equation}
\label{tensortrans}
V^{[N,N-1]} \longrightarrow V^{[N,N-1]}\left(\psi^{[1,N-1]}+i\psi^{[1,N]}\right)\left(\psi^{[2,N-1]}+i\psi^{[2,N]}\right)\cdots \left(\psi^{[N-2,N-1]}+i\psi^{[N-2,N]}\right)~.
\end{equation}
Notice also that depending on $N$ and $K$ there may in fact be no gauge invariant monopole operator in the spectrum.  Specifically, when $N$ is even and $K-\frac{c_{\mathbf{R}}}{2}$ is odd the monopole operator is odd under the center of $SO(N)$ and this charge cannot be screened by any other fields in the model.  Meanwhile, for $N$ odd, or $K-\frac{c_{\mathbf{R}}}{2}$ even, there is a gauge invariant local monopole operator.   

Assuming that a gauge invariant monopole operator exists, let us now act on it with the charge conjugation symmetry $\mathcal{C}$.  As in \eqref{cdef} we take the operator $\mathcal{C}$ to act as a minus sign on the index $1$.  From \eqref{tensortrans} we therefore see that the $\mathcal{C}$ charge of the monopole operator differs for $m_{\psi}<0$ and $m_{\psi}>0$.  An identical analysis applies to the case of $\psi^{(i,j)}$ in a symmetric traceless tensor representation.

We can phrase this result in terms of the discrete $\theta$-parameters of the section \ref{gaugegrps}.  Consider the $SO(N)$ gauge theory in the presence of a background $\mathcal{C}$ gauge field $B^{\mathcal{C}}$.  The monopole charge is measured by the  second Stiefel-Whitney class $w_{2}$.  The fact that the monopole charge differs across a tensor transition means that the positive and negative mass theories differ by the coupling $\pi \int_{X}w_{2}\cup B^{\mathcal{C}}$.  This means that the theory with vanishing $m_{\psi}$ can be described imprecisely by saying that the coupling is $\pm 1/2$.  This coupling is shifted when a tensor field is given a mass by $\text{sign}(m_{\psi})/2$, so that it is properly quantized after the massive fermions have been integrated out.  

This effect described above is even more dramatic in the $O(N)$ gauge theory, where $B^{\mathcal{C}}=w_{1}$ is dynamical.  In that case the different integral values of the coupling $\pi \int_{X}w_{2}\cup w_{1}$ label the theories $O(N)^{0}$ and $O(N)^{1}$.  Thus we see that these two gauge theories are related through a transition of a massless tensor fermion.  We denote the theory at the origin by $O(N)^{\pm1/2}.$  In particular this implies that $O(N)$ Chern-Simons theory coupled to an odd number of fermions in a two index tensor representation has a parity anomaly.

One immediate application of this analysis is that we can rephrase the discussion in \cite{Aharony:2016jvv} about the  relationship between the non-supersymmetric Chern-Simons matter dualities reviewed in section \ref{funddual} and the supersymmetric $\mathcal{N}=2$ dualities described in \cite{Aharony:2013kma}.  As discussed in section \ref{LRDsec}, level-rank duality of $SO(N)_{K}$  exchanges the $\mathbb{Z}_{2}$ symmetries $\mathcal{C}$ and $\mathcal{M}$.  This map of symmetries also holds in the boson-fermion dualities for $SO(N)_{K}$ coupled to vector matter.

By contrast, in $\mathcal{N}=2$ supersymmetric theories there are also symmetries $\mathcal{C}_{\text{susy}}$ and $\mathcal{M}_{\text{susy}}$ however under the duality the map of symmetries is instead
\begin{equation}
 \mathcal{C}_{\text{susy}}\longleftrightarrow \mathcal{C}_{\text{susy}}~, \hspace{.5in} \mathcal{M}_{\text{susy}}\longleftrightarrow \mathcal{C}_{\text{susy}}\mathcal{M}_{\text{susy}}~.
\end{equation}

To explain the difference in the symmetry map under duality, note that the $\mathcal{N}=2$ theory has a pair of gauginos transforming in the adjoint representation under $SO(N)$.  The non-supersymmetric duality can be obtained from the supersymmetric one by giving mass to these gauginos and flowing to the infrared. Therefore, according to the discussion above, the supersymmetric theory differs from the non-supersymmetric theory by the coupling $\pi \int_{X} w_{2}\cup B^{\mathcal{C}}$.  In particular, this means that the symmetries of the supersymmetric and non-supersymmetric theories are related as
\begin{equation}
\mathcal{C}_{\text{susy}}\longrightarrow \mathcal{C}\mathcal{M}~, \hspace{.5in}\mathcal{M}_{\text{susy}}\longrightarrow\mathcal{M}~.
\end{equation}

In fact, the superymmetric theory, defined by the massless gauginos, has effective coupling $\pi \int_{X} w_{2}\cup B^{\mathcal{C}}.$  If the gauginos are given mass (of either sign) the coefficient of this term shifts to become trivial.  But at the massless point it still has non-trivial effects on the $\mathcal{C}$ charges of monopole operators.

\subsection{Dualities with Fundamental Matter}
\label{funddual}

We can apply our refined understanding of level-rank duality of orthogonal gauge theories to obtain dualities of Chern-Simons theories coupled to vector matter.  

Let us recall briefly the dualities described in \cite{Aharony:2016jvv}.  They concern $SO(N)$ Chern-Simons theories coupled to real scalars $(\phi)$ or Majorana fermions ($\psi$) in vector representations. Explicitly the duality is
\begin{equation}
SO(N)_{K}~\text{with}~N_{f} ~\phi \longleftrightarrow SO(K)_{-N+N_{f}/2}~\text{with}~N_{f}~ \psi~, \label{sovectordual}
\end{equation}
where in the above, the scalars are subject to a quartic potential and both theories are tuned to flow to a transition point in the infrared.

Level-rank duality provides an essential consistency check on \eqref{sovectordual}. Indeed, the obvious mass terms in the UV description of the theories flow to relevant operators at the fixed point that may be used to give mass to any of the scalar or fermi fields.  In particular, if we gap all the matter fields we find level-rank dual topological field theories.

We can also turn this logic around, and use the consistency of the duality \eqref{sovectordual} to provide a check on our result \eqref{lrct} for level-rank duality coupled to background fields.  

Consider the theory with scalars, now coupled to background $\mathbb{Z}_{2}$ gauge fields $B^{\mathcal{C}}$ and $B^{\mathcal{M}}$ for the charge conjugation and magnetic symmetry.  As is common, the validity of the duality \eqref{sovectordual} in the presence of background fields may require us to add counterterms for these gauge fields \cite{Closset:2012vg, Closset:2012vp}.  Without loss of generality, we couple the fermionic theory to background gauge fields without counterterms.  Then, the bosonic theory as a function of background fields is
\begin{equation}
\left(SO(N)_{K}~\text{with}~N_{f} ~\phi \right)[B^{\mathcal{C}}, B^{\mathcal{M}}]+\alpha f[B^{\mathcal{C}}]+\beta f[B^{\mathcal{M}}]+\gamma f[B^{\mathcal{C}}+B^{\mathcal{M}}]~.
\end{equation}
Here, the coefficients $\alpha,\beta, \gamma$ parameterize our ignorance.  We could use \eqref{lrct} to fix them.  Instead, we will constrain them by demanding consistency of the duality \eqref{sovectordual} in the presence of background fields.

Note that the parameters $\alpha,\beta, \gamma$ are part of the ultraviolet definition of the theory, and cannot depend on symmetry breaking patterns, i.e.\ they depend only on $N$ and $K$.  If we now give negative mass squared to $L$ scalars and positive mass squared to $N_{f}-L$ we arrive in the infrared at the topological field theory
\begin{equation}
SO(N-L)_{K}[B^{\mathcal{C}}, B^{\mathcal{M}}]+\alpha f[B^{\mathcal{C}}]+\beta f[B^{\mathcal{M}}]+\gamma f[B^{\mathcal{C}} +B^{\mathcal{M}}]~. \label{scalarback}
\end{equation}

On the fermion side of the duality, we can give positive mass to $L$ of the vectors and negative mass to $N_{f}-L.$  The gauge group is unmodified, but the Chern-Simons level is shifted according the discussion in \ref{levelshifts}.  Using \eqref{discreteshifts} we find
\begin{equation}
SO(K)_{-N+L}[B^{\mathcal{M}}, B^{\mathcal{C}}]+Lf[B^{\mathcal{M}}]~. \label{fermiback}
\end{equation}  
  
We now demand that \eqref{scalarback} and \eqref{fermiback} are dual for all $L$.  From this we determine:
\begin{equation}
\alpha(N,K)=\alpha(N-L,K)~, \hspace{.5in}\beta(N,K)-L=\beta(N-L,K)~,\hspace{.5in}\gamma(N,K)=\gamma(N-L,K)~.
\end{equation}
We can also apply spacetime parity to \eqref{scalarback} and \eqref{fermiback} and again demand a consistent duality.  This implies that $\alpha(N,K)=\beta(K,N)$ and further generates  the recursion relations:
\begin{equation}
\alpha(N,K)-L=\alpha(N,K-L)~, \hspace{.5in}\beta(N,K)=\beta(N,K-L)~,\hspace{.5in}\gamma(N,K)=\gamma(N,K-L)~.
\end{equation}
From this we conclude that
\begin{equation}
\alpha(N,K)=K-p~, \hspace{.5in}\beta(N,K)=N-p~,\hspace{.5in}\gamma(N,K)=q~, \label{ctresults}
\end{equation}
for some $N$ and $K$ independent integers $p$ and $q$.  This matches our general result \eqref{lrct} for level-rank duality, provided we fix $p=q=1$.

\subsubsection{Dualities for $Spin(N)$ and $O(N)$}
\label{specialcases}

Now that we have fixed the exact counterterms required for the duality \eqref{sovectordual} to hold in the presence of background fields, we can immediately derive new results.  Gauging $\mathcal{C}$, $\mathcal{M},$ and $\mathcal{CM}$ produces three new dualities:\footnote{For odd $N$ the group $O(N)$ is a product $\mathbb{Z}_{2}\times SO(N)$ and therefore its representation theory is factorized.  In particular there is a vector representation of $SO(N)$ that is even under the $\mathbb{Z}_{2}$ factor as well as one that is odd under the $\mathbb{Z}_{2}$ factor.  Our dualities hold for the representation that is odd under the $\mathbb{Z}_{2}$ factor. }
\begin{eqnarray}
O(N)_{K,K}^{0}~\text{with}~N_{f} ~\phi &\longleftrightarrow &Spin(K)_{-N+N_{f}/2}~\text{with}~N_{f}~ \psi~, \nonumber \\
Spin(N)_{K}~\text{with}~N_{f} ~\phi &\longleftrightarrow &O(K)^{0}_{-N+N_{f}/2, -N+N_{f}/2}~\text{with}~N_{f}~ \psi~, \\
O(N)^{1}_{K,K-1+L}~\text{with}~N_{f} ~\phi &\longleftrightarrow &O(K)^{1}_{-N+N_{f}/2, -N+N_{f}/2+1+L}~\text{with}~N_{f}~ \psi~. \nonumber
\end{eqnarray}
Note that as a special case when there is no matter ($N_{f}=0$), the above reduce to the level-rank dualities of section \ref{LRDsec}.  

Some special cases of these are worth mentioning.
For $K=1$ we find
\begin{align}
O(N)^0_{1,1}~\text{with}~N_{f} ~\phi &\quad\longleftrightarrow \quad N_{f}~{\rm free\ Majorana\ fermions}+{\rm decoupled}\ (\mathbb{Z}_2)_0~,\cr
Spin(N)_{1}~\text{with}~N_{f} ~\phi &\quad\longleftrightarrow \quad (\mathbb{Z}_2)_{-N+N_{f}/2}~\text{with}~N_{f}~ \psi~,\cr
O(N)^1_{1,L}~\text{with}~N_{f} ~\phi &\quad\longleftrightarrow\quad (\mathbb{Z}_2)_{-N+N_{f}/2+1+L}~\text{with}~N_{f}~ \psi~,
\end{align}
where in the second and last dualities the fermions are odd under the $\mathbb{Z}_2$ gauge field.
The first duality gives infinitely many scalar theories that describe the same $N_f$ free Majorana fermions with the decoupled $(\mathbb{Z}_2)_0$, which is time-reversal invariant (similar dualities appeared in \cite{Metlitski:2016dht,Aharony:2016jvv}).
The theories on the right-hand-side of the second and third dualities are also essentially free.  Specifically, they are free fermions coupled to a discrete gauge theory, which reduces the local operators to the $\mathbb{Z}_{2}$ invariant sector.  Comparing the second duality with the last duality for $L=-1$ also gives infinite many boson/boson dualities
\begin{align}
Spin(N)_{1}~\text{with}~N_{f} ~\phi \quad\longleftrightarrow \quad O(N)^1_{1,-1}~\text{with}~N_{f} ~\phi~,
\end{align}
which describe the same fixed point as that of $(\mathbb{Z}_2)_{-N+N_{f}/2}$ coupled to $N_{f}~ \psi$.

Similarly, for $N=1$ we find
\begin{align}
(\mathbb{Z}_2)_{K}~\text{with}~N_{f} ~\phi &\quad\longleftrightarrow \quad Spin(K)_{-1+N_{f}/2}~\text{with}~N_{f}~ \psi~,\cr
N_{f} ~\text{real WF scalars}+{\rm decoupled}\ (\mathbb{Z}_2)_0 &\quad\longleftrightarrow \quad O(K)^{0}_{-1+N_{f}/2, -1+N_{f}/2}~\text{with}~N_{f}~ \psi~,\cr
(\mathbb{Z}_2)_{K-1+L}~\text{with}~N_{f} ~\phi &\quad\longleftrightarrow\quad O(K)^{1}_{-1+N_{f}/2, N_{f}/2+L}~\text{with}~N_{f}~ \psi~,
\end{align}
where in the first and last dualities the scalars are odd under the $\mathbb{Z}_2$ gauge field.
The second duality gives infinitely many fermionic theories that describe the same $N_f$ real Wilson-Fisher scalars with the decoupled $(\mathbb{Z}_2)_0$.
Comparing the first duality with the last duality for $L=1$ also gives infinite many fermion/fermion dualities
\begin{align}
Spin(K)_{-1+N_{f}/2}~\text{with}~N_{f}~ \psi\quad\longleftrightarrow\quad O(K)^{1}_{N_{f}/2-1, N_{f}/2+1}~\text{with}~N_{f}~ \psi~,
\end{align}
which describe the same fixed point as that of $(\mathbb{Z}_2)_{K}$ coupled to $N_{f} ~\phi$.

For $K=2$ we use $Spin(2)_{L}\cong U(1)_{4L}$ to find
\begin{eqnarray}
\label{speciali}
O(N)^0_{2,2}~\text{with}~N_{f} ~\phi &\quad\longleftrightarrow \quad  &U(1)_{-4N+2N_{f}}~\text{with}~N_{f}~ \psi~, \nonumber \\
Spin(N)_{2}~\text{with}~N_{f} ~\phi &\quad\longleftrightarrow \quad&O(2)_{-N+N_{f}/2, -N+N_{f}/2}~\text{with}~N_{f}~ \psi~, \\
O(N)^1_{2,1+L}~\text{with}~N_{f} ~\phi &\quad\longleftrightarrow \quad&O(2)_{-N+N_{f}/2, -N+N_{f}/2+1+L}~\text{with}~N_{f}~ \psi~, \nonumber
\end{eqnarray}
where the fermion in the first duality has charge 2.  These dualities are valid only for $N_f<N$ \cite{Aharony:2016jvv}.  For other values the dualities are still valid but with a more subtle interpretation \cite{Komargodski:2017keh,Gomis:2017ixy,Cordova:2017kue}.
Comparing the second duality with the last duality for $L=-1$ gives the boson/boson duality
\begin{align}
Spin(N)_{2}~\text{with}~N_{f} ~\phi\quad\longleftrightarrow \quad O(N)^1_{2,0}~\text{with}~N_{f} ~\phi~,
\end{align}
which describes the same theory as $O(2)_{-N+N_{f}/2, -N+N_{f}/2}$ with $N_{f}~ \psi$.

Similarly, for $N=2$ we find
\begin{align}
O(2)_{K,K}~\text{with}~N_{f} ~\phi &\quad\longleftrightarrow \quad Spin(K)_{-2+N_{f}/2}~\text{with}~N_{f}~ \psi~,\cr
U(1)_{4K}~\text{with}~N_{f} ~\phi &\quad\longleftrightarrow \quad O(K)^{0}_{-2+N_{f}/2, -2+N_{f}/2}~\text{with}~N_{f}~ \psi~,\cr
O(2)_{K,K-1+L}~\text{with}~N_{f} ~\phi &\quad\longleftrightarrow\quad O(K)^{1}_{-2+N_{f}/2, -1+N_{f}/2+L}~\text{with}~N_{f}~ \psi~,
\end{align}
where the scalar in the second duality has charge 2.
Comparing the first duality with the last duality for $L=1$ gives the fermion/fermion duality
\begin{align}
Spin(K)_{2-N_{f}/2}~\text{with}~N_{f}~ \psi \quad\longleftrightarrow \quad  O(K)^{1}_{2-N_{f}/2,-N_{f}/2}~\text{with}~N_{f}~ \psi~,
\end{align}
which describes the same theory as $O(2)_{-K,-K}$ with $N_{f}~ \phi$.

If we specialize to $N=K=2$ and $N_f=1$ they become
\begin{eqnarray}
\label{specialii}
O(2)_{2,2}~\text{with}~\phi &\quad\longleftrightarrow \quad  &U(1)_{-6}~\text{with}~ \psi~, \nonumber \\
U(1)_{8}~\text{with}~\phi &\quad\longleftrightarrow \quad&O(2)_{-3/2, -3/2}~\text{with}~\psi~, \\
O(2)_{2,1+L}~\text{with}~\phi &\quad\longleftrightarrow \quad&O(2)_{-3/2, -1/2+L}~\text{with}~ \psi~, \nonumber
\end{eqnarray}
where the matter in the $U(1)$ theories has charge 2.
They can be summarized into
\begin{align}
O(2)_{2,0}~\text{with}~\phi&\quad\longleftrightarrow \quad U(1)_{8}~\text{with $\phi$ of charge two}
\cr
{\Big\updownarrow}\quad\quad&\qquad\qquad\qquad\qquad{\Big\updownarrow}\quad\qquad
\cr
O(2)_{-3/2, -3/2}~\text{with}~\psi&\quad\longleftrightarrow \quad  
{U(1)_{-6}\times (\mathbb{Z}_2)_{-2}\over \mathbb{Z}_2}~\text{with $\psi$ of $U(1)$ charge two}~.
\end{align}
Here we also deviate from our standard notation and the quotient is not of the gauge group, but denotes a gauging of the one-form global symmetry.  Specifically, the quotient uses the Wilson line of $(\mathbb{Z}_2)_{-2}$.

\subsection{Phase Diagram of Adjoint QCD}

As a final application of our level-rank duality result \eqref{lrct}, we consider the phase diagram of orthogonal gauge theories coupled to fermionic matter in a two-index tensor representation discussed in \cite{Gomis:2017ixy}.  We focus on the case of small Chern-Simons level, i.e. the range:
\begin{eqnarray}
SO(N)_{K}+\text{adjoint} ~\lambda && 0\leq K < \frac{N}{2}-2~,\\
SO(N)_{K}+\text{symmetric} ~S && 0\leq K < \frac{N}{2}~.
\end{eqnarray}
The problem is to determine the infrared behavior of these theories as a function of the mass $m$ of the fermionic matter.  A complete solution to this problem was conjectured in \cite{Gomis:2017ixy} and we review it below.

There are two obvious phases where $|m|$ is large and the fermions may be integrated out semi-classically.   These are gapped TQFTs described by $SO(N)$ Chern-Simons theory with level depending on the sign of the mass.
 
 As $|m|$ is reduced, a new quantum phase appears.  This quantum phase is also gapped and described by an $SO$ Chern-Simons theory but with a different gauge group and value of the level.
 
The passage from a semiclassical phase to the quantum phase proceeds through a transition that is weakly coupled in dual variables.  The dual theory also has a two-index tensor fermion. However, this tensor has the opposite symmetry to the original matter defining the model.  Thus, in the adjoint theory, the transition from the semiclassical to quantum phase is governed by a symmetric tensor transition, while in the theory with symmetric matter it is governed by an adjoint transition.  
 
 As we dial further into the quantum regime, we find that the adjoint theory becomes $\mathcal{N}=1$ supersymmetric at a special value of the bare mass.  Based on index calculations in \cite{Witten:1999ds}, it is expected that this theory spontaneously breaks supersymmetry and therefore in addition to the TQFT we also find a massless goldstino.  In the symmetric tensor theory, no such effect occurs.
 
\begin{figure}
  \centering
  \subfloat{\label{figSOWA}\includegraphics[width=.9\textwidth]{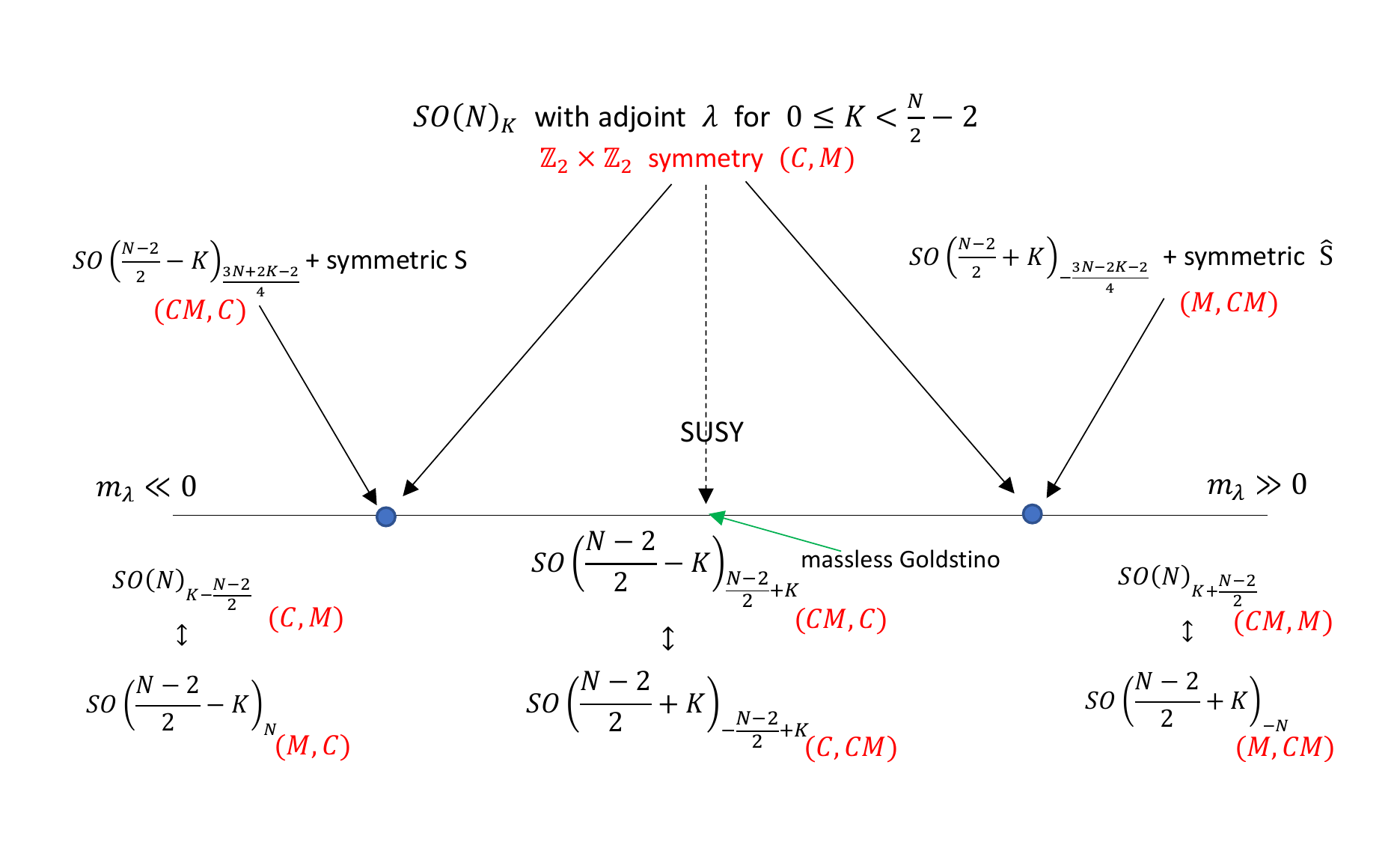}}\\         
  \subfloat{\label{figSOWS}\includegraphics[width=.9\textwidth]{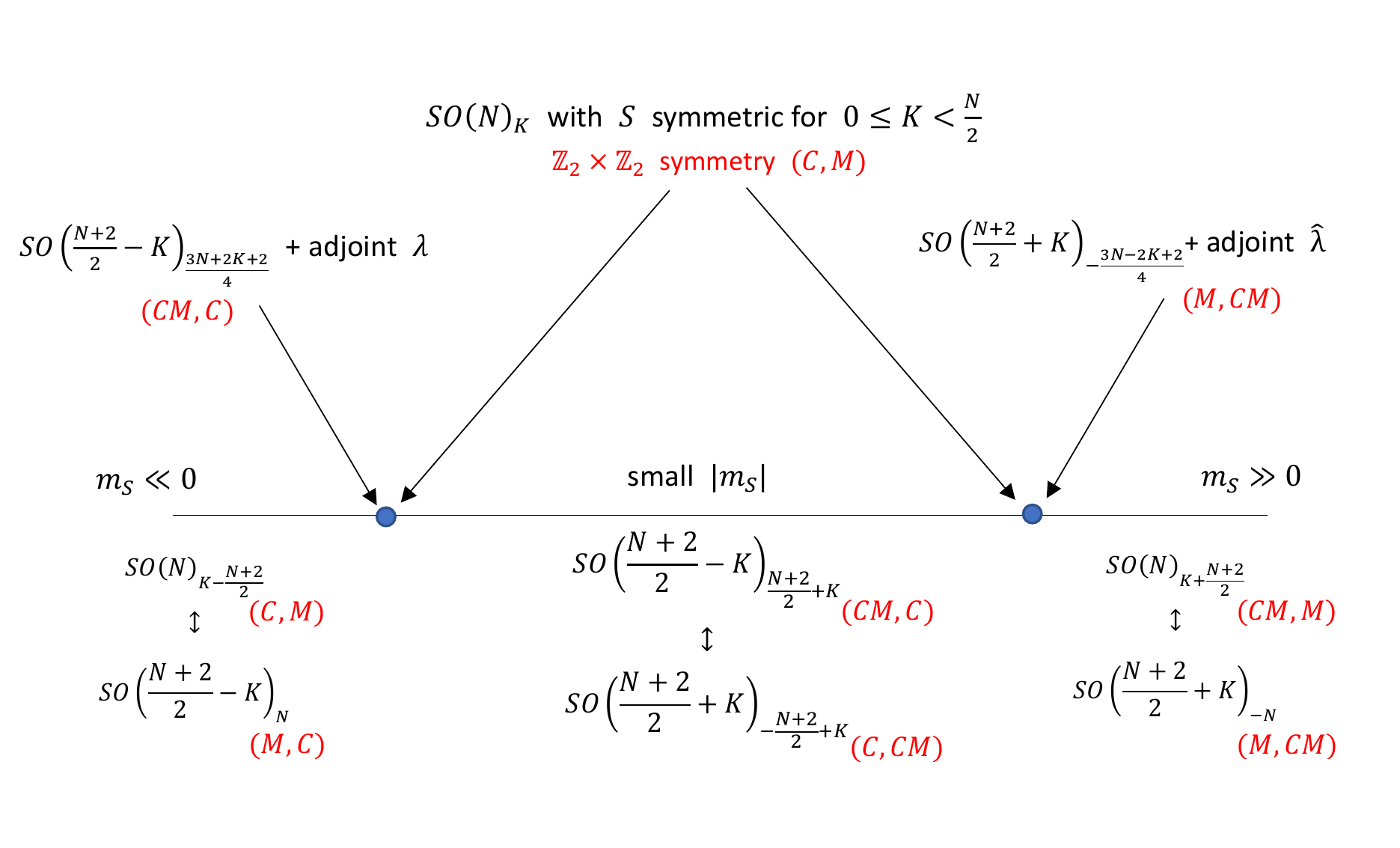}}
  \caption{The phase diagrams of $SO(N)$ gauge theory coupled tensor fermions.  The infrared TQFTs, together with relevant level-rank duals are shown along the bottom.  The blue dots indicate the transitions from the semiclassical phase to the quantum phase. This proceeds through a tensor transition described by a dual theory, which covers part of the phase diagram.  At a special value of the mass in the quantum phase of the adjoint theory, a massless goldstino appears.  These figures are identical to those in \cite{Gomis:2017ixy} except that we now add the map of the $\mathbb{Z}_{2}\times \mathbb{Z}_{2}$ global symmetry.}
  \label{figSOphase}
\end{figure}

These phase diagrams are illustrated in figure \ref{figSOphase}.  As is evident from the explicit levels shown there, level-rank duality is crucial for the consistency of the full picture.

We can use our improved understanding of level-rank duality in the presence of background gauge fields to provide a new consistency check of this conjectured picture.  As we have discussed in section \ref{LRDsec}, level-rank duality shifts the values of the $\mathbb{Z}_{2}\times \mathbb{Z}_{2}$ counterterms.  Additionally the various tensor matter transitions change the counterterms as well as shift the $\int_{X}w_{2}\cup B^{\mathcal{C}}$ counterterm as discussed in section \ref{levelshifts}.  The fact that all these discrete counterterms close on a consistent picture is a striking test of the validity of this phase diagram.

One way to present the information encoded by the counterterm consistency described above is to promote some of the background fields to be dynamical and hence obtain the phase diagram for another gauge group.  We illustrate the results for the $Spin(N)$ and $O(N)$ theories in figures \ref{figSpinphase}, \ref{figOpphase}, and \ref{figOmphase} below.

In these figures it is convenient to introduce a notation analogous to the $\mathbb{Z}_{2}$ level for theories with gauge group $Spin(N)$:
\begin{equation}
\label{spindiscretetheta}
\widetilde{Spin}(N)_{K,L}\equiv \frac{Spin(N)_{K}\times Spin(L)_{-1}}{\mathbb{Z}_{2}}\quad\longleftrightarrow\quad \frac{Spin(N)_{K}\times \left(\mathbb{Z}_{2}\right)_{L}}{\mathbb{Z}_{2}}~.
\end{equation}
The first expression is defined in terms of a Chern-Simons theory based on the group $\frac{Spin(N)\times Spin(L)}{\mathbb{Z}_{2}} $.  The third expression uses the duality \eqref{SpinZ2}.  Here the $\mathbb{Z}_{2}$ quotient is not simply a quotient of the gauge group in the numerator.  It is given by gauging the diagonal one-form symmetry of
the $\mathbb{Z}_{2}$ subgroup in the center of $Spin(N)$, whose quotient results in $SO(N)$, and the $\mathbb{Z}_2$ one-form symmetry generated by the Wilson line of $\left(\mathbb{Z}_{2}\right)_{L}$. 
Note that this quotient exists for all $L$.  An example of an allowed line in this quotient is the product of a spinor in $Spin(N)$ and an 't Hooft line in $(\mathbb{Z}_{2})_{L}$.   For $L=0$ the theory $\widetilde{Spin}(N)_{K,0}$ is the same as the standard $Spin(N)_K$ Chern-Simons theory.

\begin{figure}
  \centering
  \subfloat{\label{figSpinWA}\includegraphics[width=.9\textwidth]{SpinwA.pdf}}\\         
  \subfloat{\label{figSpinWS}\includegraphics[width=.9\textwidth]{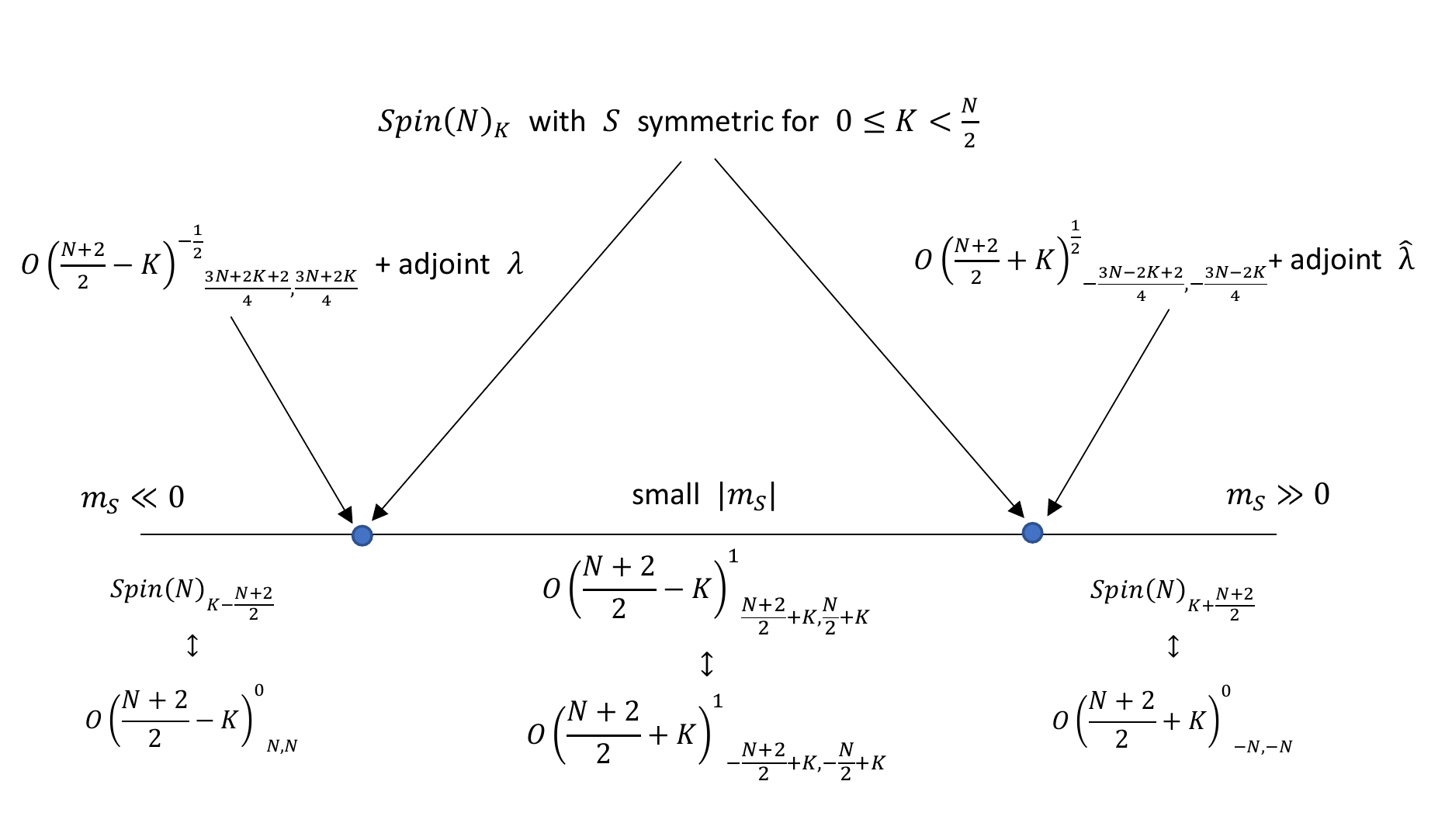}}
  \caption{The phase diagrams of $Spin(N)$ gauge theory coupled tensor fermions.  The infrared TQFTs, together with relevant level-rank duals are shown along the bottom.  The blue dots indicate the transitions from the semiclassical phase to the quantum phase. This proceeds through a tensor transition described by a dual theory, which covers part of the phase diagram.  Across these tensor transitions $O(L)^{0}$ and $O(L)^{1}$ are exchanged.  At a special value of the mass in the quantum phase of the adjoint theory, a massless goldstino appears.}
  \label{figSpinphase}
\end{figure}

\begin{figure}
  \centering
  \subfloat{\label{figOpWA}\includegraphics[width=.9\textwidth]{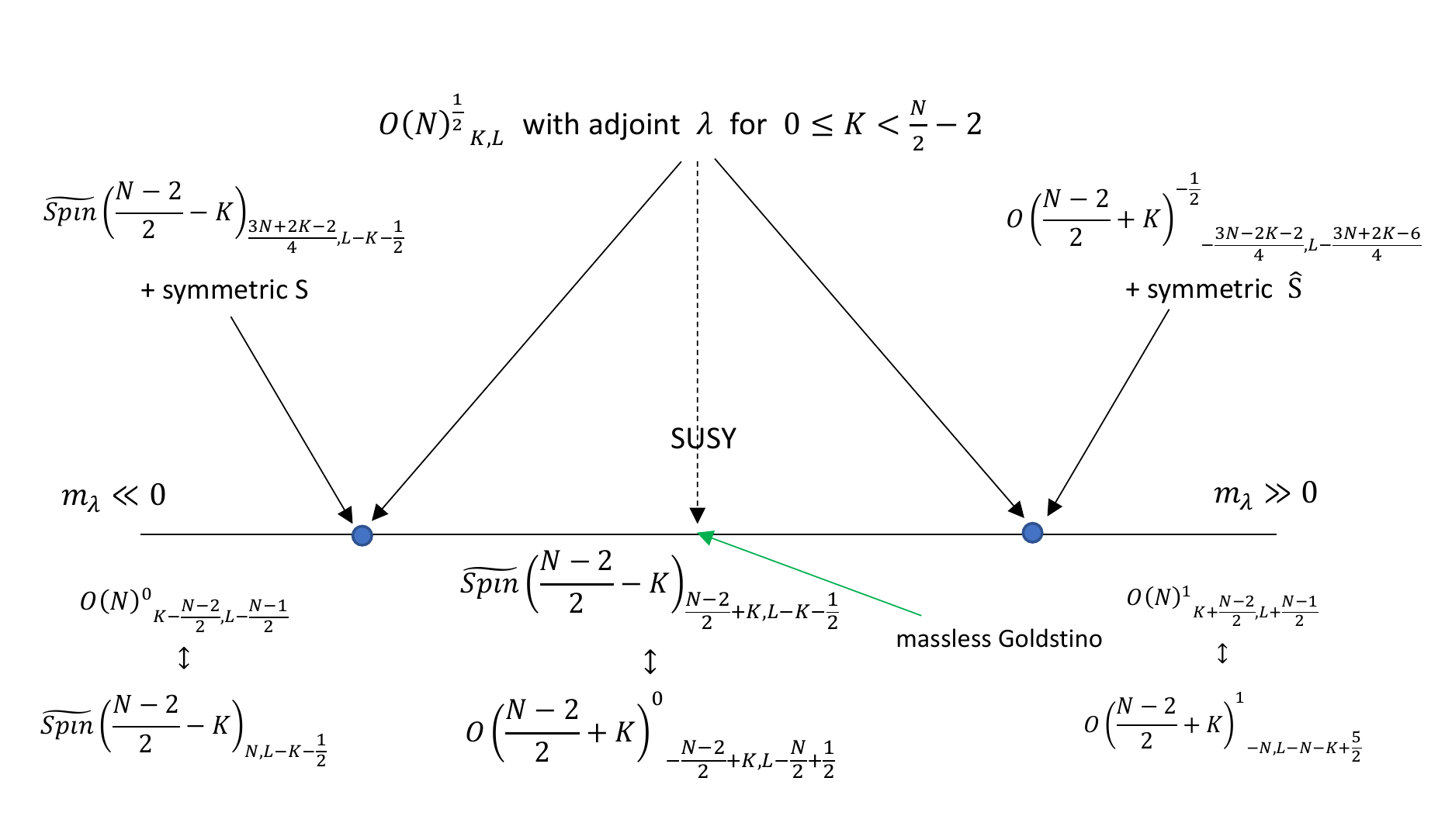}}\\         
  \subfloat{\label{figOpWS}\includegraphics[width=.9\textwidth]{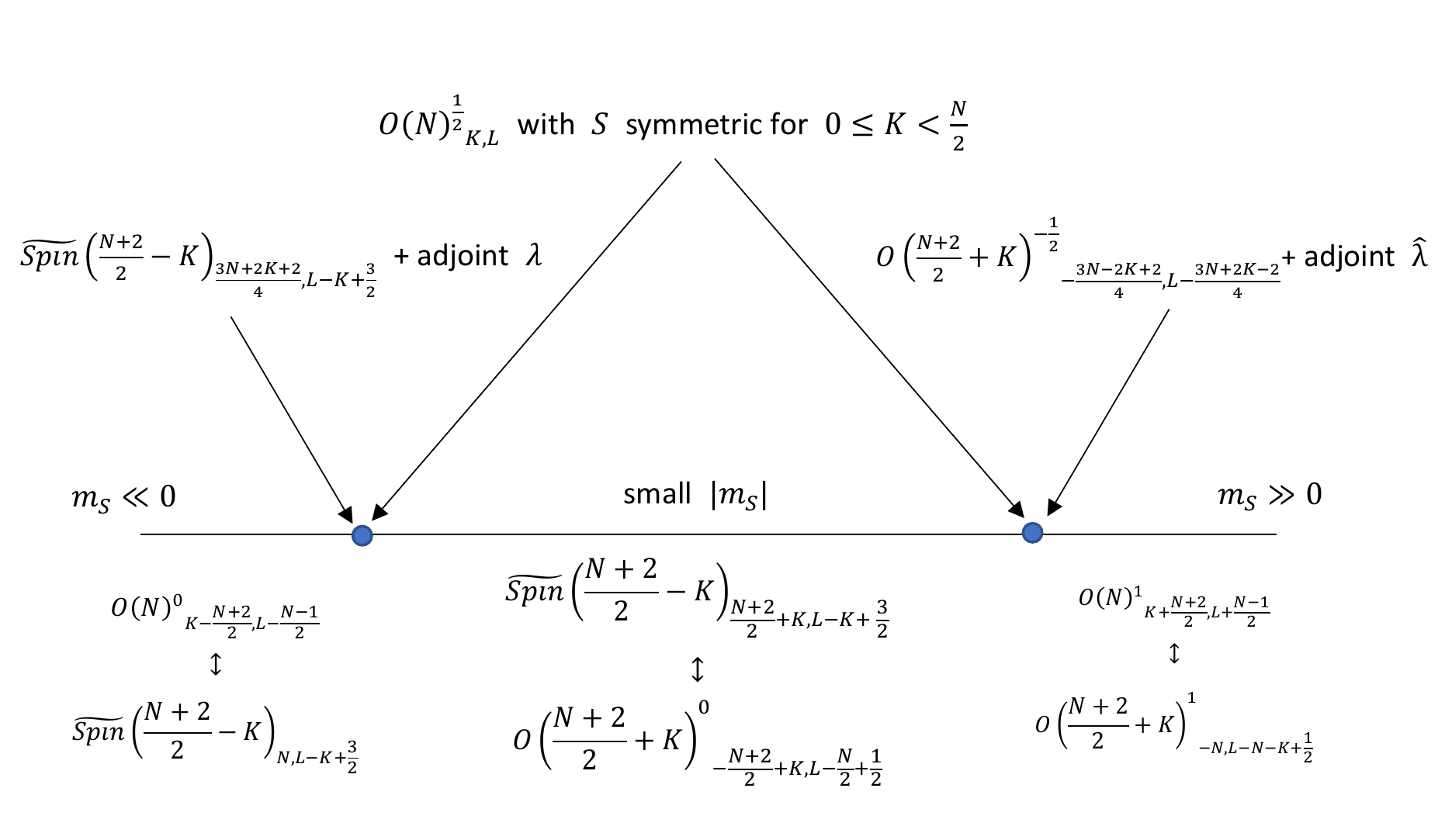}}
  \caption{The phase diagrams of $O(N)^{\frac{1}{2}}$ gauge theory coupled tensor fermions.  The infrared TQFTs, together with relevant level-rank duals are shown along the bottom.  The blue dots indicate the transitions from the semiclassical phase to the quantum phase. This proceeds through a tensor transition described by a dual theory, which covers part of the phase diagram.  Across these tensor transitions $O(L)^{0}$ and $O(L)^{1}$ are exchanged.  At a special value of the mass in the quantum phase of the adjoint theory, a massless goldstino appears. The notation $\widetilde{Spin}(N)_{K,L}$ is explained around \eqref{spindiscretetheta}.}
  \label{figOpphase}
\end{figure}

\begin{figure}
  \centering
  \subfloat{\label{figOmWA}\includegraphics[width=.9\textwidth]{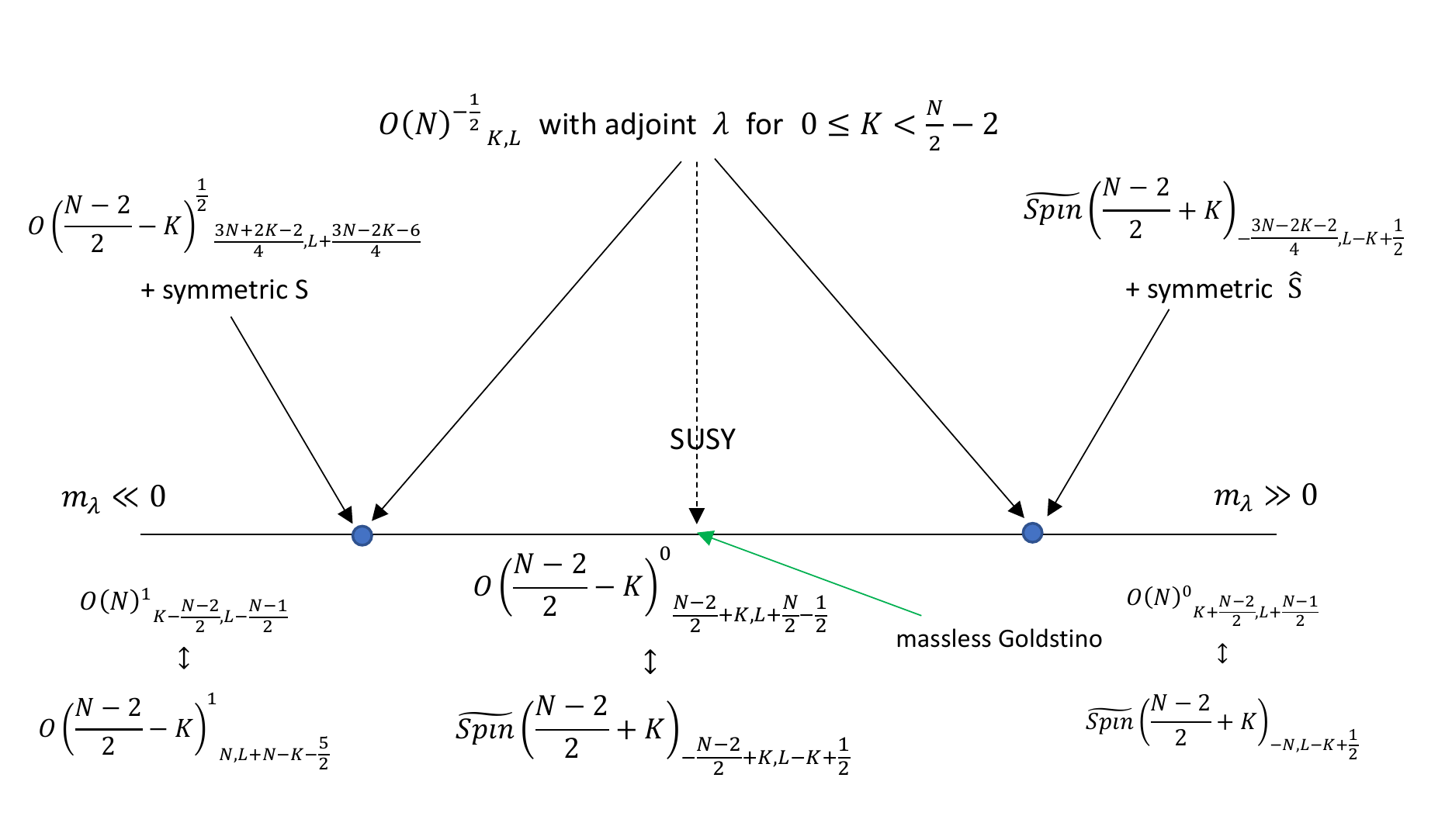}}\\        
  \subfloat{\label{figOmWS}\includegraphics[width=.9\textwidth]{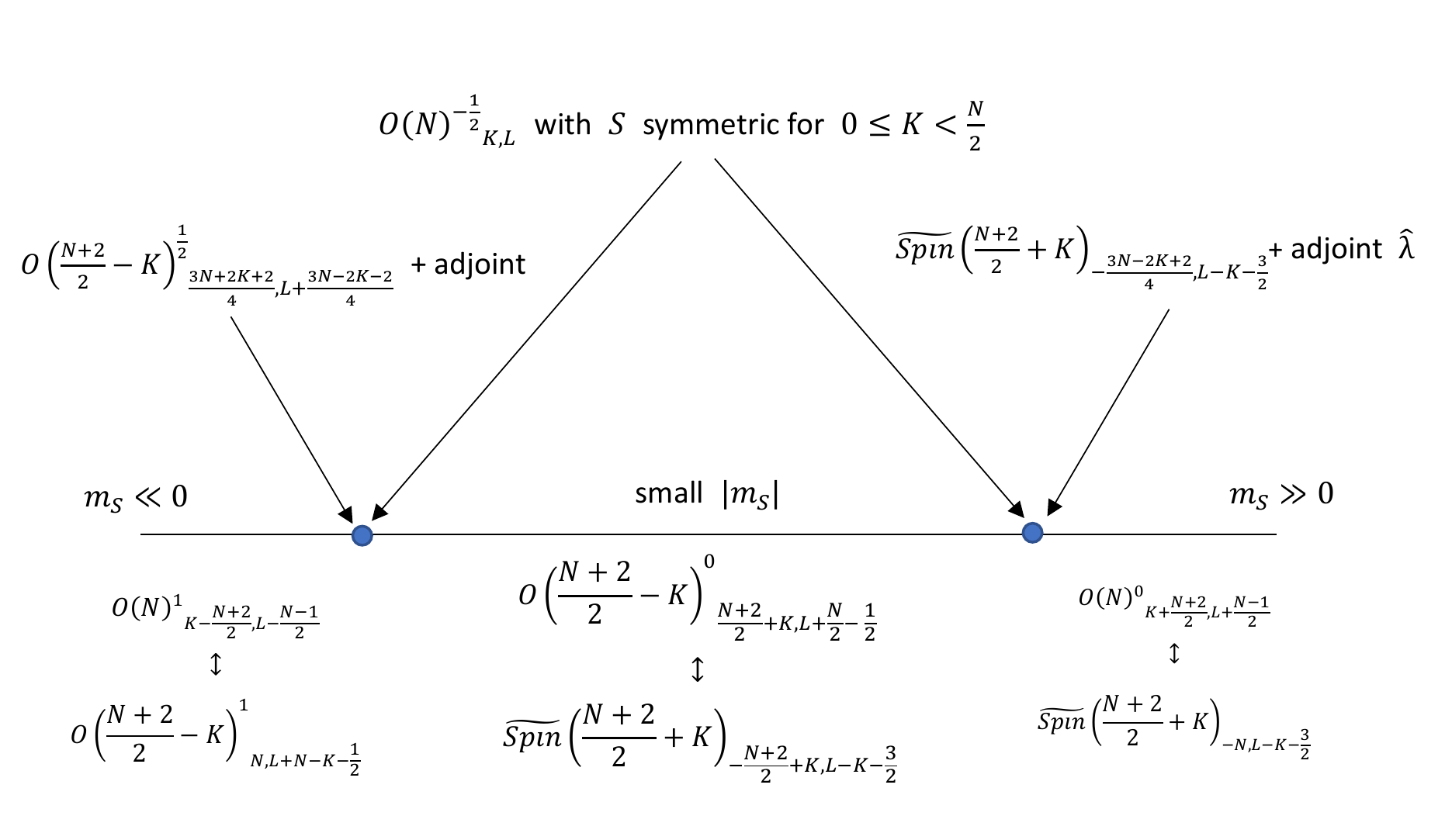}}
  \caption{The phase diagrams of $O(N)^{-\frac{1}{2}}$ gauge theory coupled tensor fermions.  The infrared TQFTs, together with relevant level-rank duals are shown along the bottom.  The blue dots indicate the transitions from the semiclassical phase to the quantum phase. This proceeds through a tensor transition described by a dual theory, which covers part of the phase diagram.  Across these tensor transitions $O(L)^{0}$ and $O(L)^{1}$ are exchanged.  At a special value of the mass in the quantum phase of the adjoint theory, a massless goldstino appears. The notation $\widetilde{Spin}(N)_{K,L}$ is explained around \eqref{spindiscretetheta}.}
  \label{figOmphase}
\end{figure}

\section*{Acknowledgements} 

We thank O. Aharony, M. Barkeshli, F. Benini, M. Cheng, D. Freed, J. Gomis, Z. Komargodski, A. Vishwanath and E. Witten for discussions. C.C. is supported by the Marvin L. Goldberger Membership at the Institute for Advanced Study, and DOE grant DE-SC0009988. The work of P.-S.H. is supported by the Department of Physics at Princeton University. The work of N.S. is supported in part by DOE grant DE-SC0009988.

\appendix

\section{Representation Theory of $\frak{so}(N)$}
\label{dynkin}

The Young tableaux is defined from Dynkin labels $\lambda_i$ by non-increasing row lengths $l_i$, $i=1,\cdots,n$ where $n$ is the rank.
\begin{eqnarray}
\frak{so}(2n+1): &&l_i={\lambda_n\over 2}+\sum_{j=i}^{n-1} \lambda_j\quad {\rm for}\  1\leq i\leq n-1;\quad
l_n={\lambda_n\over 2}~,\\
\frak{so}(2n): &&l_i={\lambda_n-\lambda_{n-1}\over 2}+\sum_{j=i}^{n-1} \lambda_j  \quad {\rm for}\ 1\leq i\leq n-1;\quad
l_n={\lambda_n-\lambda_{n-1}\over 2}~.
\end{eqnarray}

The number of boxes is $r=\sum_{i=1}^n l_i$.
Note for $\frak{so}(2n)$ the last row length comes with a sign while other row lengths are non-negative.

Tensor representations are defined by even $\lambda_n$ for $\frak{so}(2n+1)$ and even $(\lambda_n-\lambda_{n-1})$ for $\frak{so}(2n)$, which correspond to Young tableaux with integral row lengths. Otherwise they are spinor representations (we also refer to them as spinorial representations).

The conformal weight of $\frak{so}(N)_K$ representation with row lengths $\{l_i\}$ can be computed from the Casimir \cite{di1997conformal}
\begin{equation}
h={1\over 2(N+K-2)}
\left(
Nr+\sum_{i=1}^n l_i(l_i-2i)
\right)~.
\end{equation}
For example, the fundamental spinor representation of Dynkin labels $(0,\cdots,0,1)$ corresponds to the Young tableaux with row lengths $(\frac{1}{2},\frac{1}{2},\cdots,\frac{1}{2})$, with conformal weight $\frac{N(N-1)}{16(N+K-2)}$ for all $N$.

\section{$\mathbb{Z}_2$ Topological Gauge Theory in Three Dimensions}
\label{z2appendix}

In this section we describe $\mathbb{Z}_{2}$ classical gauge theories in three spacetime dimensions.  We are interested in theories that depend on the spin structure of the underlying three-manifold $X$.  
These are also known as fermionic symmetry protected topological phases (SPT) with $\mathbb{Z}_{2}$ unitary symmetry.
Such theories have been completely determined \cite{GuLevin,Kapustin:2014dxa,Wang:2016lix,Kapustin:2017jrc}, and admit a $\mathbb{Z}_{8}$ classification.

A practical way to produce these local actions is via $SO(L)_{1}$ spin Chern-Simons theory.
Since these theories depend on the spin structure, they can couple to a $\mathbb{Z}_{2}$ gauge field $B$ by shifting the spin structure with the background $B$. Equivalently, from the relation (\ref{eqn:shiftspinstruc}) they couple to $B$ by the magnetic symmetry ${\cal M}$, and we define the action $f_{L}[B]$ via the normalized partition function
\begin{equation}
e^{-if_{L}[B]}=\frac{Z(SO(L)_{1})[B]}{Z(SO(L)_{1})[0]}~.
\end{equation}
Although we have defined the action by a functional integral, it is known that the result is local since $SO(L)_1$ is an invertible field theory \cite{Freed:2004yc}.  

There are two significant properties that are useful in manipulating these counterterms.  First, the index $L$ behaves as a level (i.e.\ $f_{L}[B]$ is linear in $L$).  This follows from the simple duality
\begin{equation}
SO(L)_{1}\times SO(L')_{1} \quad\longleftrightarrow\quad SO(L+L')_{1}~.
\end{equation}
We use this linearity to simplify notation and write $f_{L}[B]=Lf[B]$ where $f[B]$ is the minimal non-trivial action.

A second useful property is that for even level $L=2n$ the action $2nf[B]$ is simply related to Abelian Chern-Simons theory.  Explicitly\footnote{
It is sufficient to show the relation for $2f[B]$. Using $SO(2)_1\cong U(1)_1$ and $w_2=c_1$ mod 2 for $SO(2)$ bundle, we can express $SO(2)_{1}[B]$ as
\begin{align}
\frac{1}{4\pi}ada+\left( \frac{da}{2\pi}+2\frac{du}{2\pi}\right)B~.
\end{align}
Integrating out $a$ gives the Abelian Chern-Simons theory $-\frac{1}{4\pi}BdB+\frac{2}{2\pi}Bdu-2CS_{\rm grav}$, where the gravitational Chern-Simons term is cancelled by $SO(2)_1[0]$.
}
\begin{equation}
\exp(i2nf[B])=
\int[\mathcal{D}a]\exp\left( \frac{2i}{2\pi} \int_{X}a dB+\frac{in}{4\pi}\int_{X}BdB\right)~, \label{acsdescription}
\end{equation}
where in the above, the $U(1)$ field $a$ is dynamical and enforces the constraint that $B$ is a $\mathbb{Z}_{2}$ gauge field \cite{Maldacena:2001ss,Banks:2010zn}.  
(For a discussion of these theories see also \cite{Kapustin:2014gua, Seiberg:2016rsg}.)
Note that $n=1$ above is the minimal allowed level in $U(1)$ spin Chern-Simons theory.  Thus $Lf[B]$ for odd $L$ cannot be expressed using continuum $U(1)$ actions.  Observe that by redefining $a\rightarrow a+B$  in \eqref{acsdescription} we change the counterterm by $8f[B]$, and hence we have the identification
\begin{align}
8f[B]\sim 0~.
\end{align}
The special cases $Lf[B]$ for $L=0,4$ mod 8 are the bosonic Dijkgraaf-Witten theories \cite{Dijkgraaf:1989pz} for $\mathbb{Z}_2$ gauge group classified by $H^3(B\mathbb{Z}_2,U(1))=\mathbb{Z}_2$. They are independent of the spin structure.  The particular case $L=4$ has the action $\pi \int_{X}B\cup B\cup B.$

If we promote the $\mathbb{Z}_2$ background gauge field $B$ to be a dynamical field $b$ we have
\begin{align}\label{eqn:defztwoL}
Lf[b]\quad\longleftrightarrow\quad Spin(L)_{-1}\times SO(L)_{1}~.
\end{align}
We will denote the left hand side by $(\mathbb{Z}_2)_L$ (with $L=0$ mod 4 defined to be bosonic).  It is a spin TQFT for $L\neq 0$ mod 4.
In particular, $(\mathbb{Z}_2)_{L}$ for odd $L$ has anyons that obey the fusion rule of the Ising TQFT with spins $0,{1\over 2},-{L\over 16}$ mod 1, tensored with $\{1,\psi\}$ where $\psi$ is the transparent spin one-half line.
The Wilson line of $(\mathbb{Z}_2)_L$ corresponds to the product of the line in the vector representation of $Spin(L)$ and $\psi$, and it has integral spin.
The basic magnetic line of $(\mathbb{Z}_2)_L$ corresponds to the line in the fundamental spinor representation of $Spin(L)$ and it has spin $-{L\over 16}$.
$(\mathbb{Z}_2)_L$ has zero framing anomaly.
The chiral algebra corresponding to $(\mathbb{Z}_2)_L$ can be read off from (\ref{eqn:defztwoL}) (also from (\ref{acsdescription}) for even $L$).

This construction of SPT phases may be generalized to any discrete group $G$ that contains a $\mathbb{Z}_{2}$ subgroup.  Consider a $G$ bundle on $X$.  A gauge field $\gamma$ may be viewed as a map to the classifying space  $\gamma:X\rightarrow BG$.  For every choice $\rho\in H^1(BG,\mathbb{Z}_2)$, we then obtain a $\mathbb{Z}_2$ gauge field $\gamma^*\rho\in H^1(X,\mathbb{Z}_2)$. 
Consider
\begin{align}\label{eqn:fsptbasic}
e^{-i  L f[\gamma^*\rho]} = \frac{Z(SO(L)_{1})[\gamma^*\rho]}{Z(SO(L)_{1})[0]}~,
\end{align}
where as before $Z(SO(L)_{1})[B]$ with $B\in H^1(X,\mathbb{Z}_2)$ is the partition function of $SO(L)_1$ coupled to $B$ by the magnetic symmetry ${\cal M}$ (equivalently it couples via shifting the spin structure by $B$ and using (\ref{eqn:shiftspinstruc})).
$f[\gamma^*\rho]$ is a local topological action of the classical $G$ gauge theory. 
For discrete symmetry $G$ it can be expressed as the Arf invariant of the symmetry defects \cite{Kapustin:2014dxa,Kapustin:2017jrc}. 
Other descriptions include \cite{Gaiotto:2015zta,Bhardwaj:2016clt} and the references there.

We can use the above construction, together with fermionic SPT phases that can be produced from Abelian Chern-Simons actions to produce a complete list of the fermionic SPT phases for $\mathbb{Z}_{2n}$ unitary symmetry.   We will reproduce the classifications in \cite{GuLevin} and \cite{Wang:2016lix} obtained from different methods.

All fermionic SPT phases for $\mathbb{Z}_{2n}$ unitary symmetry can be expressed as follows (denote the background gauge field by $B\in H^1(X,\mathbb{Z}_{2n})$)
\begin{align}
L f[(nB)] + \frac{L'}{4\pi}\int_{X}BdB+\frac{2n}{2\pi}\int_{X}adB~,
\end{align}
where in the above $a$ is a dynamical field that constrains the $U(1)$ field $B$ to be a $\mathbb{Z}_{2n}$ gauge field.

The parameters have the identification
\begin{align}
L\sim L+8,\quad L'\sim L'+4n,\quad (L,L') \sim (L+2, L' - n^2)~.
\end{align}
where the last equation uses (\ref{acsdescription}).
Thus the parameter space is $L=0,1$ and $L'=0...4n-1$ for $n>1$, while $L=0,...7, L'=0$ for $n=1$.
For even $n$, $4n$ is multiple of 8 and the classification is $\mathbb{Z}_{4n}\times\mathbb{Z}_2$ generated by $(L,L') = (1,1)$ of order $4n$ and $(1,-n^2/2)$ of order 2.
For odd $n>1$, the classification is $\mathbb{Z}_{8n}$ generated by $(L,L')=(1,1)$.
For $n=1$ the $\mathbb{Z}_8$ classification is generated by $(L,L') = (1,0)$.

For unitary symmetry $\mathbb{Z}_{2n+1}$ there is no $\mathbb{Z}_2$ subgroup and thus no coupling with $w_2(SO(L))$. 
Therefore all fermionic SPT phases for unitary symmetry $\mathbb{Z}_{2n+1},$ can be constructed from Abelian Chern-Simons actions, and have a $\mathbb{Z}_{2n+1}$ classification.\footnote{
Consider the Abelian Chern-Simons theory $(Z_N)_K\times SO(0)_1$ with odd $N$
\begin{align}
\frac{K}{4\pi}B'dB'+\frac{N}{2\pi}B'du+\frac{1}{4\pi}xdx+\frac{1}{2\pi}xdy~.
\end{align}
The change of variables $u\rightarrow u+\frac{N+1}{2}B'-y$, $z\rightarrow x+N B'$, $y\rightarrow y-N B'$ produces the identification $K\sim K+N$. For even $N$ there is no such identification.
}

\section{$Pin^-(N)$ and $O(N)^1$ from $SO(N)$}
\label{gauging}

Since an $SO(N)$ bundle is an $O(N)$ bundle with $w_1=0$, we can write the $SO(N)$ Lagrangian as
\begin{align}\label{eqn:SON}
\int_X {\cal L}[SO(N)] \quad\longleftrightarrow\quad \int_X {\cal L}[O(N)] +\pi \int_X w_1(O(N))\cup B_2 ~,
\end{align}
where the dynamical $\mathbb{Z}_2$ two-form gauge field $B_2$ constrains $w_1(O(N))=0$.
Thus turning on background $B^{\cal C}$ for ${\cal C}$ changes the Lagrangian to\footnote{In this section, as throughout the paper, $w_{i}(G)$ refers to the Stiefel-Whitney classes of a principle bundle with structure group $G$.}
\begin{align}
\label{appsoc}
\int_X {\cal L}[SO(N),B^{\cal C}] \quad\longleftrightarrow\quad \int_X {\cal L}[O(N)]+\pi \int_X \left( w_1(O(N))+B^{\cal C}\right) \cup B_2~,
\end{align}
such that making $B^{\cal C}$ dynamical without any additonal counterterm produces the $O(N)$ theory.
The dynamical multiplier $B_2$ implies that an $SO(N)$ bundle with background $B^{\cal C}$ produces $O(N)$ bundle with fixed first Stiefel-Whitney class $w_1(O(N))=B^{\cal C}$.
In addition, we can include the coupling with background $B^{\cal M}$ for ${\cal M}$
\begin{align}\label{eqn:SONCM}
\int_X {\cal L}[&SO(N),B^{\cal C},B^{\cal M}] \quad\longleftrightarrow\quad\cr &\int_X {\cal L}[O(N)]
+\pi \int_X \left( w_1(O(N))+B^{\cal C}\right) \cup B_2 +\pi\int_X w_2(O(N))\cup B^{\cal M}~.
\end{align}

We can gauge the $\mathbb{Z}_2$ symmetries by promoting the gauge fields to be dynamical.  In this way we produce figure \ref{gaugemap}.
 
\begin{itemize}
\item Gauging ${\cal C,M}$ by promoting $B^{\cal C},B^{\cal M}$ to be dynamical and adding a counterterm $S_{ct}[B^{\cal C}]$ that depends only on $B^{\cal C}$. The multiplier $B^{\cal M}$ imposes $w_2(O(N))=0$, which changes the gauge group to $Pin^+(N)$, and the counterterm becomes $S_{ct}[w_1(Pin^+(N))]$.

\item 
Gauging ${\cal C,M}$ by promoting $B^{\cal C},B^{\cal M}$ to be dynamical and adding the counterterm $S_{ct}[B^{\cal C}]+\pi\int_X B^{\cal M}\cup  B^{\cal C}\cup  B^{\cal C}$.
The equation of motion for $B_2$ imposes $B^{\cal C}=w_1(O(N))$, and the dynamical multiplier $B^{\cal M}$ imposes $w_2(O(N))+w_1(O(N))\cup w_1(O(N))=0$. Thus the theory becomes a $Pin^-(N)$ gauge theory, and the other counterterm becomes $S_{ct}[w_1(Pin^-(N))]$.

\item Gauging the diagonal ${\cal CM}$. We take $B^{\cal C}=B^{\cal M}$ (denoted by $B$) and add some counterterm $S_{ct}[B]$. 
Then we promote $B$ to be dynamical.
The dynamical multiplier $B_2$ imposes $B=w_1(O(N))$, and the theory becomes
\begin{align}
\int_X {\cal L}[O(N)]+\pi \int_X w_1(O(N))\cup w_2(O(N))+S_{ct}[w_1(O(N))]~.
\end{align}
Thus it produces an $O(N)$ gauge theory with discrete topological term $w_1\cup w_2$, namely an $O(N)^1$ theory.

\end{itemize}

We can also gauge the one-form symmetry in the $Pin^+(N)$ theory.
We will focus on one-form symmetries dual to the gauged ${\cal C,M}$ symmetries.
Denote their background two-form $\mathbb{Z}_2$ gauge fields by $B_2',B_2''$, they couple to the $Pin^+(N)$ theory by
\begin{align}
\int_X {\cal L}[&SO(N),B^{\cal C},B^{\cal M}]+\pi\int_X \left( B^{\cal C}\cup B_2' +B^{\cal M}\cup B_2''\right)~,
\end{align}
where $B^{\cal C},B^{\cal M}$ are dynamical.  As in the discussion below \eqref{appsoc}, the presence of the dynamical fields $B^{\mathcal{C}}$ and $B^{\mathcal{M}}$ means that the bundles are $Pin^{+}(N)$ bundles with non-trivial Stiefel-Whitney classes.
\begin{itemize}
\item We can gauge both $\mathbb{Z}_2$ one-form symmetries by promoting both $B_2',B_2''$ to be dynamical. 
This imposes $B^{\cal C}=B^{\cal M}=0$ and we recover the $SO(N)$ gauge theory.

\item We can gauge either one-form symmetry by promoting either $B_2'$ or $B_2''$ to be dynamical. This produces the $Spin(N)$ or $O(N)^0$ theory.

\item Gauging the diagonal one-form symmetry by promoting the diagonal $B_2'=B_2''$ to be dynamical. This imposes $B^{\cal C}=B^{\cal M}$, and we recover the $O(N)^1$ theory.
\end{itemize}

\section{The Chern-Simons Action of $O(N)$ for Odd $N$}
\label{CSOddsec}

As discussed in section \ref{sec:lagrangians}, the Lagrangian for $O(N)=SO(N)\rtimes \mathbb{Z}_2$ can have discrete couplings involving $w_1$ of the $O(N)$ bundle. We view the Stiefel-Whitney class $w_1$ as a $\mathbb{Z}_2$ gauge field for the charge conjugation ${\cal C}$ of $SO(N)$.  These discrete couplings depend on how the $\mathbb{Z}_2$ is embedded in $O(N)$.  For even $N$ there is no natural choice, and our convention is specified by (\ref{cdef}). If we change our convention and define $\mathcal{C}$ to flip the signs of more indices, the $\mathbb{Z}_2$ levels are modified.  Formulas for the change in the action can be derived following the logic below.

When $N$ is odd, the orthogonal group factorizes $O(N)=SO(N)\times \mathbb{Z}_2,$ and thus every $O(N)$ bundle is separately an $SO(N)$ bundle and a $\mathbb{Z}_2$ bundle.  One natural choice of $\mathbb{Z}_{2}$ subgroup is this $\mathbb{Z}_{2}$ factor, which commutes with $SO(N)$.  However, to be uniform with our treatment of even $N$ we continue to use the convention \eqref{cdef}, which  corresponds to a $\mathbb{Z}_2$ that is embedded non-trivially in each factor of $SO(N)\times\mathbb{Z}_2$. 

Let us derive the relation between the Lagrangians for these two choices of $\mathbb{Z}_{2}$ subgroup.  Explicitly the formula we would like to prove is:
\begin{equation}
\label{CS(O(odd))ap}
CS(O(N))=CS(SO(N))+\pi \int_X w_{2}(SO(N))\cup w_{1}+(N-1)f[w_{1}]~,
\end{equation}
where on the left the $\mathbb{Z}_2$ with gauge field $w_1$ is non-trivially embedded in $SO(N)\times\mathbb{Z}_2$ as defined by the charge conjugation (\ref{cdef}), but on the right it is factorized from $SO(N)$.

To derive \eqref{CS(O(odd))ap} observe that the charge conjugation transformation \eqref{cdef} can be expressed as $e^{\pi J}$ where $J$ is the element of the Lie algebra 
\begin{equation}
J=\left(\begin{array}{ccccc}
0 &   &   &  &\\
  & 0 & 1 &  &\\
  & -1& 0 &  &\\
  &   &   & 0 & 1\\
  &   &   & -1 & 0\\
  &   &   &    &\cdots
\end{array}
\right)
~.
\end{equation}
Therefore the dynamical $O(N)$ gauge field may be expressed as a sum of an $SO(N)$ gauge field together with a $\mathbb{Z}_{2}$ gauge field $B^{\mathcal{C}}$ which also sources $J$
\begin{equation}
A_{O(N)}=(A_{SO(N)},B^{\mathcal{C}})~. \label{onfield}
\end{equation}
For practical manipulations it is useful to view $B^{\mathcal{C}}$ as a $U(1)$ gauge field constrained to $\mathbb{Z}_{2}$ by a Lagrange multiplier field.

We now evaluate the Chern-Simons action \eqref{spincsinv}.  To evaluate the terms involving the field $B^{\mathcal{C}}$ it suffices to consider the case where the $SO(N)$ gauge field is of the form
\begin{equation}
A_{SO(N)}=\left(\begin{array}{ccccc}
0 &   &   &  &\\
  & 0 & a_1 &  &\\
  & -a_1& 0 &  &\\
  &   &   & 0 & a_2\\
  &   &   & -a_2 & 0\\
  &   &   &    &\cdots
\end{array}
\right)
\end{equation}
Evaluating the $O(N)$ Chern-Simons action on \eqref{onfield} we then find (below $B^{\mathcal{C}}$ is normalized to have periods $0$ and $\pi$):
\begin{equation}
CS(O(N))=CS(SO(N))+\int_X \sum_{i=1}^{(N-1)/2}\frac{da_{i}}{2\pi}  B^{\mathcal{C}}+\frac{N-1}{8\pi}\int_X B^{\mathcal{C}}  d B^{\mathcal{C}}~.
\end{equation}
We recognize the sum over fluxes $\sum_{i=1}^{(N-1)/2}\frac{da_{i}}{2\pi} $ is the expression for the Stiefel-Whitney class $w_{2}(SO(N)).$ Meanwhile, using the relationship between Abelian Chern-Simons theory and $\mathbb{Z}_{2}$ gauge theory discussed in appendix \ref{z2appendix}, we see that the final term above is $(N-1)f[B^{\mathcal{C}}]$.

As a consistency check, integrating out one Majorana fermion $\psi$ of positive mass in the vector representation of $O(N)$ generates the topological term (\ref{etadiscrete}):
\begin{align}\label{eqn:csoddtop1}
CS(O(N))+f[w_1]+\int_X { CS}_{\rm grav}~.
\end{align}
On the other hand, up to an $SO(N)$ gauge transformation, charge conjugation acts on the fermion in the same way as flipping the sign of $\psi^i$ for all $SO(N)$ vector indices $i$. The latter symmetry is the $\mathbb{Z}_{2}$ subgroup of $O(N)$ that commutes with $SO(N)$.   Note also that for positive vs negative mass, the charge of the monopole under this $\mathbb{Z}_{2}$ factor subgroup of $O(N)$ is changed as in section \ref{levelshifts}.  Thus, in the factorized variables, integrating out the same fermion generates the topological terms
\begin{align}\label{eqn:csoddtop2}
CS(SO(N))+ \pi\int_X  w_2(SO(N))\cup w_1+Nf[w_1]+\int_X { CS}_{\rm grav}~.
\end{align}
Equating the two expressions (\ref{eqn:csoddtop1}),(\ref{eqn:csoddtop2}) reproduces the relation (\ref{CS(O(odd))ap}).

\section{Derivation of Level-Rank Duality for Odd $N$ or $K$}
\label{derlrdualodd}

In this appendix we derive the counterterm in (\ref{lrct}) when $N$ or $K$ is odd.  As in section \ref{LRDsec}, we derive this result, by proving the spin Chern-Simons dualities \eqref{spindualspinO}-\eqref{spindualOO} for $N$ or $K$ odd:
\begin{align} \label{eqn:deroddspindual} 
O(N)^0_{K,K}&\quad\longleftrightarrow\quad Spin(K)_{-N} \cr
O(N)^1_{K,K-1} &\quad\longleftrightarrow\quad O(K)^1_{-N,-N+1} ~.
\end{align}

We will use various formulas from section \ref{ordinarysym} for the Chern-Simons action of $SO(N)_K$ with odd $N$ coupled to the backgrounds $B^{\cal C},B^{\cal M}$ for the symmetries ${\cal C,M}$.  The Chern-Simons action can be expressed as
\begin{align} \label{eqn:SOoddcoupling}
SO(N)_K[B^{\cal C},B^{\cal M}] 
&= SO(N)_K[0,B^{\cal M}+KB^{\cal C}] + 
(K-1)(N-1)f[B^{\cal C}]\cr
&\quad-(N-1)f[B^{\cal M}] + (N-1)f[B^{\cal C}+B^{\cal M}]
\quad{\rm odd}\ N~. 
\end{align}
From (\ref{eqn:SOoddcoupling}) we can obtain the $O(N)^0_{K,0}$ theory by promoting $B^{\cal C}$ to be dynamical with $B^{\cal M}=0$.
Thus
\begin{align}
O(N)^0_{K,0}=\left\{\begin{array}{cl}
SO(N)_K\times (\mathbb{Z}_2)_{K(N-1)} & {\rm odd}\ N,\,{\rm even}\ K\\
\left( Spin(N)_K\times (\mathbb{Z}_2)_{K(N-1)}\right)/\mathbb{Z}_2  
& {\rm odd}\ N,\,{\rm odd}\ K~.
\end{array}\right.\label{eqn:opodd}
\end{align}
Similarly, we can obtain the $O(N)^1_K$ theory by promoting $B^{\cal C}=B^{\cal M}$ to be dynamical\footnote{
From (\ref{eqn:opodd}) and (\ref{eqn:omodd}) one can also find the one-form global symmetry of $O(N)^0_{K,0}, O(N)^1_{K,0}$ for odd $N$. In $O(N)^0_{K,0}$ it is $\mathbb{Z}_2\times\mathbb{Z}_2$ for odd $N$ and even $K$, while $\mathbb{Z}_2$ for odd $N$ and $K$.
In $O(N)^1_{K,0}$ it is $\mathbb{Z}_2$ for odd $N$ and even $K$, $\mathbb{Z}_2\times\mathbb{Z}_2$ for odd $K$ and $N=1$ mod 4, and $\mathbb{Z}_4$ for odd $K$ and $N=3$ mod 4.
}
\begin{align}
O(N)^1_{K,0}=\left\{\begin{array}{cl}
\left(Spin(N)_K\times (\mathbb{Z}_2)_{(K-2)(N-1)} \right)/ \mathbb{Z}_2 & {\rm odd}\ N,\,{\rm even}\ K\\
SO(N)_K\times (\mathbb{Z}_2)_{(K-2)(N-1)} & {\rm odd}\ N,\,{\rm odd}\ K~.
\end{array}\right.\label{eqn:omodd}
\end{align}
In both \eqref{eqn:opodd} and \eqref{eqn:omodd}, the $\mathbb{Z}_{2}$ quotients use the product of $\chi$ in the continuous factor and the electric Wilson line of the $\mathbb{Z}_{2}$ gauge theories in the numerator, and thus the theories are not based on quotients of the gauge group that appears in the numerator.  (See the discussion around \eqref{eqn:GKL}.)

\subsection{Even $N$ and odd $K$}

We start with the conformal embedding with $NK$ real fermions
\begin{align}\label{eqn:confembevenodd}
Spin(N)_K\times SO(K)_N\subset Spin(NK)_1~,
\end{align}
where in the above the relation between the centers of the left and right chiral algebras is subject to the same discussion as that following \eqref{confembedding}.

Consider gauging $({\cal C},{\cal M})$ on the left hand side (and an outer automorphism equivalent to ${\cal C}$ on the right).
${\cal C}$ transforms the currents of $Spin(N)_K$ by changing the sign of $J_{1i}$ with $i=2,\cdots,N$.
To be compatible with the transformation the $NK$ indices are partitioned into $K$ and $NK$ as in  section \ref{sec:embed}.
Therefore we obtain the conformal embedding
\begin{align}
Pin^+(N)_K\times Spin(K)_N\subset Spin(K)_1\times Spin(NK-K)_1 ~. \label{confebede1}
\end{align}

The conformal embedding implies the duality of chiral algebras
\begin{align}
Pin^+(N)_K\quad\longleftrightarrow\quad {Spin(K)_1\times Spin(NK-K)_1\over Spin(K)_{N}}~,
\end{align}
where we again use the fact that each factor in the subalgebra in  \eqref{confebede1} acts faithfully. (See the discussion around \eqref{eqn:equivnonspinalg}.)
This implies the non-spin Chern-Simons duality
\begin{align}\label{eqn:nonspincosetevenodd}
Pin^+(N)_K\quad\longleftrightarrow\quad {Spin(K)_{-N}\times  Spin(K)_1\times Spin(NK-K)_1\over\mathbb{Z}_2}~.
\end{align}
We denote the generators of the $\mathbb{Z}_2\times\mathbb{Z}_2$ simple currents on the left-hand side by $\chi,j_C.$  They have spin of ${1\over 2}$ and $0$ respectively.  Gauging the $\mathbb{Z}_{2}$ one-form symmetry  generated by $\chi$ results in the theory $O(N)^0_K$, while gauging  the $\mathbb{Z}_{2}$ one-form symmetry  generated by $j_{C}$ results in the theory $Spin(N)_K.$ On the right-hand side $\chi$ maps to $(1,1,\chi)$ and $j_{C}$ maps to $(1,\chi,\chi).$ (See also footnote \ref{foot}.)  

Taking the $\mathbb{Z}_2$ quotient in (\ref{eqn:nonspincosetevenodd}) generated by $\chi$ on the left and $(1,1,\chi)$ on the right produces the spin duality
\begin{align}\label{eqn:OSpinevenodd}
O(N)^0_{K,0}\quad\longleftrightarrow\quad {Spin(K)_{-N}\times  Spin(K)_1\over\mathbb{Z}_2}~,
\end{align}
which implies the spin Chern-Simons duality $O(N)^0_{K,0}\leftrightarrow \widetilde{Spin}(K)_{-N,-K}$.  By adding a $(\mathbb{Z}_{2})_{K}$ theory on both sides and gauging a one-form symmetry as in appendix \ref{dualityoneform}, the above implies
\begin{align}
O(N)^0_{K,K}\quad\longleftrightarrow\quad Spin(K)_{-N}~.
\end{align}
For the special case $K=1$ the spin duality (\ref{eqn:OSpinevenodd}) reproduces $O(N)^0_{1,0}\leftrightarrow Spin(1)_1$ with $Spin(1)_N\equiv (\mathbb{Z}_2)_0$ for even $N$.

Consider next the $\mathbb{Z}_2$ quotient in (\ref{eqn:nonspincosetevenodd}), which on the left brings $Pin^+(N)_K$ to $Spin(N)_K$, and on the right it is generated by $(1,\chi,\chi).$  We find
\begin{align}
Spin(N)_K\quad\longleftrightarrow\quad SO(K)_{-N} \times  Spin(NK)_1~.
\end{align}
Comparing the right-hand side with (\ref{eqn:opodd}), which defines the $O(K)$ Chern-Simons theory for odd $K$, gives the spin Chern-Simons duality
\begin{align}
Spin(N)_K\quad\longleftrightarrow\quad O(K)^0_{-N,-N}~.
\end{align}

Similarly consider the $\mathbb{Z}_2$ quotient in (\ref{eqn:nonspincosetevenodd}), which on the left brings $Pin^+(N)_K$ to $O(N)^1_K$, and on the right it is generated by $(1,\chi,1):$
\begin{align}
O(N)^1_K\quad\longleftrightarrow\quad {Spin(K)_{-N}\times  Spin(NK-K)_1\over\mathbb{Z}_2} ~.
\end{align}
Comparing with (\ref{eqn:omodd}) gives the spin Chern-Simons duality
\begin{align}
O(N)^1_{K,K-1}\quad\longleftrightarrow\quad O(K)^1_{-N,-N+1}~.
\end{align}

\subsection{Odd $N,K$}

We start from the conformal embedding for odd $N,K$
\begin{align}\label{eqn:confembodd}
Spin(N)_K\times Spin(K)_N\subset Spin(NK)_1~.
\end{align}
Each factor of the subalgebra above acts faithfully and thus we obtain the following duality of chiral algebras
\begin{align}
Spin(N)_K\quad\longleftrightarrow\quad {Spin(NK)_1\over Spin(K)_N}~.
\end{align}
The corresponding non-spin Chern-Simons duality is
\begin{align}\label{eqn:lroddspinspin}
Spin(N)_K\quad\longleftrightarrow\quad {Spin(K)_{-N}\times Spin(NK)_1\over \mathbb{Z}_2}~.
\end{align}

Promoting to spin theories and comparing the right-hand side with (\ref{eqn:opodd}) gives the spin Chern-Simons duality
\begin{align}
Spin(N)_K\quad\longleftrightarrow\quad O(K)^0_{-N,-N}~.
\end{align}
Note also that since $N$ and $K$ are both odd we also have the duality \eqref{oddduality}.  Combining with the above, this implies
\begin{equation}
O(N)^{0}_{K,K-NK}\quad\longleftrightarrow\quad O(K)^0_{-N,-N}~.
\end{equation}

Taking the $\mathbb{Z}_2$ quotient in (\ref{eqn:lroddspinspin}) generated by $\chi$ of $Spin(N)_K,Spin(K)_{-N}$ gives 
\begin{align}
SO(N)_K \quad\longleftrightarrow\quad SO(K)_{-N}~.
\end{align}
Comparing with (\ref{eqn:omodd})  produces the spin Chern-Simons duality
\begin{align}
O(N)^1_{K,K-1}\quad\longleftrightarrow\quad O(K)^1_{-N,-N+1}~.
\end{align}

Therefore for $N$ or $K$ odd we establish the same dualities (\ref{eqn:deroddspindual}), and (\ref{lrct}) follows.

\section{Low Rank Chiral Algebras and Level-Rank Duality}
\label{so2so4sec}

In this appendix we present some details of familiar chiral algebras and their relation to level-rank duality.

\subsection{Chiral Algebras Related to $\frak{so}(2)$}

Let us describe the chiral algebras related to the Lie algebra $\frak{so}(2).$\footnote{
For simplicity we will not discuss $Pin^-(2)_k$.
}  We consider
\begin{equation}
SO(2)_k=U(1)_k~, \hspace{.5in}Spin(2)_k=U(1)_{4k}~, \hspace{.5in}O(2)_{k,0}~, \hspace{.5in}Pin^+(2)_k\leftrightarrow  O(2)_{4k,0}~. \label{so2exapp}
\end{equation}

We start with $Pin^+(2)_{k,0}$, its chiral algebra is the charge conjugation orbifold of $Spin(2)_k=U(1)_{4k}$ that takes charge $Q$ to $-Q$ mod $4k$.
This chiral algebra was constructed in \cite{Dijkgraaf:1989hb}.  The spectrum is built from $U(1)_{4k}$ as:
\begin{itemize}
\item the identity in $U(1)_{4k}$ is split into the invariant part 1 and $j=i\partial\phi$. They have conformal dimensions 0,1.

\item the original primary of charge $2k$ splits into a symmetric part $\phi_{2k}^1$ and an antisymmetric part $\phi^2_{2k}$. They have conformal dimensions $k/2$.

\item the original primary of charge $q$ is identified with charge $4k-q$, thus we have the states $\phi_q$ with $q=1,2,\cdots,2k-1$. 
The conformal dimension is $q^2/(8k)$.

\item the twisted sectors of the original identity and primary of charge $2k$.
Denote them by $\sigma^1,\sigma^2$ of conformal dimension ${1\over 16}$ and $\tau^1,\tau^2$ of conformal dimension ${9\over 16}$.
\end{itemize}

The modular $S$ matrix is summarized in table \ref{tab:Ssotwo}  \cite{Dijkgraaf:1989hb}.
\begin{table}[h!]
\begin{center}
\begin{tabular}{ c|cccccc } 
  &  1 & $j$ & $\phi^{a}_{2k}$ & $\phi_q$ & $\sigma^a$ & $\tau^a$ \\
 \hline
1 & 1 & 1 & 1 & 2 & $\sqrt{2k}$ & $\sqrt{2k}$ \\
$j$ & 1 & 1 & 1 & 2 & $-\sqrt{2k}$ & $-\sqrt{2k}$ \\
$\phi_{2k}^{b}$ & 1 & 1 & 1 & $2(-1)^q$ & $(-1)^{a+b}\sqrt{2k}$ & $(-1)^{a+b}\sqrt{2k}$ \\
$\phi_{q'}$ & 2 & 2 & $2(-1)^{q'}$ & $4\cos{\pi qq'\over 4k}$ & $0$ & $0$ \\
$\sigma^{b}$ & $\sqrt{2k}$ & $-\sqrt{2k}$ & $(-1)^{a+b}\sqrt{2k}$ & 0 & $\delta_{a,b}\sqrt{2k}$ & $-\delta_{a,b}\sqrt{2k}$ \\
$\tau^b$ & $\sqrt{2k}$ & $-\sqrt{2k}$ & $(-1)^{a+b}\sqrt{2k}$ & 0 & $-\delta_{a,b}\sqrt{2k}$ & $\delta_{a,b}\sqrt{2k}$ 
\end{tabular}
\end{center}
\caption{Modular $S$ matrix for $Pin^+(2)_k$, normalized by $S_{1,1}=1/(4\sqrt{k})$.
The labels are $a,b=1,2$, $q,q'=1,\cdots 2k-1$.}
\label{tab:Ssotwo}
\end{table}

The fusion rules can be obtained from the modular $S$ matrix (see \cite{Dijkgraaf:1989hb}). In particular there are $\mathbb{Z}_2\times\mathbb{Z}_2$ simple currents $1,j,\phi^1_{2k},\phi^2_{2k}$. They have conformal weights $0,1,k/2,k/2$ respectively.  These currents are the Abelian anyons generating the one-form global symmetry.  In particular by extending the chiral algebra by these currents we can obtain the other algebras in \eqref{so2exapp}.
\begin{itemize}
\item Extending the chiral algebra $Pin^+(2)_k$ by $j$.
From the modular $S$ matrix we find that $\sigma^a,\tau^a$ are projected out, and from the fusion rules we find that $\phi^1_{2k},\phi^2_{2k}$ are identified. Thus the resulting chiral algebra is $U(1)_{4k}=Spin(2)_k$.

\item 
Extending the chiral algebras $Pin^+(2)_k$ by $\phi_{2k}^1$.
From the modular $S$ matrix we find that $\sigma^2,\tau^2$ and $\phi_q$ with odd $q$ are projected out.

For even $k$ the extending primary $\phi_{2k}^1$ has integral conformal weight, and
from the fusion rules we find that $j,\phi^2_{2k}$ are identified, $\phi_q,\phi_{2k-q}$ are identified (for $q\neq k$), and $\phi_k,\sigma^1,\tau^1$ are doubled. Thus the spectrum in the new theory is
$1,j,\phi_k^\pm,\sigma_\pm^1,\tau^1_\pm$ and $\phi_2,\phi_4\cdots\phi_{k-2}$ i.e. $k/2+7$ primaries. Thus the chiral algebra is $O(2)_{k,0}$.

For odd $k$ the extending primary $\phi_{2k}^1$ has half-integral conformal weight thus there is no doubling and the primaries differed by fusing with $\phi_{2k}^1$ are not identified. 
Thus the spectrum in the new theory is $1,j,\phi_{2k}^1,\phi_{2k}^2,\sigma^1,\tau^1$ and $\phi_2,\phi_4,\cdots,\phi_{2k-2}$ together $k+5$ primaries.
The resulting chiral algebra is $O(2)_{k,0}$.
This corresponds to a spin TQFT.

\item
Extending the chiral algebras $Pin^+(2)_k$ with $\phi_{2k}^2$.
Since the modular $S$ matrix respects $\phi_{2k}^1\leftrightarrow\phi_{2k}^2,\sigma^1\leftrightarrow \sigma^2,\tau^1\leftrightarrow \tau^2$, we find that the extended chiral algebra is also $O(2)_{k,0}$.  This is consistent with the fact that the discrete $\theta$-parameter differentiating $O(N)^{1}$ from $O(N)^{0}$ does not exist for $N=2$, and further agrees with the classification of bosonic topological gauge theory with $O(2)$ gauge group \cite{Dijkgraaf:1989pz,Chen:2011pg,Wen:2014zga}.

\end{itemize}

\subsection{Chiral Algebras Related to $\frak{so}(4)$}

Next we discuss chiral algebras related to $\frak{so}(4)$.  The simplest is $Spin(4)_k$ which is simply the product $SU(2)_k\times SU(2)_k.$

The chiral algebra $Pin^+(4)_{k,0}$ is the charge conjugation orbifold of $Spin(4)_k$ that acts on representations by $(j_1,j_2)\leftrightarrow (j_2,j_1)$ where $j_a$ are $SU(2)\times SU(2)$ spins. The spectrum is (see e.g. \cite{Borisov:1997nc}):

\begin{itemize}
\item Off-diagonal fields $(j_1,j_2)$ with $j_1\neq j_2$. There are $[(k+1)^2-(k+1)]/2=k(k+1)/2$ of them.
Their conformal weight is
\begin{equation}
\label{hoffdiag}
h=\frac{j_1(j_1+1)}{k+2}+\frac{j_2(j_2+1)}{ k+2}~.
\end{equation}

\item
Diagonal fields $j_1=j_2=j$ denoted by $(j,s)$ where $s=0,1$ denotes two states in the (two dimensional) untwisted sector.
There are $(k+1)$ of them.
Their conformal weight is
\begin{equation}
\label{hdiaguntwist}
h=\frac{2j(j+1)}{k+2}~.
\end{equation}

\item
Twisted diagonal fields $j_1=j_2=j$ denoted by $\widehat{(j,s)}$, where $s=0,1$ denotes two states in the (two-dimensional) twisted sector.
Their conformal weight is
\begin{equation}
\label{hdiaguntwist}
h={j(j+1)\over 2(k+2)}+{3k\over 16(k+2)}+{s\over 2}~.
\end{equation}

\end{itemize}
 
The modular $S$ matrix is given in \cite{Borisov:1997nc}. We list some of the entries
\begin{eqnarray}
S_{(i,j),(i',j')}&=&S^{SU(2)}_{i,i'}S^{SU(2)}_{j,j'}+S^{SU(2)}_{i,j'}S^{SU(2)}_{j,i'}~, \nonumber\\
S_{(i,j),(i',s)}&=&S^{SU(2)}_{i,i'}S_{j,i'}^{SU(2)}~,\nonumber \\
S_{(i,j),\widehat{(i',s)}}&=&0~,  \label{SmatrixPinfour} \\
S_{(i,s),(i',s')}&=&{1\over 2}\left( S^{SU(2)}_{i,i'}\right)^2~,\nonumber\\
S_{(i,s),\widehat{(i',s')}} &=&{1\over 2}(-1)^s S^{SU(2)}_{i,i'}~.\nonumber
\end{eqnarray}
Denote by $1,J,\sigma$ and $J\sigma$ the $\mathbb{Z}_2\times\mathbb{Z}_2$ simple currents $(j=0,s=0),(j=0,s=1),(j={k\over 2},s=0)$ and $(j={k\over 2},s=1)$ of conformal weights $0,1,{k\over 2},{k\over 2}$ respectively.
In particular, extending the chiral algebra $Pin^+(4)_{k,0}$ with $J$ produces $Spin(4)_k$ by projecting out the twisted sectors, identifying $s=0,1$ for the untwisted diagonal fields and doubling the off-diagonal fields.

Consider extending the chiral algebra $Pin^+(4)_{k,0}$ with $\sigma$.
The resulting theory is $O(4)^0_{k,0}$. For odd $k$ there are no fixed modules under fusion with $\sigma$, and therefore the spectrum is the subset of primaries in $Pin^+(4)_{k,0}$ that are mutually local with $\sigma.$  Namely, we exclude the off-diagonal fields $(j_1,j_2)$ for $j_1,j_2$ not both $SU(2)$ tensor or both spinor, and the diagonal field $\widehat{(j',s)}$ for $SU(2)$ spinor $j'$.
For even $k$ we need to take into account the identification and doubling, and the spectrum of $O(4)^0_{k,0}$ theory is
\begin{itemize}
\item
Off-diagonal fields $(j_1,j_2)\oplus (j_2,j_1)\sim ({k\over 2}-j_1,{k\over 2}-j_2)\oplus ({k\over 2}-j_2,{k\over 2}-j_1)$ with $j_1,j_2$ both either $SU(2)$ tensor or spinor, and $j_1\neq j_2$ or ${k\over 2}-j_2$. There are doubled representations $[(j,{k\over 2}-j)\oplus({k\over 2}-j,j)]_\pm$ for $j<{k\over 4}$.

\item
Diagonal fields $(j,s)\oplus ({k\over 2}-j,s)$ for $j<{k\over 4}$ and the doubled representation $({k\over 4},s)_\pm$.

\item
Twisted diagonal fields $\widehat{(j,s)}_\pm$ with $SU(2)$ tensor $j=0,1,\cdots {k\over 2}$ (all of them are doubled).

\end{itemize}

Instead, consider extending the chiral algebra $Pin^+(4)_k$ with $\sigma J$.
The resulting theory is $O(4)^1_{k,0}$.
For odd $k$ the spectrum is the original spectrum without the primaries that are projected out, namely without the off-diagonal fields $(j_1,j_2)$ for $j_1,j_2$ not both $SU(2)$ tensor or both spinor, and without the diagonal field $\widehat{(j',s)}$ for $SU(2)$ tensor $j'$.
For even $k$ we need to take into account the identification and doubling, and the spectrum of $O(4)^1_k$ theory is
\begin{itemize}
\item
Off-diagonal fields the same as that of $O(4)^0_k$.

\item
Diagonal fields $(j,s=0)\oplus({k\over 2}-j,s=1)$ with $j=0,{1\over 2},1,\cdots {k\over 2}$.

\item
Twisted diagonal fields $\widehat{(j,s)}_\pm$ with $SU(2)$ spinor $j={1\over 2},{3\over 2},\cdots {k-1\over 2}$ (all of them are doubled).

\end{itemize}
Note that unlike the previous section, which discussed $O(2)_{k}$ chiral algebras, here we see that $O(4)^0_k,O(4)^1_k$ have different twisted sectors and are distinct theories.

As a check,  the spins match in the spin Chern-Simons duality
\begin{equation}
\label{omomfour}
O(4)^1_{4,3} \quad\longleftrightarrow\quad O(4)^1_{-4,-3}~.
\end{equation}
Here, the $\mathbb{Z}_2$ quotient in $O(4)^1_{4,3}$ is generated by the product of $J$ and the basic Wilson line in $(\mathbb{Z}_{2})_{3}$. 
The quotient pairs the twisted sector with the magnetic lines of $(\mathbb{Z}_{2})_{3}$.
In particular the lowest-dimension twisted-sector primary matches in the two theories: the conformal weight of $\widehat{(1/2,0)}$ is ${3\over 16}$, which is cancelled up to an integer by the basic magnetic line of $(\mathbb{Z}_{2})_{3}$.  Thus this primary maps to itself under the duality \eqref{omomfour}.

\section{Projective Representations in $SO(4)_4$ with ${\cal CM}\neq {\cal MC}$}
\label{ProjectiveReps}

The anyons of $SO(4)_4$ can be labelled by spin $(j_1,j_2)$ of $Spin(4)=SU(2)\times SU(2)$, with the extending representation $\sigma= (2,2)$.  It acts as by fusion as $\sigma\cdot (j_1,j_2)=(2-j_{1},2-j_{2})$.
Fixed points under fusion with $\sigma$ lead to a pair of distinct representations of the extended chiral algebra, denoted by $(j_1,j_2)_\pm$.
Meanwhile, for $(j_1,j_2)\neq \sigma \cdot (j_1,j_2)$ the representation of the extended chiral algebra is $(j_1,j_2)\oplus \sigma \cdot(j_1,j_2)$.

The resulting list of anyons of $SO(4)_4,$ together with their conformal weights $h=\frac{1}{6}\left(j_1(j_1+1)+j_2(j_2+1)\right),$ are summarized in table \ref{tabso44}.
\begin{table}[h!]
\begin{center}
\begin{tabular}{|c|c|c|c|c|c|c|c|c|}
\hline
$(j_{1}, j_{2})$ & $(0,0)$& $(0,1)$& $(1,0)$& $(1,1)_{+}$& $(1,1)_{-}$& $(0,2)$& $(1/2,1/2)$& $(1/2,3/2)$\\
\hline
$h$ & $0$ &$1/3$ &$1/3$ &$2/3$ &$2/3$ &$1$ &$1/4$ &$3/4$ \\
\hline
\end{tabular}
\end{center}
\caption{Anyons and their conformal dimensions for the theory $SO(4)_{4}.$}
\label{tabso44}
\end{table}

The chiral algebra has $\mathbb{Z}_2\times\mathbb{Z}_2$ anyon permutation symmetry ${\cal C}$, ${\cal M}$ defined by
\begin{align}
{\cal C}:\quad (0,1)&\;\longleftrightarrow\; (1,0)\cr
{\cal M}:\quad (1,1)_+&\;\longleftrightarrow\; (1,1)_-~.
\end{align}
In addition the theory has time-reversal symmetry that permutes the anyons as $(0,1)\leftrightarrow (1,1)_+$, 
$(1,0)\leftrightarrow (1,1)_-$ and $(\frac{1}{2},\frac{1}{2})\leftrightarrow (\frac{1}{2},\frac{3}{2})$. 
We will focus on the unitary symmetries ${\cal C,M}$.

Consider anyons that are not permuted by ${\cal C,M}$.
${\cal M}$ assigns $\pm1$ to the two states in the representation $(j_1,j_2)\oplus \sigma\cdot (j_1,j_2)$ respectively.
${\cal C}$ is inherited from the symmetry that exchanges $(j_1,j_2)$ as $Spin(4)$ representations.
${\cal C}$ may not preserve the value assigned by ${\cal M}$.  This happens when
$\sigma\cdot(j_1,j_2)={\cal C}(j_1,j_2)=(j_2,j_1)$, namely the anyon $(j_1,j_2)$ is invariant under ${\cal C}$ only up to fusion with the extending representation $\sigma$.
This occurs for the primaries $(j_1,j_2)=(0,2),(\frac{1}{2},\frac{3}{2})$, and they realize the symmetry projectively with $({\cal CM})^2=-1$.

The value of $({\cal CM})^2$ for each anyon can be changed by a sign that preserves the fusion rules and anyon permutations.
The sign comes from braiding the anyon with the Abelian anyon $(0,2)$, which is not permuted by ${\cal C,M}$. Thus the values of $({\cal CM})^2$ for
 $({1\over 2},{1\over 2}),({1\over 2},{3\over 2})$ can be $+1,-1$ or $-1,+1$, while the values for $(0,0),(0,2)$ stay unchanged.

Consider the following two orbits of three-punctured spheres under the permutation ${\cal C,M}$:
\begin{align*}
&\left\{ |(0,1),(0,1),(0,0)\rangle,\;|(1,0),(1,0),(0,0)\rangle\right\}\quad{\rm and}\cr
&\left\{|(0,1),(0,1),(0,2)\rangle,\;|(1,0),(1,0),(2,0)\rangle\right\}~.
\end{align*}
Since $({\cal CM})^2=+1$ for $(0,0)$ and $({\cal CM})^2=-1$ for $(0,2)$, the two orbits of three-punctured spheres cannot both be in a linear representation of ${\cal C,M}$. 
Thus, the three-punctured spheres in general are in projective representation of ${\cal C,M}.$\footnote{
We thank M. Barkeshli and M. Cheng for pointing out the same result can be derived from the method of \cite{Barkeshli:2014cna} for $SO(4)_4$ as an abstract TQFT.}

This does not occur in every orbit of three-punctured spheres (in particular, the sphere without anyons is in the trivial linear representation of the symmetry), and it does not imply the symmetries ${\cal C,M}$ have an `t Hooft anomaly. 
In fact gauging ${\cal C,M}$ without counterterm produces the well-defined $Pin^+(4)_4$ theory.

\section{$O(2)_{2,L}$ as Family of Pfaffian States}
\label{sec:OtwoPfaffian}

In this appendix we show that the Pfaffian and T-Pfaffian theories are special cases of the spin Chern-Simons theory $O(2)_{2,L}$ for odd $L$. For time-reversal invariant theories, we also analyze the mixed anomaly between the time-reversal and $U(1)$ symmetries.

The theories $O(2)_{2,L}$ as spin Chern-Simons theories can be expressed as
\begin{align}\label{eqn:otwotwo}
O(2)_{2,L}\quad\longleftrightarrow\quad \frac{U(1)_8\times Spin(L)_{-1}}{\mathbb{Z}_2}~,
\end{align}
where we used $O(2)_{2,0}\leftrightarrow U(1)_8$ and (\ref{eqn:defztwoL}).
For odd $L$ the theory has 12 lines, with non-Abelian fusion algebra that can be read off from the right hand side.

The Moore-Read Pfaffian theory \cite{MOORE1991362} can be expressed as the spin Chern-Simons theory \cite{Seiberg:2016rsg}
\begin{align}
\frac{SU(2)_2\times U(1)_{-4}\times U(1)_8}{\mathbb{Z}_2\times\mathbb{Z}_2}~,
\end{align}
where the first $\mathbb{Z}_2$ quotient acts on $SU(2)_2\times U(1)_{-4}$ and the second quotient acts on  $U(1)_{-4}\times U(1)_8$.
To simplify the quotients we can use the duality $U(1)_{-4}\times U(1)_8\leftrightarrow U(1)_{4}\times U(1)_{-8}$ (which can be proven by a change of variables as in \cite{Seiberg:2016rsg}).
In the new variables, the first $\mathbb{Z}_2$ quotient acts diagonally on $SU(2)_2\times U(1)_{4}\times U(1)_{-8}$, while
the second $\mathbb{Z}_2$ quotient acts on $U(1)_4$ and turns it into the trivial spin TQFT $U(1)_1$.
Thus the quotient simplifies to
\begin{align}
\text{Moore-Read}\quad=\quad\frac{SU(2)_2\times U(1)_{-8}}{\mathbb{Z}_2}\times U(1)_1~.
\end{align}
Since $SU(2)_2\cong Spin(3)_1$, the Moore-Read theory is equivalent to $O(2)_{-2,-3}\leftrightarrow O(2)_{2,7}$ (up to a gravitational Chern-Simons term, which we ignore), where we used the second duality in \eqref{lrgaugedi}.
The spins of the 12 lines in the theory modulo integers are $\{0,0,\frac{1}{2},\frac{1}{2},\pm\frac{1}{4},\pm\frac{1}{4}\}$ and $\{\frac{1}{8},\frac{1}{8},\frac{5}{8},\frac{5}{8}\}$, and thus the theory is not time-reversal invariant.  The time-reversal of the Moore-Read Pfaffian theory is also called the anti-Pfaffian theory, and it is dual to $O(2)_{2,3}$.

Both $O(2)_{2,1}$ and $O(2)_{2,5}$ are time-reversal invariant\footnote{
In fact, from the spins of lines one can conclude that $O(2)_{2,L}$ for other integers $L\neq 1$ mod 4 are not time-reversal invariant.
}
 by the level-rank duality (\ref{lrgaugedi}):
\begin{align}\label{eqn:OtwoTsym}
O(2)_{2,1}&\quad\longleftrightarrow\quad O(2)_{-2,-1}\cr
O(2)_{2,5}&\quad\longleftrightarrow\quad O(2)_{-2,-5}~.
\end{align}
The theory $O(2)_{2,1}$ is equivalent to the T-Pfaffian theory \cite{Bonderson:2013pla,Chen:2013jha}, which can be expressed as $\left(U(1)_{8}\times {\overline{\rm Ising}}\right)/\mathbb{Z}_2$.
The theory $O(2)_{2,5}$ has the same fusion algebra as $O(2)_{2,1}$, but the spin of the twist field in the Ising anyons is different (spin $-\frac{5}{16}$ instead of $-\frac{1}{16}$).
In particular, both $O(2)_{2,1}$ and $O(2)_{2,5}$ have $\mathbb{Z}_4$ one-form symmetry.
They are related by
\begin{align}
O(2)_{2,5} \quad\longleftrightarrow\quad { O(2)_{2,1}\times (\mathbb{Z}_2)_4\over \mathbb{Z}_2}~,
\end{align}
where the quotient on $(\mathbb{Z}_2)_4$ uses the Wilson line (of integral spin). In particular, this means that the theory is not defined by a quotient of the gauge group in the numerator.  (See the discussion around \eqref{eqn:GKL}.)  Note $(\mathbb{Z}_2)_4\leftrightarrow U(1)_2\times U(1)_{-2}$ is the semion-antisemion theory, and it is time-reversal invariant.
Since $O(2)_{2,5}= O(2)_{2,-3}$ and $(\mathbb{Z}_2)_{-3}\leftrightarrow Spin(3)_1\cong SU(2)_2$, the spin Chern-Simons theories $O(2)_{2,5}$ and $U(2)_{2,4}$ are dual, where $U(2)_{2,4}$ is time-reversal invariant by the level-rank duality (\ref{SUlevelranki}).

We can couple the theories  $O(2)_{2,1},O(2)_{2,5}$ to a background $U(1)$ gauge field $A$ and investigate the mixed anomaly between the $U(1)$ and the time-reversal symmetries. The $O(2)$ Chern-Simons theory does not have $U(1)$ magnetic symmetry, but it can couple to $U(1)$ symmetry using its one-form symmetry.\footnote{
This is an example of coupling a TQFT to an ordinary global symmetry using the Abelian anyons as in \cite{Barkeshli:2014cna} for discrete symmetries. Here we couple $O(2)_{2,L}$ to the continuous $U(1)$ symmetry that does not permute anyons.
The $O(2)_{2,L}$ theory also has an intrinsic $\mathbb{Z}_2$ magnetic symmetry that permutes anyons (it is not the $\mathbb{Z}_2$ subgroup of the $U(1)$ symmetry).
}

Both $O(2)_{2,1}$ and $O(2)_{2,5}$ have $\mathbb{Z}_4$ one-form symmetry, and the generating line has spin $\frac{1}{4}$.
Consider coupling the dualities (\ref{eqn:OtwoTsym}) to the two-form background $\mathbb{Z}_4$ gauge field $B_2$ (with the period in $(2\pi/ 4)\mathbb{Z}$) for the one-form symmetry:
\begin{align}
O(2)_{2,1}[B_2]+\frac{8}{4\pi}\int_{\rm bulk} B_2B_2&\quad\longleftrightarrow\quad O(2)_{-2,-1}[B_2]-\frac{8}{4\pi}\int_{\rm bulk} B_2B_2\cr
O(2)_{2,5}[B_2]+\frac{8}{4\pi}\int_{\rm bulk} B_2B_2&\quad\longleftrightarrow\quad O(2)_{-2,-5}[B_2]-\frac{8}{4\pi}\int_{\rm bulk} B_2B_2~,
\end{align}
where the bulk terms $\pm\frac{8}{4\pi}\int B_2B_2$ are equivalent on a closed four-manifold as expected from the anomaly matching for the dualities.
By substituting the value for the background $B_2=\frac{1}{4}dA$ we find the mixed anomaly 
$\frac{1}{8\pi}\int dAdA$ between the time-reversal symmetry and the $U(1)$ symmetry.\footnote{
The analysis of the mixed anomaly can be generalized to the theories $SO(N)_N$, $O(N)^1_{N,N-1}$ and $O(N)^1_{N,N+3}$, which are time-reversal invariant by the level-rank dualities (\ref{SOlevelranki}),(\ref{lrgaugedi}).  When $N=2$ mod 4 they have one-form symmetry with non-trivial `t Hooft anomaly.
}
This is the same parity anomaly as that of one massless Dirac fermion.
The lines can carry fractional $U(1)$ charges determined by their charges under the one-form symmetry.

\section{Duality Via One-form Symmetry}
\label{dualityoneform}

Consider a TQFT ${\cal T}$ with an Abelian anyon $a$ of spin $L/(2N)$ mod 1 that generates a $\mathbb{Z}_N$ one-form symmetry (we assume $NL$ is even for simplicity).\footnote{
For example, ${\cal T}$ can be the Chern-Simons theory with gauge algebra $\frak{so}(N)$, with the one-form symmetry discussed in section \ref{reviewso}.
}
Then we can prove the following duality\footnote{Notice the intentional change of font below.  The theory $(\mathbb{Z}_{2})_{L}$ is not the same as the theory $(Z_{N})_{L}$ when $N=2.$ They are related by (\ref{acsdescription}) (i.e. $(Z_2)_{L}\leftrightarrow (\mathbb{Z}_2)_{2L}$). }
\begin{align}\label{eqn:oneformdual}
{\rm Even}\ L:&\quad {\cal T}\quad\longleftrightarrow\quad {{\cal T}\times (Z_N)_{0}\over \mathbb{Z}_N}\cr
{\rm Odd}\ L:&\quad {\cal T}\quad\longleftrightarrow\quad {{\cal T}\times (Z_N)_{N}\over \mathbb{Z}_N}~,
\end{align}
where $(Z_N)_{k}$ denotes the Abelian Chern-Simons theory $\frac{k}{4\pi}xdx+\frac{N}{2\pi}xdy$ with $U(1)$ gauge fields $x,y$.
The $\mathbb{Z}_N$ quotient on the $(Z_N)_0$ and $(Z_N)_N$ theories are generated by the line $\exp\left(i\oint y-i{L\over 2}\oint x\right)$ and $\exp\left(i\oint y-i{L-1\over 2}\oint x\right)$ with even and odd $L$ respectively.
(Note that $L$ is defined mod $2N$.)

To prove \eqref{eqn:oneformdual} denote the $\mathbb{Z}_N$ two-form gauge field of the gauged one-form symmetry on the right by $B_2$ and couple it to
 $(Z_N)_{0}$ and $(Z_N)_N$ for even and odd $L$ as
\begin{align}\label{eqn:ZNcoupl}
{\rm Even}\ L:\quad &\int_{X=\partial M_4}\left( {N\over 2\pi}xdy + {N\over 2\pi}B_2(y-{L\over 2}x)\right) -{NL\over 4\pi}\int_{M_4} B_2B_2\cr
{\rm Odd}\ L:\quad&\int_{X=\partial M_4}\left( {N\over 4\pi}xdx+{N\over 2\pi}xdy + {N\over 2\pi}B_2(y-{L-1\over 2}x)\right) -{NL\over 4\pi}\int_{M_4} B_2B_2~,
\end{align}
and the one-form gauge transformation is
\begin{align}
{\rm Even}\ L:\quad &B_2\rightarrow B_2-d\lambda,\; x\rightarrow x+\lambda,\; y\rightarrow y-{L\over 2}\lambda\cr
{\rm Odd}\ L:\quad &B_2\rightarrow B_2-d\lambda,\; x\rightarrow x+\lambda,\; y\rightarrow y-{L+1\over 2}\lambda~.
\end{align}
In (\ref{eqn:ZNcoupl}) the last terms $B_2B_2$ can be ignored, since they will be cancelled by the `t Hooft anomaly of the $\mathbb{Z}_N$ one-form symmetry in ${\cal T}$.  In this step, we use the fact that the `t Hooft anomaly of a one-form symmetry is given by the self-braiding of the generating line\cite {Kapustin:2014gua, Gaiotto:2014kfa,GomisKS:unpublish}.

To simplify (\ref{eqn:ZNcoupl}) we can perform the one-form gauge transformation with $\lambda=-x$, which change the gauge fields $(B_2,x,y)$ to $(B_2+dx,0,y+Lx/2)$ for even $L$ and $(B_2+dx,0,y+(L+1)x/2)$ for odd $L$. 
Then (\ref{eqn:ZNcoupl}) becomes $\int_X N(B_2+dx)\tilde y/(2\pi)$ where $\tilde y=y+Lx/2$ for even $L$ and $\tilde y=y+(L+1)x/2$ for odd $L$.
Integrating out $\tilde y$ constrains $B_2$ to be a trivial two-form $\mathbb{Z}_N$ gauge field and removes the quotient.  Therefore the theory on the right of (\ref{eqn:oneformdual}) is the original theory ${\cal T}$.

A simple example illustrating the duality (\ref{eqn:oneformdual}) is the relationship between the $SU(N)$ Chern-Simons theory, and $U(N)$ Chern-Simons theory coupled to a $U(1)$ multiplier that constrains the $U(N)$ gauge field \cite{Kapustin:2014gua,Hsin:2016blu}.
Consider $SU(N)_K\times (Z_N)_{NK}$ with the Lagrangian
\begin{align}\label{eqn:oneformdualeg}
{K\over 4\pi}{\rm Tr}\left[bdb+\frac{2}{3}b^3\right] + \frac{NK}{4\pi}xdx +\frac{N}{2\pi}xdy~,
\end{align}
where $b$ is an $SU(N)$ gauge field, and $x,y$ are $U(1)$ gauge fields.
For simplicity we will take $K$ to be even.
Next we perform a $\mathbb{Z}_N$ quotient such that $b,x$ are not properly quantized but the $U(N)$ gauge field $u=b + {\mathbf 1}_N x$ is well-defined. 
The Lagrangian in the new variables is
\begin{align}
{K\over 4\pi}{\rm Tr}\left[udu+\frac{2}{3}u^3\right] + \frac{1}{2\pi}({\rm Tr}\; u)dy~,
\end{align}
which is the $U(N)_{K,K}$ Chern-Simons action constrained by a $U(1)$ multiplier $y$, and we recognize it as the $SU(N)_K$ Chern-Simons action \cite{Hsin:2016blu}.
Thus we find that for even $K$,
\begin{align}
SU(N)_K\quad\longleftrightarrow\quad \frac{SU(N)_K\times (Z_N)_{NK}}{\mathbb{Z}_N}~,
\end{align}
where the quotient on $(Z_N)_{NK}$ leads to selection rule on Wilson line of $x$ but not for $y$, namely it is generated by $\exp(i\oint y+iK\oint x)$.
For even $K$ we can use $(Z_N)_{NK}\cong (Z_N)_0$ by the redefinition $y\rightarrow y-{K\over 2}x$, then the quotient on $(Z_N)_0$ is generated by the line $\exp(i\oint y+i{K\over 2}\oint x)$. This agrees with the duality (\ref{eqn:oneformdual}), where the generating line of the $\mathbb{Z}_N$ one-form symmetry in $SU(N)_K$ has spin $-K/(2N)$ and thus $L=-K$.


\bibliographystyle{utphys}
\bibliography{SO(N)LevelRank}{}

\providecommand{\href}[2]{#2}\begingroup\raggedright\begin{thebibliography}{10}

\bibitem{Kapustin:2014gua}
A.~Kapustin and N.~Seiberg, ``{Coupling a QFT to a TQFT and Duality},''
  \href{http://dx.doi.org/10.1007/JHEP04(2014)001}{{\em JHEP} {\bfseries 04}
  (2014) 001},
\href{http://arxiv.org/abs/1401.0740}{{\ttfamily arXiv:1401.0740 [hep-th]}}.

\bibitem{Gaiotto:2014kfa}
D.~Gaiotto, A.~Kapustin, N.~Seiberg, and B.~Willett, ``{Generalized Global
  Symmetries},'' \href{http://dx.doi.org/10.1007/JHEP02(2015)172}{{\em JHEP}
  {\bfseries 02} (2015) 172},
\href{http://arxiv.org/abs/1412.5148}{{\ttfamily arXiv:1412.5148 [hep-th]}}.

\bibitem{Aharony:2013hda}
O.~Aharony, N.~Seiberg, and Y.~Tachikawa, ``{Reading between the lines of
  four-dimensional gauge theories},''
  \href{http://dx.doi.org/10.1007/JHEP08(2013)115}{{\em JHEP} {\bfseries 08}
  (2013) 115},
\href{http://arxiv.org/abs/1305.0318}{{\ttfamily arXiv:1305.0318 [hep-th]}}.

\bibitem{Aharony:2013kma}
O.~Aharony, S.~S. Razamat, N.~Seiberg, and B.~Willett, ``{3$d$ dualities from
  4$d$ dualities for orthogonal groups},''
  \href{http://dx.doi.org/10.1007/JHEP08(2013)099}{{\em JHEP} {\bfseries 08}
  (2013) 099},
\href{http://arxiv.org/abs/1307.0511}{{\ttfamily arXiv:1307.0511 [hep-th]}}.

\bibitem{Komargodski:2017keh}
Z.~Komargodski and N.~Seiberg, ``{A symmetry breaking scenario for
  QCD$_{3}$},'' \href{http://dx.doi.org/10.1007/JHEP01(2018)109}{{\em JHEP}
  {\bfseries 01} (2018) 109},
\href{http://arxiv.org/abs/1706.08755}{{\ttfamily arXiv:1706.08755 [hep-th]}}.

\bibitem{Gomis:2017ixy}
J.~Gomis, Z.~Komargodski, and N.~Seiberg, ``{Phases Of Adjoint QCD$_3$ And
  Dualities},''
\href{http://arxiv.org/abs/1710.03258}{{\ttfamily arXiv:1710.03258 [hep-th]}}.

\bibitem{Cordova:2017kue}
C.~Cordova, P.-S. Hsin, and N.~Seiberg, ``{Time-Reversal Symmetry, Anomalies,
  and Dualities in (2+1)$d$},''
\href{http://arxiv.org/abs/1712.08639}{{\ttfamily arXiv:1712.08639
  [cond-mat.str-el]}}.

\bibitem{Closset:2012vg}
C.~Closset, T.~T. Dumitrescu, G.~Festuccia, Z.~Komargodski, and N.~Seiberg,
  ``{Contact Terms, Unitarity, and F-Maximization in Three-Dimensional
  Superconformal Theories},''
  \href{http://dx.doi.org/10.1007/JHEP10(2012)053}{{\em JHEP} {\bfseries 10}
  (2012) 053},
\href{http://arxiv.org/abs/1205.4142}{{\ttfamily arXiv:1205.4142 [hep-th]}}.

\bibitem{Closset:2012vp}
C.~Closset, T.~T. Dumitrescu, G.~Festuccia, Z.~Komargodski, and N.~Seiberg,
  ``{Comments on Chern-Simons Contact Terms in Three Dimensions},''
  \href{http://dx.doi.org/10.1007/JHEP09(2012)091}{{\em JHEP} {\bfseries 09}
  (2012) 091},
\href{http://arxiv.org/abs/1206.5218}{{\ttfamily arXiv:1206.5218 [hep-th]}}.

\bibitem{Freed:2004yc}
D.~S. Freed and G.~W. Moore, ``{Setting the quantum integrand of M-theory},''
  \href{http://dx.doi.org/10.1007/s00220-005-1482-7}{{\em Commun. Math. Phys.}
  {\bfseries 263} (2006) 89--132},
\href{http://arxiv.org/abs/hep-th/0409135}{{\ttfamily arXiv:hep-th/0409135
  [hep-th]}}.

\bibitem{Seiberg:2010qd}
N.~Seiberg, ``{Modifying the Sum Over Topological Sectors and Constraints on
  Supergravity},'' \href{http://dx.doi.org/10.1007/JHEP07(2010)070}{{\em JHEP}
  {\bfseries 07} (2010) 070},
\href{http://arxiv.org/abs/1005.0002}{{\ttfamily arXiv:1005.0002 [hep-th]}}.

\bibitem{Freed:2017rlk}
D.~S. Freed, Z.~Komargodski, and N.~Seiberg, ``{The Sum Over Topological
  Sectors and $\theta$ in the 2+1-Dimensional $\mathbb{C}\mathbb{P}^1$
  $\sigma$-Model},''
\href{http://arxiv.org/abs/1707.05448}{{\ttfamily arXiv:1707.05448
  [cond-mat.str-el]}}.

\bibitem{EtingofNOM2009}
P.~Etingof, D.~Nikshych, V.~Ostrik, and E.~Meir, ``{Fusion Categories and
  Homotopy Theory},'' \href{http://dx.doi.org/10.4171/QT/6}{{\em Quantum
  Topology} {\bfseries 1} (2010) 209},
  \href{http://arxiv.org/abs/0909.3140}{{\ttfamily 0909.3140}}.

\bibitem{Barkeshli:2014cna}
M.~Barkeshli, P.~Bonderson, M.~Cheng, and Z.~Wang, ``{Symmetry, Defects, and
  Gauging of Topological Phases},''
\href{http://arxiv.org/abs/1410.4540}{{\ttfamily arXiv:1410.4540
  [cond-mat.str-el]}}.

\bibitem{Naculich:1990pa}
S.~G. Naculich, H.~A. Riggs, and H.~J. Schnitzer, ``{Group Level Duality in
  {WZW} Models and {Chern-Simons} Theory},''
\href{http://dx.doi.org/10.1016/0370-2693(90)90623-E}{{\em Phys. Lett.}
  {\bfseries B246} (1990) 417--422}.

\bibitem{Mlawer:1990uv}
E.~J. Mlawer, S.~G. Naculich, H.~A. Riggs, and H.~J. Schnitzer, ``{Group level
  duality of WZW fusion coefficients and Chern-Simons link observables},''
\href{http://dx.doi.org/10.1016/0550-3213(91)90110-J}{{\em Nucl. Phys.}
  {\bfseries B352} (1991) 863--896}.

\bibitem{Witten:1993xi}
E.~Witten, ``{The Verlinde algebra and the cohomology of the Grassmannian},''
\href{http://arxiv.org/abs/hep-th/9312104}{{\ttfamily arXiv:hep-th/9312104
  [hep-th]}}.

\bibitem{Douglas:1994ex}
M.~R. Douglas, ``{Chern-Simons-Witten theory as a topological Fermi liquid},''
\href{http://arxiv.org/abs/hep-th/9403119}{{\ttfamily arXiv:hep-th/9403119
  [hep-th]}}.

\bibitem{Naculich:2007nc}
S.~G. Naculich and H.~J. Schnitzer, ``{Level-rank duality of the U(N) WZW
  model, Chern-Simons theory, and 2-D qYM theory},''
  \href{http://dx.doi.org/10.1088/1126-6708/2007/06/023}{{\em JHEP} {\bfseries
  06} (2007) 023},
\href{http://arxiv.org/abs/hep-th/0703089}{{\ttfamily arXiv:hep-th/0703089
  [HEP-TH]}}.

\bibitem{Hsin:2016blu}
P.-S. Hsin and N.~Seiberg, ``{Level/rank Duality and Chern-Simons-Matter
  Theories},'' \href{http://dx.doi.org/10.1007/JHEP09(2016)095}{{\em JHEP}
  {\bfseries 09} (2016) 095},
\href{http://arxiv.org/abs/1607.07457}{{\ttfamily arXiv:1607.07457 [hep-th]}}.

\bibitem{Moore:1988ss}
G.~W. Moore and N.~Seiberg, ``{Naturality in Conformal Field Theory},''
\href{http://dx.doi.org/10.1016/0550-3213(89)90511-7}{{\em Nucl. Phys.}
  {\bfseries B313} (1989) 16--40}.

\bibitem{Moore:1989yh}
G.~W. Moore and N.~Seiberg, ``{Taming the Conformal Zoo},''
\href{http://dx.doi.org/10.1016/0370-2693(89)90897-6}{{\em Phys. Lett.}
  {\bfseries B220} (1989) 422--430}.

\bibitem{Bais:2008ni}
F.~A. Bais and J.~K. Slingerland, ``{Condensate induced transitions between
  topologically ordered phases},''
  \href{http://dx.doi.org/10.1103/PhysRevB.79.045316}{{\em Phys. Rev.}
  {\bfseries B79} (2009) 045316},
\href{http://arxiv.org/abs/0808.0627}{{\ttfamily arXiv:0808.0627
  [cond-mat.mes-hall]}}.

\bibitem{Hasegawa:1989741}
K.~Hasegawa, ``{Spin Module Versions of Weyl's Reciprocity Theorem for
  Classical Kac-Moody Lie Algebras \textemdash An Application to Branching Rule
  Duality},'' \href{http://dx.doi.org/10.2977/prims/1195172705}{{\em
  Publications of the Research Institute for Mathematical Sciences} {\bfseries
  25} no.~5, (1989) 741--828}.

\bibitem{Verstegen:1990at}
D.~Verstegen, ``{Conformal embeddings, rank level duality and exceptional
  modular invariants},''
\href{http://dx.doi.org/10.1007/BF02100278}{{\em Commun. Math. Phys.}
  {\bfseries 137} (1991) 567--586}.

\bibitem{Aharony:2016jvv}
O.~Aharony, F.~Benini, P.-S. Hsin, and N.~Seiberg, ``{Chern-Simons-matter
  dualities with $SO$ and $USp$ gauge groups},''
  \href{http://dx.doi.org/10.1007/JHEP02(2017)072}{{\em JHEP} {\bfseries 02}
  (2017) 072},
\href{http://arxiv.org/abs/1611.07874}{{\ttfamily arXiv:1611.07874
  [cond-mat.str-el]}}.

\bibitem{Seiberg:2016rsg}
N.~Seiberg and E.~Witten, ``{Gapped Boundary Phases of Topological Insulators
  via Weak Coupling},'' \href{http://dx.doi.org/10.1093/ptep/ptw083}{{\em PTEP}
  {\bfseries 2016} no.~12, (2016) 12C101},
\href{http://arxiv.org/abs/1602.04251}{{\ttfamily arXiv:1602.04251
  [cond-mat.str-el]}}.

\bibitem{Benini:2017dus}
F.~Benini, P.-S. Hsin, and N.~Seiberg, ``{Comments on global symmetries,
  anomalies, and duality in (2 + 1)d},''
  \href{http://dx.doi.org/10.1007/JHEP04(2017)135}{{\em JHEP} {\bfseries 04}
  (2017) 135},
\href{http://arxiv.org/abs/1702.07035}{{\ttfamily arXiv:1702.07035
  [cond-mat.str-el]}}.

\bibitem{Bonderson:2013pla}
P.~Bonderson, C.~Nayak, and X.-L. Qi, ``{A time-reversal invariant topological
  phase at the surface of a 3D topological insulator},''
  \href{http://dx.doi.org/10.1088/1742-5468/2013/09/P09016}{{\em J. Stat.
  Mech.} {\bfseries 1309} (2013) P09016},
\href{http://arxiv.org/abs/1306.3230}{{\ttfamily arXiv:1306.3230
  [cond-mat.str-el]}}.

\bibitem{Chen:2013jha}
X.~Chen, L.~Fidkowski, and A.~Vishwanath, ``{Symmetry Enforced Non-Abelian
  Topological Order at the Surface of a Topological Insulator},''
  \href{http://dx.doi.org/10.1103/PhysRevB.89.165132}{{\em Phys. Rev.}
  {\bfseries B89} no.~16, (2014) 165132},
\href{http://arxiv.org/abs/1306.3250}{{\ttfamily arXiv:1306.3250
  [cond-mat.str-el]}}.

\bibitem{Chen:2011pg}
X.~Chen, Z.-C. Gu, Z.-X. Liu, and X.-G. Wen, ``{Symmetry protected topological
  orders and the group cohomology of their symmetry group},''
  \href{http://dx.doi.org/10.1103/PhysRevB.87.155114}{{\em Phys. Rev.}
  {\bfseries B87} no.~15, (2013) 155114},
\href{http://arxiv.org/abs/1106.4772}{{\ttfamily arXiv:1106.4772
  [cond-mat.str-el]}}.

\bibitem{Wen:2014zga}
X.-G. Wen, ``{Construction of bosonic symmetry-protected-trivial states and
  their topological invariants via $G\times SO(\infty)$ non-linear
  $\sigma$-models},'' \href{http://dx.doi.org/10.1103/PhysRevB.91.205101}{{\em
  Phys. Rev.} {\bfseries B91} (2015) 205101},
\href{http://arxiv.org/abs/1410.8477}{{\ttfamily arXiv:1410.8477
  [cond-mat.str-el]}}.

\bibitem{Metlitski:2016dht}
M.~A. Metlitski, A.~Vishwanath, and C.~Xu, ``{Duality and bosonization of (2+1)
  -dimensional Majorana fermions},''
  \href{http://dx.doi.org/10.1103/PhysRevB.95.205137}{{\em Phys. Rev.}
  {\bfseries B95} no.~20, (2017) 205137},
\href{http://arxiv.org/abs/1611.05049}{{\ttfamily arXiv:1611.05049
  [cond-mat.str-el]}}.

\bibitem{Giveon:2008zn}
A.~Giveon and D.~Kutasov, ``{Seiberg Duality in Chern-Simons Theory},''
  \href{http://dx.doi.org/10.1016/j.nuclphysb.2008.09.045}{{\em Nucl. Phys.}
  {\bfseries B812} (2009) 1--11},
\href{http://arxiv.org/abs/0808.0360}{{\ttfamily arXiv:0808.0360 [hep-th]}}.

\bibitem{Kapustin:2011gh}
A.~Kapustin, ``{Seiberg-like duality in three dimensions for orthogonal gauge
  groups},''
\href{http://arxiv.org/abs/1104.0466}{{\ttfamily arXiv:1104.0466 [hep-th]}}.

\bibitem{Aharony:2013dha}
O.~Aharony, S.~S. Razamat, N.~Seiberg, and B.~Willett, ``{3d dualities from 4d
  dualities},'' \href{http://dx.doi.org/10.1007/JHEP07(2013)149}{{\em JHEP}
  {\bfseries 07} (2013) 149},
\href{http://arxiv.org/abs/1305.3924}{{\ttfamily arXiv:1305.3924 [hep-th]}}.

\bibitem{Witten:1999ds}
E.~Witten, ``{Supersymmetric index of three-dimensional gauge theory},''
\href{http://arxiv.org/abs/hep-th/9903005}{{\ttfamily arXiv:hep-th/9903005
  [hep-th]}}.

\bibitem{lawson1989spin}
H.~Lawson and M.~Michelsohn, {\em Spin Geometry}.
\newblock Princeton mathematical series. Princeton University Press, 1989.
\newblock \url{https://books.google.com/books?id=3d9JkN8w3X8C}.

\bibitem{kirby_taylor_1991}
R.~Kirby and L.~Taylor, {\em Pin structures on low-dimensional manifolds},
  vol.~2 of {\em London Mathematical Society Lecture Note Series},
  \href{http://dx.doi.org/10.1017/CBO9780511629341.015}{pp.~177--242}.
\newblock Cambridge University Press, 1991.

\bibitem{Berg2001}
M.~Berg, C.~DeWitt-Morette, S.~Gwo, and E.~Kramer, ``{The Pin Groups in
  Physics: C, P and T},''
  \href{http://dx.doi.org/10.1142/S0129055X01000922}{{\em Reviews in
  Mathematical Physics} {\bfseries 13} no.~08, (2001) 953--1034}.
  \url{http://www.worldscientific.com/doi/abs/10.1142/S0129055X01000922}.

\bibitem{Varlamov2004}
V.~V. Varlamov, ``Universal coverings of orthogonal groups,''
  \href{http://dx.doi.org/10.1007/s00006-004-0006-4}{{\em Advances in Applied
  Clifford Algebras} {\bfseries 14} no.~1, (Mar, 2004) 81--168}.
  \url{https://doi.org/10.1007/s00006-004-0006-4}.

\bibitem{Moore2013pin}
G.~W. Moore, ``{Quantum Symmetries and Compatible Hamiltonians},''.
  \url{http://www.physics.rutgers.edu/~gmoore/695Fall2013/CHAPTER1-QUANTUMSYMMETRY-OCT5.pdf}.

\bibitem{steenrod1962cohomology}
N.~Steenrod, {\em Cohomology Operations}.
\newblock Annals of mathematics studies. Princeton University Press, 1962.
\newblock \url{https://books.google.com/books?id=4isPAAAAIAAJ}.

\bibitem{Hatcher:2001}
A.~Hatcher, {\em {Algebraic topology}}.
\newblock Cambridge Univ. Press, Cambridge, 2001.
\newblock
  \url{https://books.google.com/books/about/Algebraic_Topology.html?id=BjKs86kosqgC}.

\bibitem{GuLevin}
Z.-C. Gu and M.~Levin, ``Effect of interactions on two-dimensional fermionic
  symmetry-protected topological phases with ${Z}_{2}$ symmetry,''
  \href{http://dx.doi.org/10.1103/PhysRevB.89.201113}{{\em Phys. Rev. B}
  {\bfseries 89} (May, 2014) 201113}.
  \url{https://link.aps.org/doi/10.1103/PhysRevB.89.201113}.

\bibitem{Kapustin:2014dxa}
A.~Kapustin, R.~Thorngren, A.~Turzillo, and Z.~Wang, ``{Fermionic Symmetry
  Protected Topological Phases and Cobordisms},''
  \href{http://dx.doi.org/10.1007/JHEP12(2015)052}{{\em JHEP} {\bfseries 12}
  (2015) 052}, \href{http://arxiv.org/abs/1406.7329}{{\ttfamily arXiv:1406.7329
  [cond-mat.str-el]}}.
[JHEP12,052(2015)].

\bibitem{Wang:2016lix}
C.~Wang, C.-H. Lin, and Z.-C. Gu, ``{Interacting fermionic symmetry-protected
  topological phases in two dimensions},''
  \href{http://dx.doi.org/10.1103/PhysRevB.95.195147}{{\em Phys. Rev.}
  {\bfseries B95} no.~19, (2017) 195147},
\href{http://arxiv.org/abs/1610.08478}{{\ttfamily arXiv:1610.08478
  [cond-mat.str-el]}}.

\bibitem{Kapustin:2017jrc}
A.~Kapustin and R.~Thorngren, ``{Fermionic SPT phases in higher dimensions and
  bosonization},'' \href{http://dx.doi.org/10.1007/JHEP10(2017)080}{{\em JHEP}
  {\bfseries 10} (2017) 080},
\href{http://arxiv.org/abs/1701.08264}{{\ttfamily arXiv:1701.08264
  [cond-mat.str-el]}}.

\bibitem{Dijkgraaf:1989pz}
R.~Dijkgraaf and E.~Witten, ``{Topological Gauge Theories and Group
  Cohomology},''
\href{http://dx.doi.org/10.1007/BF02096988}{{\em Commun. Math. Phys.}
  {\bfseries 129} (1990) 393}.

\bibitem{Jenquin2005CS}
J.~A. Jenquin, ``{Classical Chern-Simons on manifolds with spin structure},''
  \href{http://arxiv.org/abs/math.DG/0504524}{{\ttfamily arXiv:math.DG/0504524
  [math.DG]}}.

\bibitem{Gaiotto:2017yup}
D.~Gaiotto, A.~Kapustin, Z.~Komargodski, and N.~Seiberg, ``{Theta, Time
  Reversal, and Temperature},''
  \href{http://dx.doi.org/10.1007/JHEP05(2017)091}{{\em JHEP} {\bfseries 05}
  (2017) 091},
\href{http://arxiv.org/abs/1703.00501}{{\ttfamily arXiv:1703.00501 [hep-th]}}.

\bibitem{Dijkgraaf:1989hb}
R.~Dijkgraaf, C.~Vafa, E.~P. Verlinde, and H.~L. Verlinde, ``{The Operator
  Algebra of Orbifold Models},''
\href{http://dx.doi.org/10.1007/BF01238812}{{\em Commun. Math. Phys.}
  {\bfseries 123} (1989) 485}.

\bibitem{Ginsparg:1988ui}
P.~H. Ginsparg, ``{Applied Conformal Field Theory},'' in {\em {Les Houches
  Summer School in Theoretical Physics: Fields, Strings, Critical Phenomena Les
  Houches, France, June 28-August 5, 1988}}, pp.~1--168.
\newblock 1988.
\newblock \href{http://arxiv.org/abs/hep-th/9108028}{{\ttfamily
  arXiv:hep-th/9108028 [hep-th]}}.
\newblock
\url{https://inspirehep.net/record/265020/files/arXiv:hep-th_9108028.pdf}.
\newblock

\bibitem{Ganor:2002ya}
O.~Ganor, M.~B. Halpern, C.~Helfgott, and N.~A. Obers, ``{The Outer automorphic
  WZW orbifolds on so(2n), including five triality orbifolds on so(8)},''
  \href{http://dx.doi.org/10.1088/1126-6708/2002/12/019}{{\em JHEP} {\bfseries
  12} (2002) 019},
\href{http://arxiv.org/abs/hep-th/0211003}{{\ttfamily arXiv:hep-th/0211003
  [hep-th]}}.

\bibitem{di1997conformal}
P.~Di~Francesco, P.~Mathieu, and D.~S{\'e}n{\'e}chal, {\em Conformal Field
  Theory}.
\newblock Graduate Texts in Contemporary Physics. Springer, 1997.
\newblock \url{https://books.google.com/books?id=keUrdME5rhIC}.

\bibitem{Witten:2015aba}
E.~Witten, ``{Fermion Path Integrals And Topological Phases},''
  \href{http://dx.doi.org/10.1103/RevModPhys.88.035001,
  10.1103/RevModPhys.88.35001}{{\em Rev. Mod. Phys.} {\bfseries 88} no.~3,
  (2016) 035001},
\href{http://arxiv.org/abs/1508.04715}{{\ttfamily arXiv:1508.04715
  [cond-mat.mes-hall]}}.

\bibitem{Atiyah:1975jf}
M.~F. Atiyah, V.~K. Patodi, and I.~M. Singer, ``{Spectral asymmetry and
  Riemannian Geometry 1},''
\href{http://dx.doi.org/10.1017/S0305004100049410}{{\em Math. Proc. Cambridge
  Phil. Soc.} {\bfseries 77} (1975) 43}.

\bibitem{Maldacena:2001ss}
J.~M. Maldacena, G.~W. Moore, and N.~Seiberg, ``{D-brane charges in five-brane
  backgrounds},'' \href{http://dx.doi.org/10.1088/1126-6708/2001/10/005}{{\em
  JHEP} {\bfseries 10} (2001) 005},
\href{http://arxiv.org/abs/hep-th/0108152}{{\ttfamily arXiv:hep-th/0108152
  [hep-th]}}.

\bibitem{Banks:2010zn}
T.~Banks and N.~Seiberg, ``{Symmetries and Strings in Field Theory and
  Gravity},'' \href{http://dx.doi.org/10.1103/PhysRevD.83.084019}{{\em Phys.
  Rev.} {\bfseries D83} (2011) 084019},
\href{http://arxiv.org/abs/1011.5120}{{\ttfamily arXiv:1011.5120 [hep-th]}}.

\bibitem{Gaiotto:2015zta}
D.~Gaiotto and A.~Kapustin, ``{Spin TQFTs and fermionic phases of matter},''
  \href{http://dx.doi.org/10.1142/S0217751X16450445}{{\em Int. J. Mod. Phys.}
  {\bfseries A31} no.~28n29, (2016) 1645044},
\href{http://arxiv.org/abs/1505.05856}{{\ttfamily arXiv:1505.05856
  [cond-mat.str-el]}}.

\bibitem{Bhardwaj:2016clt}
L.~Bhardwaj, D.~Gaiotto, and A.~Kapustin, ``{State sum constructions of
  spin-TFTs and string net constructions of fermionic phases of matter},''
  \href{http://dx.doi.org/10.1007/JHEP04(2017)096}{{\em JHEP} {\bfseries 04}
  (2017) 096},
\href{http://arxiv.org/abs/1605.01640}{{\ttfamily arXiv:1605.01640
  [cond-mat.str-el]}}.

\bibitem{Borisov:1997nc}
L.~Borisov, M.~B. Halpern, and C.~Schweigert, ``{Systematic approach to cyclic
  orbifolds},'' \href{http://dx.doi.org/10.1142/S0217751X98000044}{{\em Int. J.
  Mod. Phys.} {\bfseries A13} (1998) 125--168},
\href{http://arxiv.org/abs/hep-th/9701061}{{\ttfamily arXiv:hep-th/9701061
  [hep-th]}}.

\bibitem{MOORE1991362}
G.~Moore and N.~Read, ``Nonabelions in the fractional quantum hall effect,''
  \href{http://dx.doi.org/https://doi.org/10.1016/0550-3213(91)90407-O}{{\em
  Nuclear Physics B} {\bfseries 360} no.~2, (1991) 362 -- 396}.
  \url{http://www.sciencedirect.com/science/article/pii/055032139190407O}.

\bibitem{GomisKS:unpublish}
J.~Gomis, Z.~Komargodski, and N.~Seiberg. Unpublished.

\end{thebibliography}\endgroup

\end{document}